\documentclass[%
 reprint,
 nofootinbib,
 amsmath,amssymb,
 aps,
floatfix]{revtex4-2}

\usepackage{graphicx}
\usepackage{lipsum}
\usepackage{hyperref}
\hypersetup{
    colorlinks=true,
    allcolors = {blue}
}
\usepackage[dvipsnames]{xcolor}
\usepackage{dcolumn}
\usepackage{bm}
\usepackage[mathlines]{lineno}

\usepackage{silence}
\usepackage{ulem}
\usepackage{color}
\usepackage{float}
\usepackage{bbold}
\usepackage{adjustbox}
\usepackage{pgfplots}
\usepackage{dirtytalk}
\usepackage{multirow}
\usepackage{isotope}
\pgfplotsset{compat=1.18}

\newcommand*{\Lcorner}{%
    \mathchoice%
        {\mathrel{\makebox[7pt][c]{\rule{.4pt}{7.5pt}\rule{5pt}{.4pt}}}}%
        {\mathrel{\makebox[7pt][c]{\rule{.4pt}{7.5pt}\rule{5pt}{.4pt}}}}%
        {\mathrel{\makebox[5.5pt][c]{\rule{.4pt}{5.25pt}\rule{3.5pt}{.4pt}}}}%
        {\mathrel{\makebox[4pt][c]{\rule{.4pt}{3.75pt}\rule{2.5pt}{.4pt}}}}%
}
\usepackage{lineno}

\begin{document}
\preprint{APS/123-QED}

\title{New multinucleon knockout model in NuWro Monte Carlo generator}

\author{Hemant Prasad}
\email{contact author hemant.prasad@uwr.edu.pl}
\author{Jan T. Sobczyk}%
 \email{contact author jan.sobczyk@uwr.edu.pl}
\affiliation{%
 Institute of Theoretical Physics\\
 University of Wroc{\l}aw, Plac Maxa Borna 9, 50-204 Wroc{\l}aw, Poland
}%

\author{Artur M. Ankowski}
\affiliation{%
 Institute of Theoretical Physics\\
 University of Wroc{\l}aw, Plac Maxa Borna 9, 50-204 Wroc{\l}aw, Poland
}
\author{J. Luis Bonilla}
\affiliation{%
 Institute of Theoretical Physics\\
 University of Wroc{\l}aw, Plac Maxa Borna 9, 50-204 Wroc{\l}aw, Poland
}

\author{Rwik Dharmapal Banerjee}
\affiliation{%
 Institute of Theoretical Physics\\
 University of Wroc{\l}aw, Plac Maxa Borna 9, 50-204 Wroc{\l}aw, Poland
}

\author{Krzysztof M. Graczyk}
\affiliation{%
 Institute of Theoretical Physics\\
 University of Wroc{\l}aw, Plac Maxa Borna 9, 50-204 Wroc{\l}aw, Poland
}

\author{Beata E. Kowal}
\affiliation{%
 Institute of Theoretical Physics\\
 University of Wroc{\l}aw, Plac Maxa Borna 9, 50-204 Wroc{\l}aw, Poland
}



\begin{abstract}
We present the implementation and results of a new model for the 
n-particle n-hole (\textit{np-nh}) contribution in the NuWro event generator, grounded in the theoretical framework established by the Valencia group in 2020. For the \textit{2p2h} component, we introduce a novel nucleon sampling function with tunable parameters to approximate correlations in the momenta of outgoing nucleons. These parameters are calibrated by comparing our results to those of the Valencia model across a range of incoming neutrino energies. In addition, our model incorporates a distinct contribution from the \textit{3p3h} mechanism. We discuss the differences between the new NuWro implementation, the original Valencia model, and the previous NuWro version, focusing on the distribution of outgoing nucleon momenta. Finally, we assess the impact of the hadronic model on experimental analyses involving hadronic observables.

\end{abstract}

\maketitle


\section{\label{sec:Introduction}INTRODUCTION}

A new generation of long-baseline neutrino oscillation experiments, including Hyper-Kamiokande~\cite{Hyper-KamiokandeProto-:2015xww} and DUNE~\cite{DUNE:2016hlj}, requires highly accurate modeling of neutrino-nucleus interactions to meet performance demands~\cite{NuSTEC:2017hzk}. The probability for a neutrino of one flavor to oscillate into another depends on its energy. However, unlike in electron scattering experiments, the energy of a neutrino beam is not precisely known due to the broad spectrum with a significant tail, even in off-axis configurations. For example, in the T2K experiment with a 2.5-degree off-axis configuration, the energy spectrum has a peak at around 650 MeV with a spread of approximately 400 MeV. Therefore, in experiments, the only way to infer oscillation parameters is by comparing measured particle distributions with predictions from Monte Carlo (MC) event generators~\cite{GallagherHayato}. These predictions depend on the oscillation parameters governed by the PMNS mixing matrix~\cite{Gonzalez-GarciaYokoyama}. Early T2K analyses examined simple distributions such as muon energy and scattering angle, but current analyses, which include additional particles (e.g., pions in SuperKamiokande, neutrons detected thanks to gadolinium doping in SK~\cite{Super-Kamiokande:2021the}, and low-energy protons in DUNE’s liquid argon detector), increase sensitivity to oscillation parameters, provided that MC predictions are sufficiently accurate.

Over the last ~15 years, intensive experimental and theoretical studies have shown that modeling even the simplest neutrino-nucleon scattering process, such as the charge-current quasi-elastic (CCQE) process~\cite{LlewellynSmith:1971uhs},
\begin{equation}
\nu_l + \ n\rightarrow l^- + \ p
\label{eqn:ccqe}
\end{equation}
where \( l \) represents the neutrino flavor and \( n \), \( p \) represent a neutron and proton, respectively, if occurring on a nuclear target is challenging due to difficulties in determining the axial form factor \cite{ETM2021} and nuclear effects. Experimentally, CCQE is not a well-defined observable, as CCQE events can easily be misidentified with pion production followed by absorption in the target nucleus or multi-nucleon ejection. In water Cherenkov detectors, knocked-out protons are often undetected what makes a separation of multinucleon contribution impossible. To address this, experimentalists have proposed the CC\(0\pi\) signal, defined by the absence of pions in the final state~\cite{MiniBooNE:2010bsu}. Significant experimental efforts have focused on measuring CC\(0\pi\) cross sections for various neutrino beams and targets \cite{McFarland2020}. While most CC\(0\pi\) events are attributed to the CCQE mechanism on bound nucleons, understanding the size and characteristics of the multi-nucleon knock-out component is critical to prevent biases in interpreting measured particle distributions.

Since the multi-nucleon mechanism was first discussed in the context of neutrino interactions~\cite{Marteau:1999kt} it has gradually become a topic of extensive theoretical study. Until recently, theoretical models primarily provided predictions for the final-state lepton only~\cite{Martini:2009uj, Nieves:2011pp, Megias:2016fjk}.
Describing the final-state lepton can be efficiently done by tabulating five two-dimensional response functions, with energy and momentum transfer as inputs. This formalism is suitable for straightforward incorporation into MC generators. The outgoing nucleons arising from the multi-nucleon knock-out mechanism has previously been modeled using the factorization scheme proposed in~\cite{Sobczyk:2012ms}. Shortcomings of this approach must be compensated by large systematic uncertainties, which ultimately contribute to the overall uncertainty in measured oscillation parameters.


In this paper, we present a method to model the isospin and momentum of outgoing nucleons from the multi-nucleon knock-out mechanism (referred to here also as np-nh or meson exchange current, MEC) in MC generators. We focus on NuWro, the first MC generator to incorporate MEC dynamics, with various treatment options. However, our results are generalizable to other MC generators, including those directly used in neutrino oscillation studies, such as NEUT~\cite{Hayato:2009zz} and GENIE~\cite{andreopoulos2009genie}. Our model leverages theoretical advancements from the Valencia group~\cite{Sobczyk:2020dkn}, providing the first detailed results on the isospin composition and momentum distribution of knocked-out nucleons. For our study, we used the numerical code provided by the authors of~\cite{Sobczyk:2020dkn}\footnote{Note that this work does not incorporate the latest calculations from the Valencia group~\cite{Sobczyk:2024ecl}}.\\

Our paper is organized as follows: in Sect.~\ref{sec:NuWro}, we present the general features of the NuWro Monte Carlo generator. Sect.~\ref{sec:MEC_MC} 
introduces a MEC model developed in 2020 by the Valencia group and describes a scheme for integrating this model into MC frameworks. 
Sect.~\ref{sec:results} presents results from the new MC implementation, including comparisons to the original theoretical model and experimental results from the MINER\ensuremath{\nu}A experiment. We conclude with some remarks in Sect.~\ref{sec:outlook}. Appendices A, B contain technical details of our study.

\section{\label{sec:NuWro} NUWRO MONTE CARLO GENERATOR}
\subsection{General structure}
NuWro is an MC neutrino event generator developed at the University of Wrocław since 2004~\cite{Juszczak:2005zs, Golan:2012wx}. It covers the energy range from $\sim 100$~MeV to $\sim 100$~GeV. All basic dynamics for the neutrino-nucleon/nucleus interaction process are included in NuWro for both charge current (CC) and neutral current (NC) reactions. For neutrino-nucleon scattering, these are:
\begin{itemize}
    \item Charge current quasi elastic  (CCQE) scattering, (see Eq.\eqref{eqn:ccqe}) and its NC counterpart
    \item Resonant production (RES) containing mostly single pion production, typically through $\Delta (1232)$ resonance excitation and its decay into nucleon-pion pair
    \begin{equation}
        \nu_l + N\ \rightarrow\ l^- + \Delta\ \rightarrow\ l^- +  N' + \pi
        \label{eqn:res}
    \end{equation}
    In NuWro RES labels all the inelastic reactions for which invariant hadronic mass $W$ satisfies $W\leq 1.6$~GeV
    \item DIS: inelastic  reactions with $W>1.6$~GeV
    \item HYP: quasi-elastic antineutrino hyperon production
    \begin{equation}
        \bar\nu_l + N \rightarrow l^+ + \Lambda/\Sigma
        \label{eqn:hyp}
    \end{equation}
\end{itemize}
Neutrino-nucleus scattering is described in the impulse approximation as a two-step process where primary neutrino-bound nucleon reaction is followed by final state interactions (FSI), hadronic re-interactions inside the nucleus. For nuclear target reactions two  new interaction modes are available:
\begin{itemize}
    \item COH: coherent pion production on nucleus $A$
    \begin{equation}
        \nu_l\ + A \rightarrow l^- + \pi^+ +  A
        \label{eqn:coh}
    \end{equation}
    and its NC counterpart.
    \item MEC: scattering on nucleon pairs correlated by exchange of virtual pion or $\rho$ meson.
\end{itemize}
NuWro includes also 
\begin{itemize}
    \item EL: contribution from neutrino-electron scattering.
\end{itemize}

In NuWro there is a variety of options to describe target nucleon immersed in the nuclear matter. These are global Fermi gas, local Fermi gas, hole spectral function~\cite{Benhar:1999bg}, effective spectral function~\cite{Ankowski:2005wi}, and effective momentum dependent potential~\cite{Juszczak:2005wk}.

NuWro's FSI is described by a custom-made intranuclear cascade model \cite{Yariv2007}. The hadrons transported inside nuclei are nucleons, pions, and hyperons. The basic scheme of the NuWro FSI model follows the seminal papers by Metropolis et al \cite{Metropolis:1958wvo, Metropolis:1958sb}. However, many quantum effects are included~\cite{Pandharipande:1992zz, Salcedo:1987md} making microscopic hadron-nucleon interactions realistic. The performance of the NuWro FSI compared to other MC event generators was investigated in detail in Ref.~\cite{Dytman:2021ohr}.\\ 

One can think of NuWro
Monte Carlo generator as a code that calculates the total cross section for neutrino scattering using the MC algorithm. The integration is done over the available phase space i.e. over a space of all the possible momenta (in practically all the situations a summation over spins is assumed to be previously performed) of final state particles. Every point in the phase space corresponds to a potentially observed configuration of the final state resulting from neutrino interaction. This allows us to assign a certain number, called {\it weight}, to every final state.  The weights are defined so that the average weight is equal to the total cross section:
\begin{equation}
    \sigma_{total} = \frac{1}{N}\sum_{j}^{N}w_{j}.
    \label{eqn:average_weight}
\end{equation}
where $w_j$ denotes the weights of individual events.
Computation of weights requires a knowledge of multidimensional differential cross sections. This information is typically not available and the calculation of weights $w_j$ is done applying various approximations. Often, the only information provided by the theoretical model is lepton inclusive differential cross section $\displaystyle d^2\sigma\over\displaystyle d\Omega(\hat{\mathbf{k'}})dE'_{l} $ where $\Omega(\hat{\mathbf{k'}})$ and $E'_{l}$ refer to final state lepton spherical angle and energy. In this situation, the cross section is calculated by sampling over the  leptonic phase space $V$ and computing weights of individual events as
\begin{equation}
    \label{eqn:lepton_inclusive_diff_xsec_nuwro}
    w = \frac{d^2\sigma}{d\Omega(\hat{\mathbf{k'}})dE'_{l}}
\end{equation}
It is then necessary to develop tools to {\it complete} events for given values of $\Omega(\hat{\mathbf{k'}})$ and $E'_{l}$ (or alternatively, the values of energy and momentum transfer i.e. $\omega$, $|\mathbf{q}|$) by assigning outgoing nucleons ispospin and momentum. At the very end, FSI effects are included as well by propagating hadrons through the nucleus under the assumption that the event's weight remains unchanged. From a little abstract perspective, FSI can be thought of as a unitary transformation in the space of hadronic final states.

\subsection{\label{subsec:Methodology}MEC models in NuWro}
In MEC reactions typically more than one nucleon is knocked out from the nucleus. In the language of nuclear theory, one often speaks about $n$-particles and $n$-holes (\textit{np-nh}) processes. MEC scattering is especially relevant in the kinematic region between quasi-elastic and $\Delta(1232)$ excitation peaks. While many theoretical models of the MEC dynamics were already present from the era of electron-nucleus scattering experiments, it was the excess of CC0$\pi$ events observed by MiniBooNE collaboration~\cite{MiniBooNE:2010bsu} that motivated the inclusion of MEC dynamics in modeling neutrino-nucleus interactions as well. Since MiniBooNE only measured the outgoing leptons and charged mesons, and did not track the final state nucleons, at that time there was no need to develop precise models for outgoing nucleons.

Monte Carlo generators describe MEC events in a three-step procedure:
\begin{enumerate}
    \item{ Generation of kinematics for outgoing lepton using information about neutrino energy and nuclear response functions. This part is referred to as the \textit{inclusive part}} of the model.
    
    \item {Generation of nucleons outgoing from primary interaction by modeling their isospin and momenta This part is referred to as the \textit{hadronic part}  of the model.}
    \item {Processing nucleons obtained in step (2) through the FSI module.}
\end{enumerate}

\subsubsection{\label{subsubsec:leptonic-part}Inclusive part}
 For the inclusive reaction on a nucleus $A$
\begin{eqnarray}
    \nu_{l}(k) + A \rightarrow l^{-}(k') + X
    \label{eqn: inclusive_nuclear_reaction}
\end{eqnarray}
with the remnant hadronic system denoted as $X$, the double differential cross section, concerning the outgoing lepton kinematical variables is given by a general expression
\begin{eqnarray}
    \label{eqn:inclusive-cross-section}
    \frac{d^2\sigma_{\nu l}}{d\Omega(\hat{\mathbf{k'}})dE_{l}'} = \frac{ G_F^2\cos^2\theta_c }{4\pi^2}\frac{|\mathbf{k}'|}{|\mathbf{k}|}  L_{\alpha\beta}W^{\alpha\beta}
\end{eqnarray}
with $\mathbf{k}$ and $\mathbf{k'}$ the incoming and outgoing lepton momenta respectively in the LAB frame, $\Omega(\hat{\mathbf{k'}})$ is the spherical angle of outgoing lepton, $E'_{l} = (\mathbf{k'}^2 + m_{l}^2)^{1/2}$ and $m_{l}$ the energy and mass of the outgoing lepton, $G_F$ is the Fermi constant, $\theta_c$ is the Cabibbo angle, and $L_{\alpha\beta}$ and $W^{\alpha\beta}$ are the leptonic and hadronic tensors, respectively. After some algebra, the expression for the cross-section in Eq.\eqref{eqn:inclusive-cross-section} can be written down as 
\begin{widetext}
\begin{eqnarray}
    \label{eqn:nuwro_double_diff_expression}
    \frac{d^2\sigma}{dE'_{l}\: d\cos\theta_{l}}=&&\frac{2G_F^2 \cos^2\theta_c  E'_{l}|\mathbf{k'}|}{\pi}\Bigg\{2W_1\sin^2\frac{\theta_{l}}{2}+W_2\cos^2\frac{\theta_{l}}{2} \pm W_3(E+ E'_{l})\sin^2\frac{\theta_{l}}{2}\nonumber\\
    && + \frac{m_{\mu}^2}{(E'_{l} + |\mathbf{k'}|)E'_{l}}\cdot\Bigg[ W_{1}\cos\theta_{l}  - \frac{W_2}{2}\cos\theta_{l} \pm \frac{W_3}{2}\left(E'_{l}(1-\cos\theta_{l}) + |\mathbf{k'}| - E\cos\theta_{l} \right) \nonumber\\
    && + \frac{W_4}{2}(m_{\mu}^2\cos\theta_{l}+ 2E'_{l}(E'_{l}+|\mathbf{k'}|)\sin^2\theta_{l}) - \frac{W_5}{5}(E'_{l}+|\mathbf{k'}|)\Bigg]\Bigg\}
\end{eqnarray}
\end{widetext} 
where $\theta_{\mathbf{k'}}$ is the scattering angle of the outgoing lepton. The five response functions, \( W_1, W_2, W_3, W_4, W_5 \), are functions of the energy and momentum transfer \((\omega, |\mathbf{q}|)\), and are linear combinations of the independent components of the hadronic tensor: \( W^{00}, W^{03}, W^{11}, W^{12}, W^{33} \). The response function $W_3$ in the $W_j$ basis flips the sign from positive for neutrinos to negative for antineutrino interactions.

In the NuWro version 21.09 several inclusive models for MEC are available. These include Valencia model~\cite{Nieves:2011pp}, Marteau model~\cite{Marteau:1999kt}, SuSav2 model~\cite{Megias:2016fjk} and transverse enhancement (TE) model~\cite{Bodek:2011ps}. An important feature of the TE model is that it is can be also used to describe neutral current MEC reaction which was essential in the studies in Refs.~\cite{Golan:2013jtj, KamLAND:2022ptk}. Inclusive models differ in the kinematic region they cover. The original Valencia model was supplemented in Ref.~\cite{Gran:2013kda} by a condition that the magnitude of momentum transfer $|\mathbf{q}|<1.2$~GeV/c. SuSav2 MEC model covers a region $|\mathbf{q}|<2$~GeV/c. Implementations of the Valencia and SuSav2 models are done by tabularization of the five response functions $W_j$. This approach is very effective because the same set of tables allows for a computation of inclusive cross section for arbitrary (anti)neutrino flavor and also at arbitrary energy. For any arbitrary pair $(\omega, |\mathbf{q}|)$ which lies within a valid leptonic kinematic region, NuWro then uses the bi-linear interpolation technique to evaluate the double differential cross section at that specific point. TE model is implemented with analytic formulas. Marteau model is implemented with tables in the region of energy and momentum transfer $w<3$~GeV and $|\mathbf{q}|<3.6$~GeV, and with analytic formulas outside this region. To generate lepton kinematics it is sufficient to know an absolute value of momentum transfer $|\mathbf{q}|$. It is always assumed that the interaction plane defined by neutrino and final lepton momentum vectors is arbitrary and as a consequence, is selected at random. Once the choice is done, the direction of momentum transfer ($\mathbf{\hat{q}}$) is known, which then will be used in the computations of the hadronic part of the model discussed below.\\

\subsubsection{\label{subsubsec:hadronic-part}Hadronic part}

The individual components $W^{ij}$ describe inclusive electroweak nuclear responses on a pair of correlated nucleons.  $W^{ij}$ do not contain information regarding correlations among the outgoing nucleon momenta and their corresponding isospin. However, in MC generators one has to explicitly model the kinematics for the outgoing nucleon states. When theoretical model information is missing, the challenge is to satisfy simultaneous constraints from energy and momentum conservation. One option to proceed is to use the hadronic model proposed in Ref.~\cite{Sobczyk:2012ms} with the basic idea that energy and momentum are automatically conserved if one assigns momentum (and energy) of outgoing nucleons in the hadronic center-of-mass frame. This approach is universal and can be applied independently of which inclusive model is used. Due to the feasibility and simplicity of the algorithm, the model~\cite{Sobczyk:2012ms} is adopted in all the MC generators as a standard approach to modeling outgoing nucleons.

\subsubsection{\label{subsubsec:FSI}FSI}
The hadronic model describes nucleons after primary interaction but before FSI.
The nucleons arising from the primary interactions must be propagated through the nucleus. For carbon, there is $\sim 40\%$ probability (see e.g.~\cite{garrow2001nuclear}) that each one of them interacts at least once. As a result, a very {\it topology} of the event is changed with more particles being knocked out (including pions). Also ejected nucleons have typically smaller kinetic energy.

\section{\label{sec:MEC_MC} New Monte Carlo MEC model}
\subsection{\label{sec:Valencia_model} The 2020 Valencia Model}
The new features of the 2020 Valencia  model~\cite{Sobczyk:2020dkn} include separation of \textit{2p2h} and \textit{3p3h} contributions to the inclusive cross-section and access to information about isospin and momenta of final state nucleons. For clarity, by 'final state nucleon pair' or 'final state three-nucleon configuration,' in the following text we refer to outgoing nucleon pair or a configuration of three outgoing nucleons before FSI, respectively. \\

Thanks to the work of the authors of Ref.~\cite{Sobczyk:2020dkn}, the numerical approximations applied to the imaginary part of the $\Delta$ self-energy have been eliminated for the \textit{2p2h}  contribution compared to the previous version of the model. The \textit{2p2h} part can now be decomposed into components based on the isospin of the final nucleon pairs: $pp$, $np$, and $pn$, representing two protons, a neutron-proton pair, and a proton-neutron pair, respectively (the last two are kinematically distinct). However, for the \textit{3p3h} mechanism, the numerical approximation remains, and no information is provided on the isospin decomposition of the total cross-section across different configurations of three nucleons in the final state. In all cases, the 2020 Valencia model imposes a constraint on the momentum transfer $|\mathbf{q}| < 1.2$ GeV, consistent with the old model.\\

The code provided by the authors of Ref. \cite{Sobczyk:2020dkn} computes the individual components of the hadronic tensor $W^{\mu\nu}$, namely ${W^{00}, W^{03}, W^{11}, W^{12}, W^{33}}$, for the entire \textit{3p3h} mechanism. For the \textit{2p2h} mechanism three sets each consisting of five individual components of tensor are computed separately based on the nucleon pairs in the final state i.e $pp$, $np$, and $pn$. The four components correspond to distinct final states and can be consistently summed up as cross sections.\\

\subsection{\label{sec:New_MEC_model}  New scheme for MEC implementation in NuWro} 

As explained in Sect.\ref{subsubsec:leptonic-part}, the inclusive contribution in NuWro implementation of the new MEC models consists of four sub-parts. Each sub-part is first used to calculate the corresponding contributions to the overall event's weight. The total weight is obtained by summing all four contributions: 
\begin{equation}
    w = \underbrace{w_{\text{\textit{2p2h}}}}_{ w_{pp}+w_{np}+w_{pn} }  +\ \   w_{\text{\textit{3p3h}}} 
\end{equation}

The code for the \textit{hadronic part} first runs a sub-routine called \textit{isospin model} which decides the \textit{event topology} whether the event is of the type $pp$, $pn$ or $np$ within \textit{2p2h}, or \textit{3p3h}. For given values of $(\omega, |\mathbf{q}|)$ probabilities to choose topologies \textit{pp}, \textit{np}, \textit{pn}, \textit{3p3h} are proportional to $w_{pp}$, $w_{np}$, $w_{pn}$ and $w_{3p3h}$ and decision is taken with the Monte Carlo algorithm. For the MEC interaction of antineutrinos, we proceed in the same way and the only difference is that individual contributions to the overall weight are different due to different sign in Eq.~\ref{eqn:inclusive-cross-section}. For \textit{3p3h}, further decisions must be taken about the isospin of the final state. With no information in this respect, it is decided based on a combinatorial model by calculating a number of possible sets of three nucleons in the final state with different isospin i.e $ppp, ppn, pnn$ in the case of $\nu$-interaction and $nnn, nnp, nnn$ in the case of $\bar{\nu}$-interaction. Specifically, for neutrino interaction
\begin{eqnarray}
    ppp\ \leftrightarrow \quad & N_{ppp} = \binom Z3 &\nonumber \\
    ppn\  \leftrightarrow \quad & N_{ppn} = \binom{Z}{2} \binom{A-Z}{1} & \\
    pnn\  \leftrightarrow \quad & N_{pnn} = \binom{Z}{1} \binom{A-Z}{2} &\nonumber
\end{eqnarray}
where $Z$ and $A$ are the atomic number and atomic mass of the nucleus respectively. Probabilities to choose a final state to be $ppp$, $ppn$ or $pnn$ are set to be proportional to $N_{ppp}$, $N_{ppn}$ and $N_{pnn}$ and the decision is taken with the Monte Carlo algorithm.

Once we decide the isospin of the nucleons in the final state, we move on to model momenta of the outgoing nucleons. We divide our next discussion into two sub-sections, namely \textit{2p2h} and \textit{3p3h}.

\subsubsection{\label{sec:2p2h}2p2h}

The algorithm is defined as follows.\\

For given values of energy $\omega$ and momentum transfer $\mathbf{q}$: 
\begin{enumerate}
    \item{Select interaction point using nuclear density profile as a probability density function.}
    
    \item{Calculate Fermi momentum at the interaction point in the local Fermi gas paradigm as\begin{equation}p_{F} = (3\pi^2\rho(r))^{1/3}\end{equation}}

    \item {Set the magnitude of the momenta of two nucleons in the initial state from the quadratic probability density function     
    \begin{eqnarray*}f(p) = \frac{3}{p_F^3}\Theta (p_F-p) p^2\end{eqnarray*}}

    \item {Set the direction of each initial nucleon uniformly chosen within the unit sphere}

    \item{Form a hadronic system with 4-momentum $p_{\text{sys}}$ equal to the sum of 4-momentum of initial nucleons and 4-momentum transfer $q^\mu=(\omega,\mathbf{q})$}
    
    \item {Move to the hadronic center-of-mass frame}
    \begin{enumerate}
    
        \item {Check whether the hadronic condition is met for a possibility of ejection of two nucleons: 
        \begin{equation}
            \label{eqn:newMEC_hadronic_conditon1}
            p^2_{\text{sys}} \geq (m_{3} + m_{4})^2
        \end{equation}
        If it's not the case, steps (3-5) are repeated.
        Here, $m_{3}$ and $m_{4}$ are the rest masses of the final-state nucleons. At this stage, both values are determined.}
        
        \item {Produce two outgoing nucleons in back-to-back configuration and set the isospins as already decided by the \textit{isospin model} for \textit{2p2h} mechanism }
    
        \item {Divide the total energy of the hadronic system to the outgoing nucleons
        \begin{eqnarray*}
            E_{3}^* = \frac{ p_{\text{sys}}^2 + m_3^2 - m_4^2}{2 \sqrt{p_{\text{sys}}^2  }}\\
E_{4}^* = \frac{ p_{\text{sys}}^2 + m_4^2 - m_3^2}{2 \sqrt{p_{\text{sys}}^2  }}
        \end{eqnarray*} where $E_3^*$ and $E_4^*$ are energies of outgoing nucleons in the hadronic centre-of-mass frame satisfying $(E_3^*+E_4^*)^2=p_{\text{sys}}^2$}

         \item{Calculate the allowed range on the scattering angle $\theta^*$ of the outgoing nucleon pair with respect to the direction of boost  for the \textit{backward} nucleon (the condition originates from he Pauli blocking)

         \begin{itemize}
            \item{Nucleon associated with $\cos\theta^* < 0$ is denoted as \textit{backward} nucleon because $\theta^* \in [\frac{\pi}{2}, \frac{3\pi}{2}]$ and so the outgoing nucleon will have anti-parallel components defined w.r.t the direction of boost. Its back-to-back partner associated with the angle $\theta^*+\pi$ has $\cos(\theta^*+\pi) >0$ and is referred as \textit{forward} nucleon.}
            \item {After boosting the nucleon pair back to the lab frame, the \textit{forward} nucleon becomes more energetic because it had the parallel component of its momentum aligned with the direction of the boost from the lab frame to the center-of-mass frame. Energy of the backward nucleon must satisfy:}
        \end{itemize}

        \begin{eqnarray}
            \label{eqn:Pauli_blocking_condition}
            E^{\text{lab}}_{b} && > m_{b} + E_{F} \implies \cos\theta^*\in [-\kappa, \kappa] \nonumber \\
            &&\left\{ \kappa: \kappa = \min\left[1,  \frac{\gamma E_b^* - m_b - E_{F}}{\beta\gamma p_b^*}\right] \right\}
        \end{eqnarray}
        here $E^{\text{lab}}_b$ is the energy of \textit{backward} nucleon in the lab frame, $E_F$ is the Fermi kinetic energy, $\beta,\gamma$ are the boost parameters, $E^*_b$ is the energy and $p^*_b$ is the magnitude of the momentum of the \textit{backward} nucleon in the centre-of-mass frame. A derivation of Eq.~\eqref{eqn:Pauli_blocking_condition} is given in the Appendix~\ref{sec:derivation}. } 
      
        \item {Sample the scattering angle of the outgoing nucleon $\cos\theta^*$ from the probability distribution function $f(\cos\theta^*)$ described in Appendix~\ref{sec:sampling} with $\cos\theta^*\in [-\kappa, \kappa]$. The azimuthal angle is sampled uniformly in the range $[0, 2\pi]$. }

    \end{enumerate}
    \item {Boost back to the lab frame. At this stage outgoing nucleons are assumed to be on-shell. Later on, the NuWro FSI module accounts for the fact that they are immersed in a nuclear potential.}
\end{enumerate}
The method we use to assign momenta to final state nucleons is an approximation of the correlations defined by the 2020 Valencia model. The choice of the approximation is motivated by the numerical efficiency of the theoretical code and the limitations of the FSI model. Exact computations require a lot of computer time. From the physics perspective, an exact model is not essential, as nucleons produced in MEC interactions propagate within the nucleus and undergo FSI, which carries significant uncertainty \cite{Niewczas:2019fro}. The hadronic model only needs to be precise enough to exceed the precision of the FSI. This will be discussed in Sect.~\ref{sec:results}.

\subsubsection{\label{sec:3p3h} 3p3h}

If a particular isospin configuration within \textit{3p3h} mechanism is selected, then modeling of final state nucleons is done using a three-body phase space model for the kinematics of the outgoing nucleons. At the initial steps, we proceed in exactly the same way as in Sect~\ref{sec:2p2h} (steps (1-5) differ only in the sense that three nucleons are sampled in the initial state). The next steps are as follows:\\

\begin{enumerate}
\setcounter{enumi}{5}
    \item {Move to the center-of-mass frame of the hadronic system}
   \begin{enumerate}
        \item {Check whether the hadronic condition are met:
        \begin{equation}
            \label{eqn:hadronic_condition_3p3h}
            p_{\text{sys}}^2 \geq (m_4+m_5+m_6)^2.
        \end{equation}
        If it is not a case, steps (3-5) are repeated. Here $m_4$, $m_5$ and $m_6$ are the rest masses of the final-state nucleons. At this stage, all three values are determined.}
    
        \item{Assign energies and momenta of three outgoing nucleons using the three body phase-space model.}
    \end{enumerate}
    \item {Boost back all the nucleons to the lab frame.}
    \item {Check whether the nucleons satisfy the Pauli blocking condition. If not, the steps 6b and 7 are repeated. At this stage, outgoing nucleons are assumed to be on-shell. Later on, NuWro FSI module accounts for the fact that they are immersed in a nuclear potential.}
\end{enumerate}

 \subsection{Choice of nucleon sampling function for \textit{2p-2h}}
When modeling the kinematics of outgoing nucleons in the \textit{2p2h} mechanism within the \textit{hadronic part} of the new MC MEC model, NuWro assigns distinct sets of parameters \((P, l)\) for each of the three final state nucleon pairs: \(pp\), \(np\), and \(pn\). In NuWro we assume that the forward nucleon in the $np\: (pn)$ contribution is neutron (proton). We limit the parameter space for all nucleon pairs to $l \in (1,2,...,10)$. The parameter sets for the $pp$, $np$, and $pn$ outgoing pairs are represented as $(P, l)_{pp}$, $(P, l)_{np}$, and $(P, l)_{pn}$.

We found best-fit values of the parameters $(P, l)_{pp}$, $(P, l)_{np}$, and $(P, l)_{pn}$ for two parameter nucleon sampling function (see Appendix~\ref{subsec:two_parameter_nucleon_sampling_function} for details). The optimization was done by demanding that two-dimensional momentum distributions of outgoing nucleons for \textit{2p2h} in NuWro are made as similar as possible to the corresponding distributions in the 2020 Valencia model. We refer to this two-dimensional distribution of momenta of outgoing nucleons as the either proton-proton or proton-neutron \textit{nucleon phase space} of the outgoing nucleon pair. The 2020 Valencia model strongly prefers a more energetic proton in the $pn$ pair and a more energetic neutron in the $np$ pair. However, the $pn$ and $np$ outgoing nucleon pairs must be analyzed together, as a small fraction of events do not follow this pattern.\\

We achieve similarity between the nucleon phase spaces of NuWro and those of the 2020 Valencia model by adjusting the $(P, l)_{pp}$ parameter set in the $pp$ nucleon phase space, and $(P, l)_{np}$ and $(P, l)_{pn}$ simultaneously in the $np+pn$ nucleon phase space, separately for selected neutrino energy values. To save computational time for each neutrino energy value, we use the reweighting technique (a general description is given in Ref.~\cite{Pickering_2017}) to generate the $pp$ and $np+pn$ nucleon phase spaces for different values of $(P, l)$ in their respective parameter spaces.\\

Firstly, we generate sufficiently large samples of $pp$, $np$ and $pn$ events with default parameter values $(P, l)_{\{pp, np, pn\}} \equiv (0, 1)$. At this stage, any value of $\cos\theta^*$ in the allowed region is selected with equal probability. Using the reweighting technique, the probability of occurrence of an event characterized by values $(\cos\theta^*, \kappa)$ can then be modified to match the new configuration $(P', l')$.\\

We chose $E_{\nu} \in \{0.5, 0.7, 1.0, 2.0, 5.0\}$ GeV for our analysis. Our choice of energy values is guided by the fact that the distribution of energy transfer in the new Valencia model has two peaks: one at $\omega \simeq 0.15$ GeV and another at $\omega \simeq 0.45$ GeV, arising from two different mechanisms that dominate the total cross-section. These two peaks contribute to the cross-section with different strengths depending on the value of the neutrino energy. We produced the $pp$ and $np+pn$ nucleon phase spaces for each $E_{\nu}$. For the analysis, we select only those bins from the 2020 Valencia model where NuWro has a non-zero contribution. To measure the differences between the nucleon phase spaces from NuWro and the new Valencia model for any given values of $(P, l)_{pp}$, $(P, l)_{np}$, and $(P, l)_{pn}$, we define a quantity denoted as $\tilde{\chi}^2$ for each neutrino energy separately.

\begin{eqnarray}
    \label{eqn:chisquare}
    \tilde{\chi}^2 = {\displaystyle 1\over\displaystyle 2N_{\text{bins}}}\sum_{i,j}^{N_{\text{bins}}} \frac{(N_{ij}^{\text{NuWro}} - N_{ij}^{\text{Valencia}})^2}
    {(N_{ij}^{\text{NuWro}} + N_{ij}^{\text{Valencia}})}
\end{eqnarray}
where $N_{ij}^{\text{NuWro}}$ is the number of events predicted by NuWro in the bin $(i,j)$ within the nucleon phase space, and $N_{ij}^{\text{Valencia}}$ is the number of events produced in the bin $(i,j)$ within the 2020 Valencia model. The overall number of events from both models are the same and we are comparing the shapes of two distributions.\\

To find the optimal values $(\hat{P}, \hat{l})_{pp}$, $(\hat{P}, \hat{l})_{np}$, and $(\hat{P}, \hat{l})_{pn}$, we minimize the sum of $\tilde{\chi}^2_{pp}$ and $\tilde{\chi}^2_{np+pn}$ from different neutrino energies over the available parameter space as follows: 

\begin{eqnarray}
    \label{eqn:sum_chisquare1}
    &&\hat{\tilde{\chi}}^2_{pp} = 
    \min\left\{\sum_{E_{\nu}} \tilde{\chi}^{2}_{pp} (P, l)_{pp}\right\}  \\
    \label{eqn:sum_chisquare2}
    &&\hat{\tilde{\chi}}^2_{np+pn} = \min\left\{\sum_{E_{\nu}} \tilde{\chi}^{2}_{np+pn}\left( (P, l)_{np}, (P, l)_{pn} \right) \right\} 
\end{eqnarray}

The best-fit values obtained are:
\begin{equation}
\label{eqn:global_configuration_pp}
    (\hat{P},\hat{l})_{pp}^{\text{Global-fit}} \equiv (0.77,4)
\end{equation} 
and
\begin{equation}
    ((\hat{P},\hat{l})_{np}, (\hat{P},\hat{l})_{pn})^{\text{Global-fit}}\equiv((0.7, 3), (0.8,4))
\end{equation}

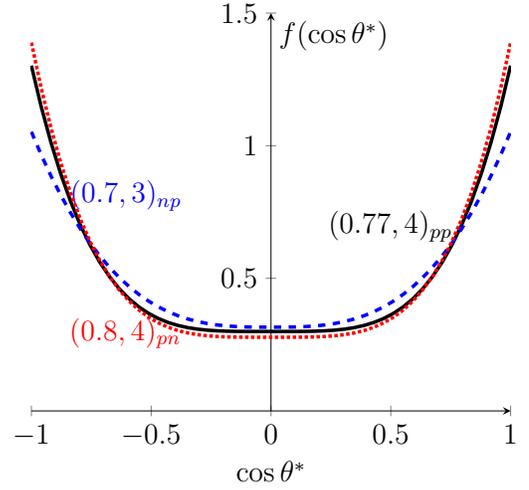
\begin{figure}[htb]
\centering
    \begin{adjustbox}{width=0.8\linewidth}
    \begin{tikzpicture}
    \tikzstyle{every node}=[font=\large]
        \begin{axis}[xmin=-1, xmax=1, ymin=0, ymax=1.5,
                    axis x line=bottom,
                    axis y line=middle,
                    xlabel=$\cos\theta^*$,
                    ylabel=$f(\cos\theta^*)$]
            \addplot[line width=0.5mm, color=black, samples=50, domain=-1:1]{(0.23 + 0.77*(abs(x))^4)/0.768}
            node[left, pos=0.8]{$(0.77,4)_{pp}$};
            \addplot[line width=0.5mm, color=blue, dashed, samples=50, domain=-1:1]{(0.3 + 0.7*(abs(x))^3)/0.95}
            node[right, pos=0.1]{$(0.7,3)_{np}$};
            \addplot[line width=0.5mm, color=red, densely dotted, samples=50, domain=-1:1]{(0.2 + 0.8*(abs(x))^4)/0.72}
            node[left, pos=0.4]{$(0.8, 4)_{pn}$};
        \end{axis}
    \end{tikzpicture}
    \end{adjustbox}
    \caption{Nucleon-sampling function (normalized to unit area) as represented in Eq.~\eqref{eqn:nucleon-sampling-function}, obtained for the optimized values of the parameters $(\hat{P}, \hat{l})$ at $\kappa=1$. The \textbf{black-solid} curve represents $f(\cos\theta^*)$ for the outgoing $pp$ nucleon pair. The \textbf{blue-dashed} and \textbf{red-dotted} curves represent $f(\cos\theta^*)$ for the outgoing $np$ and $pn$ pairs, respectively.}
    \label{fig:nucleon-sampling-function2}
\end{figure}

Fig.~\ref{fig:nucleon-sampling-function2} shows the best-fit nucleon sampling functions for $pp$, $pn$, and $np$. They are similar to each other.
The deviation of the values $(P, l)$ from $(0, 1)$, corresponding to the NuWro model of Ref.~\cite{Sobczyk:2012ms}, to positive values of $P$ for all three cases indicate that the distribution in nucleon phase space is strongly affected compared to the uniform one. An important assumption in our approach is that there exist a universal nucleon sampling functions for the $pp$, $pn$, and $np$ contributions. In general, the sampling function depends on the values of $(\omega, |\mathbf{q}|)$. Motivated by this observation, we attempted to obtain a more sophisticated sampling function by splitting the $(\omega, |\mathbf{q}|)$ domain into two sub-regions, each containing one peak in the double-differential cross-section. We chose the following condition for a boundary of two regions:

\begin{equation}
\omega [\text{GeV}] = \frac{0.65}{1.2} \cdot |\mathbf{q}| [\text{GeV}/c].
\label{eqn:split-line}
\end{equation}

Events within the nucleon phase space corresponding to the peak at lower energy transfer (dominated by the $N\Delta$ mechanism) were assigned one set of parameters, $(P, l)_{\text{reg. I}}$. Alternatively, events from higher energy transfer (dominated by the $\Delta\Delta$ mechanism) were assigned another set of parameters, denoted as $(P, l)_{\text{reg. II}}$. In this way, we had a total of four free parameters to fit the $pp$ nucleon phase space. We simultaneously varied both parameter sets $\tilde{\chi}^2_{pp}$ however, no significant improvement was found for the global $\tilde{\chi}^2_{pp}$.

\section{\label{sec:results}PERFORMANCE OF THE NEW NuWro MODEL}
We present results for the \textit{inclusive part} and \textit{hadronic model} of the NuWro implementation of the 2020 Valencia model through the new MC MEC model. First, we compare NuWro's performance in generating the total MEC cross section, which is governed by the \textit{inclusive part}. Next, we benchmark the new \textit{hadronic model} by comparing its predictions with those from the 2020 Valencia model. We show also results of the 2020 Valencia inclusive model combined with the old hadronic model. We also show the distribution of maximal proton momentum before and after applying the FSI model. Finally, we benchmark NuWro's performance using experimental data, evaluating the importance of incorporating information on correlations in the momenta of final-state nucleons.

 \subsection{\label{sec:Total-cross section} Total cross section}
\begin{figure}[htbp!]
    \centering
    \includegraphics[width=1.1\linewidth]{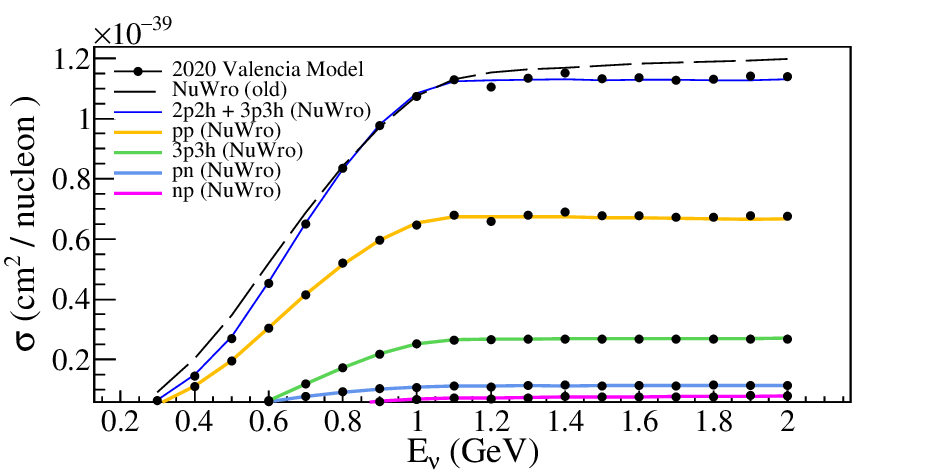}
    \caption{MEC cross section on $\isotope[12][6]{C}$ as a function of incoming neutrino energy for distinct components of the 2020 Valencia model and the corresponding NuWro implementation. For a comparison, we also show results from the old NuWro Valencia model.}
    \label{fig:inclusive_cros_section}
\end{figure}

 In Fig.~\ref{fig:inclusive_cros_section}, the performance of the \textit{inclusive part} from the NuWro implementation of the 2020 Valencia model is shown. 
 The solid curves represent NuWro prediction of the total cross section as well as individual contributions from $pp, np$, and $pn$ within \textit{2p2h} and \textit{3p3h} mechanisms using the new hadronic tables produced from the code provided by the authors of Ref.~\cite{Sobczyk:2020dkn}. The black-dotted markers show the calculation for the total cross section as well as the individual contributions done by the code of authors of Ref.~\cite{Sobczyk:2020dkn}. We also show the total cross section from the NuWro implementation of the original Valencia model~\cite{Nieves:2011pp}. About $\sim 20\%$ of the total cross section comes from \textit{3p3h} mechanism for energies $E_{\nu} \gtrsim 1.1$ GeV. All the curves in Fig.~\ref{fig:inclusive_cros_section} have a similar neutrino energy dependence, with a plateau above $E_{\nu} \sim 1.1$ GeV. This behavior can be explained by the fact that the Valencia model imposes the cut of momentum transfer $|\mathbf{q}| < 1.2$ GeV. 

\subsection{\label{sec:nucleon_phase_space} Outgoing nucleon phase space}
We present the performance of the NuWro implementation of the 2020 Valencia model for the nucleon phase spaces. For comparisons, we analyze separately two protons and neutron-proton cases in the final state. In the discussed comparisons FSI effects are not included.

\subsubsection{\label{sec:pp_nucleon_phase_space} $pp$ nucleon phase space}

\begin{figure*}[htbp!]
    \begin{minipage}[b]{0.3\linewidth}
        \begin{adjustbox}{width=\linewidth}
            \includegraphics[width=\linewidth]{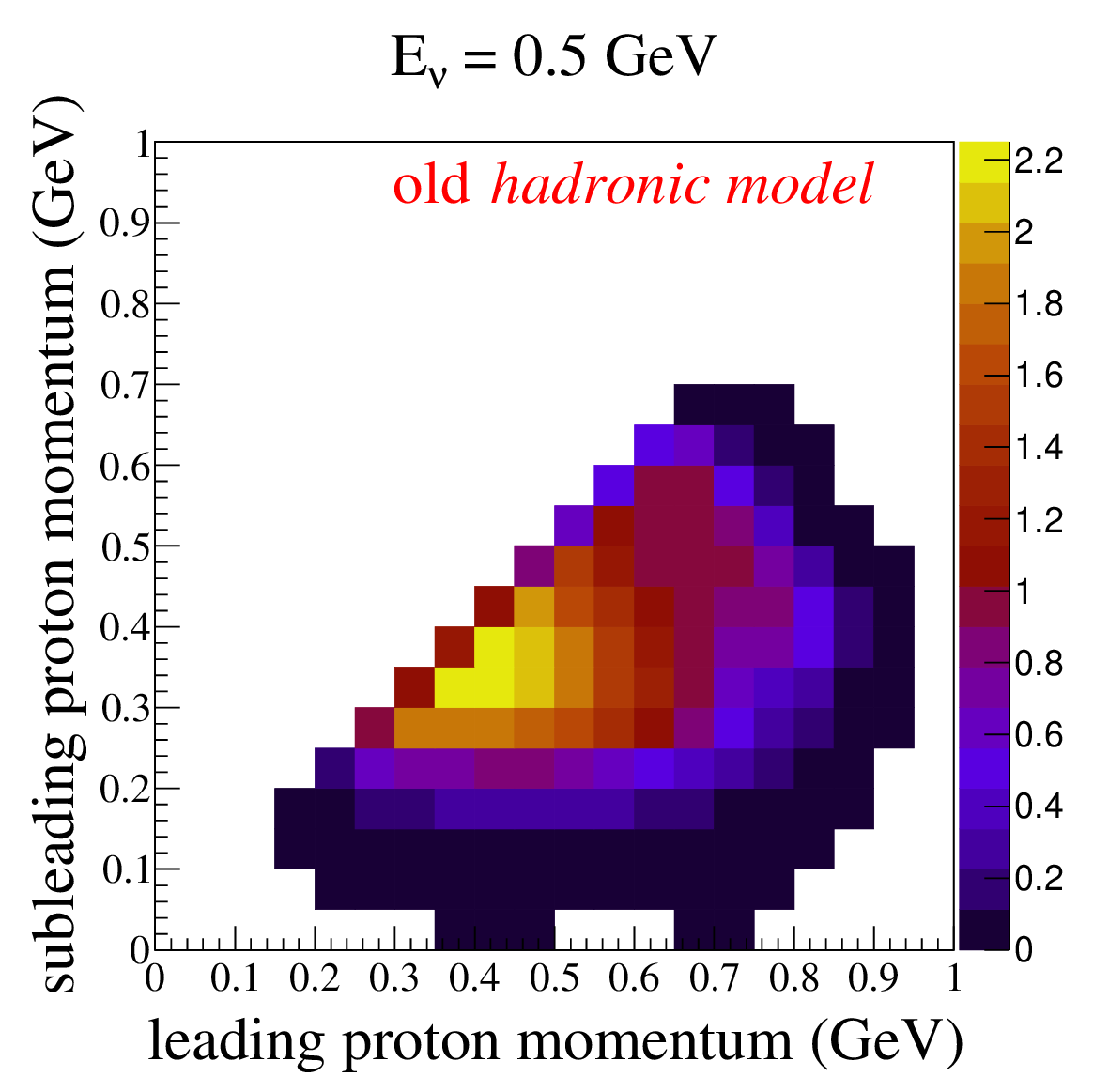}
        \end{adjustbox}
    \end{minipage}
    \begin{minipage}[b]{0.3\linewidth}
        \begin{adjustbox}{width=\linewidth}
            \includegraphics[width=\linewidth]{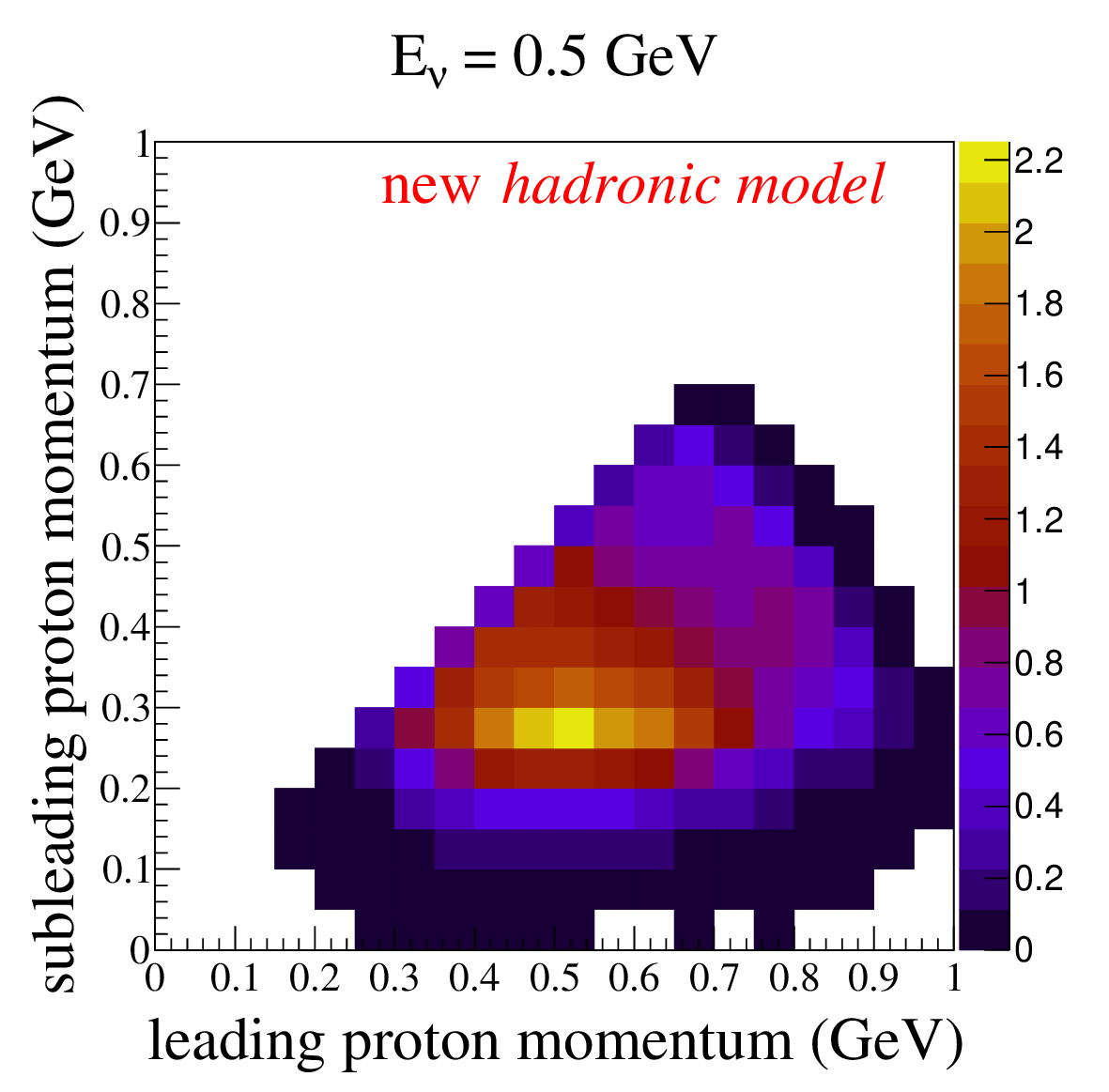}
        \end{adjustbox}
    \end{minipage}
    \begin{minipage}[b]{0.3\linewidth}
        \begin{adjustbox}{width=\linewidth}
            \includegraphics[width=\linewidth]{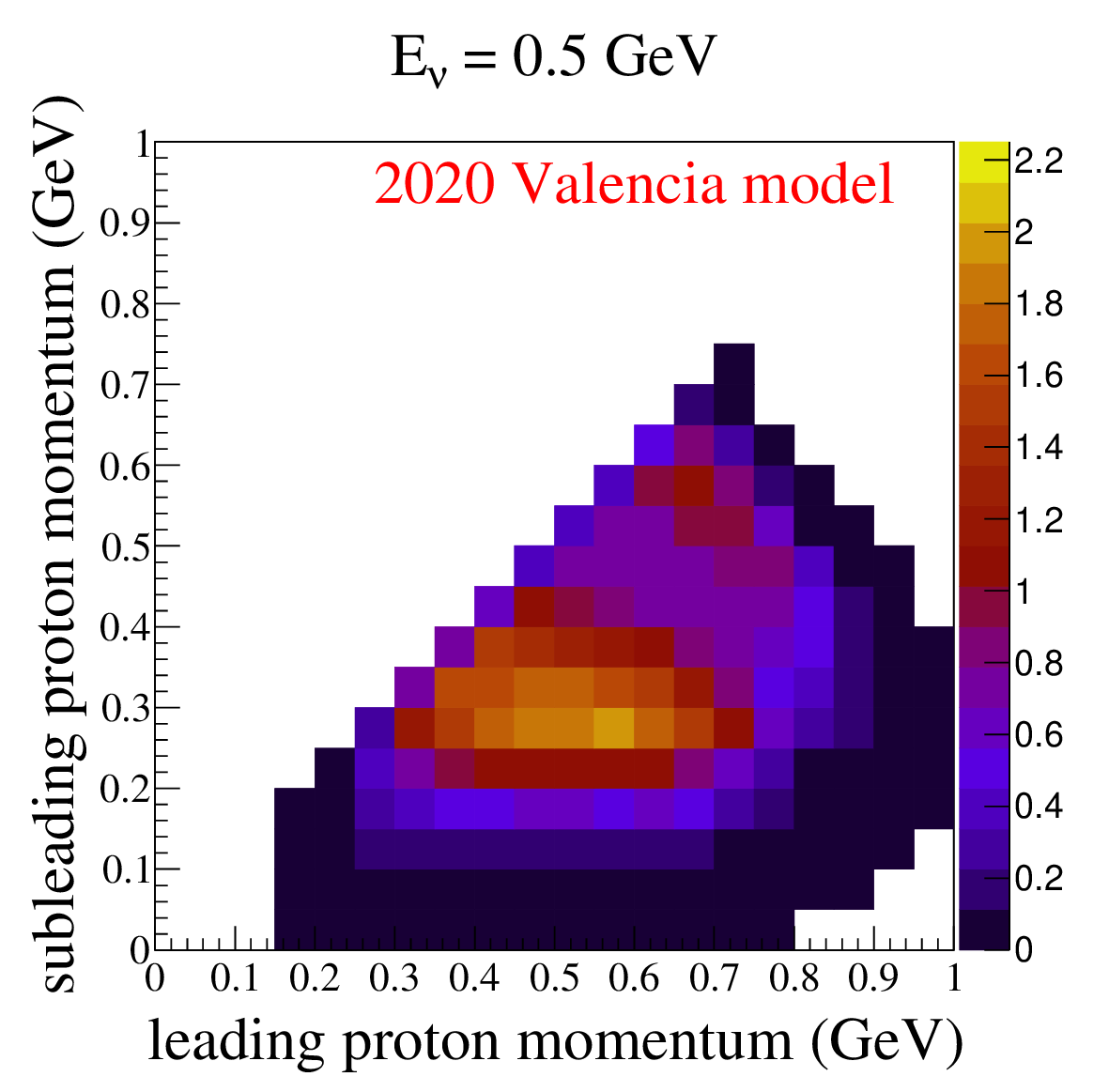}
        \end{adjustbox}
    \end{minipage}
    \begin{minipage}[b]{0.3\linewidth}
        \begin{adjustbox}{width=\linewidth}
            \includegraphics[width=\linewidth]{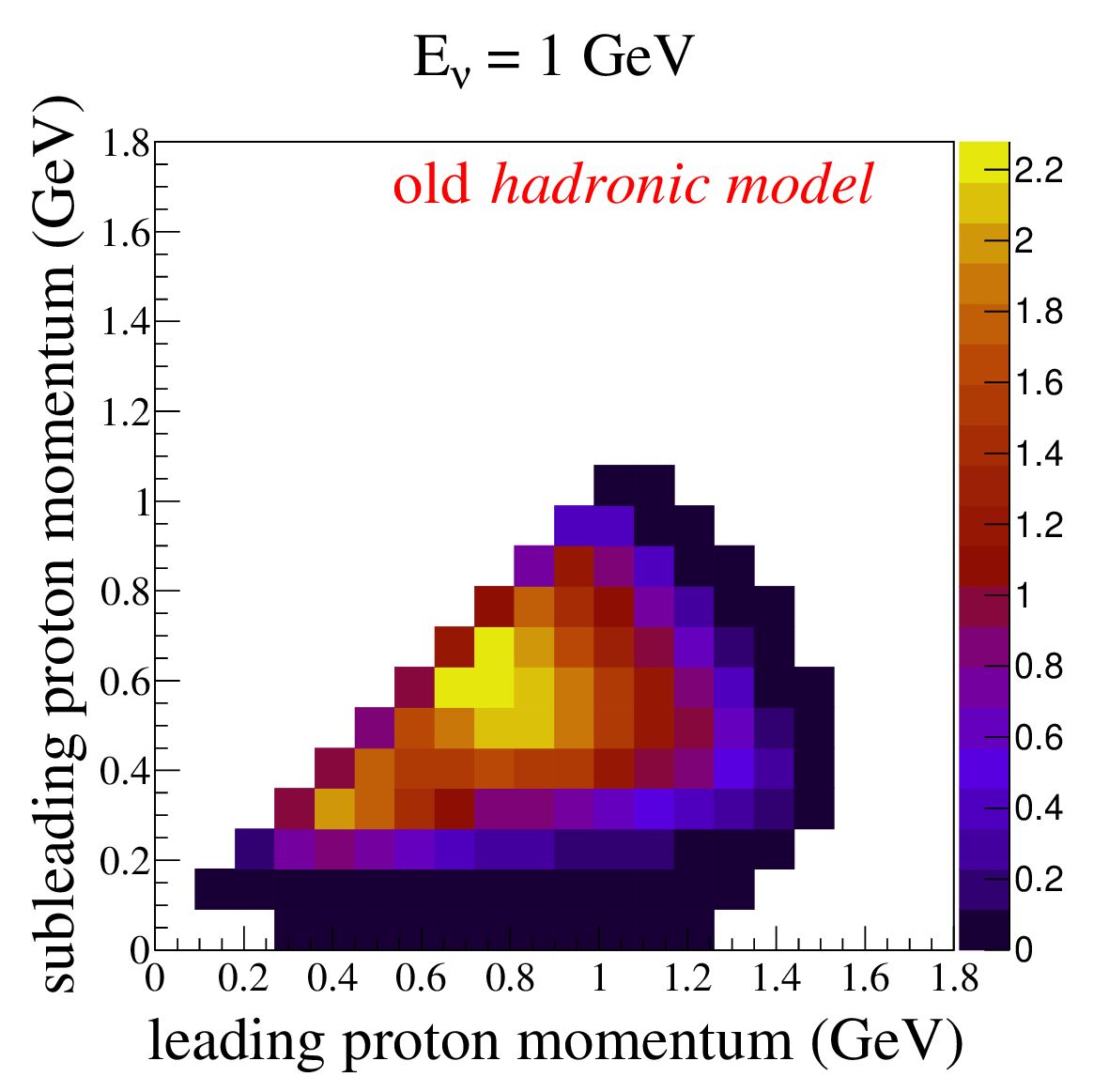}
        \end{adjustbox}
    \end{minipage}
    \begin{minipage}[b]{0.3\linewidth}
        \begin{adjustbox}{width=\linewidth}
            \includegraphics[width=\linewidth]{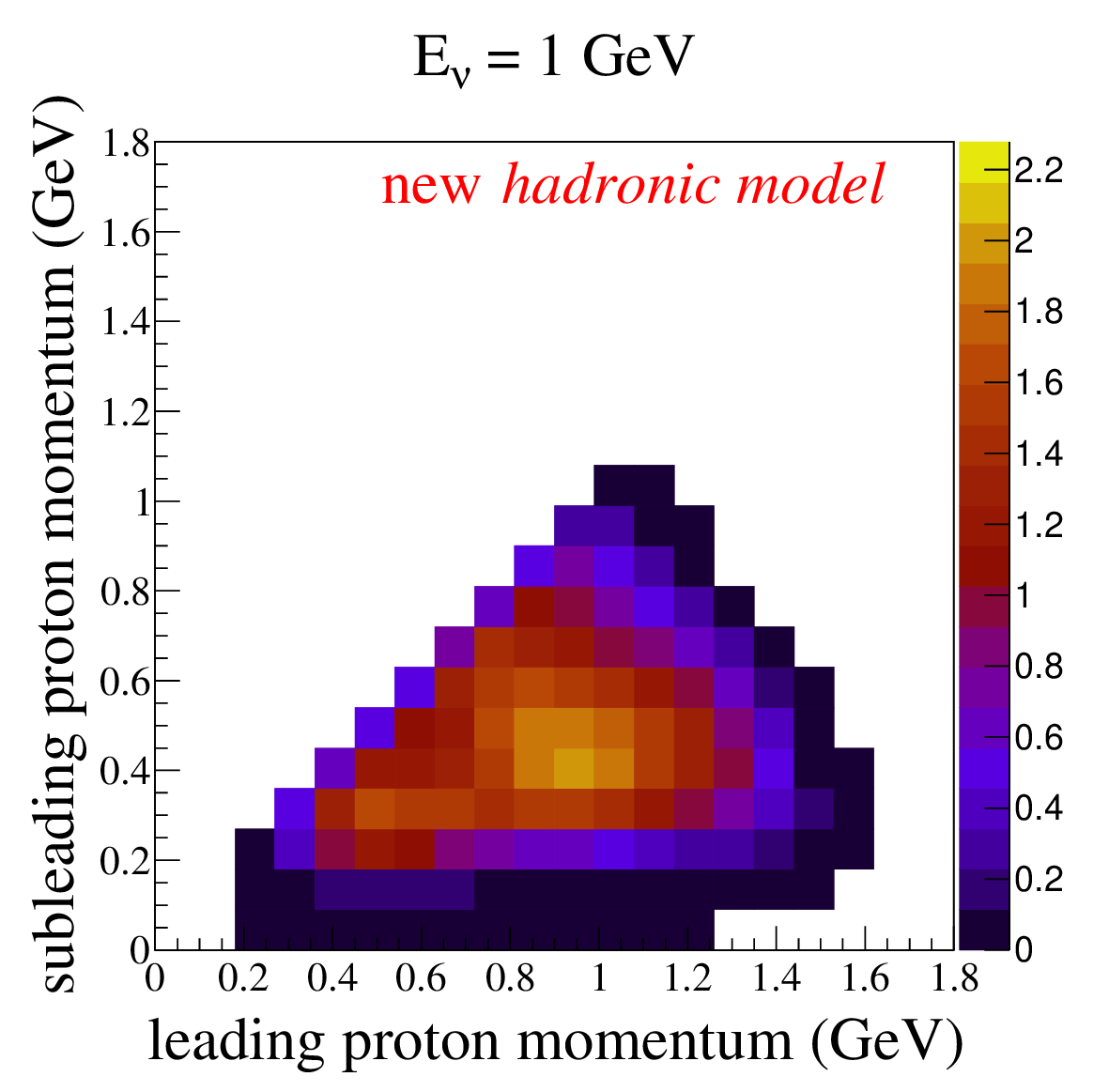}
        \end{adjustbox}
    \end{minipage}
    \begin{minipage}[b]{0.3\linewidth}
        \begin{adjustbox}{width=\linewidth}
            \includegraphics[width=\linewidth]{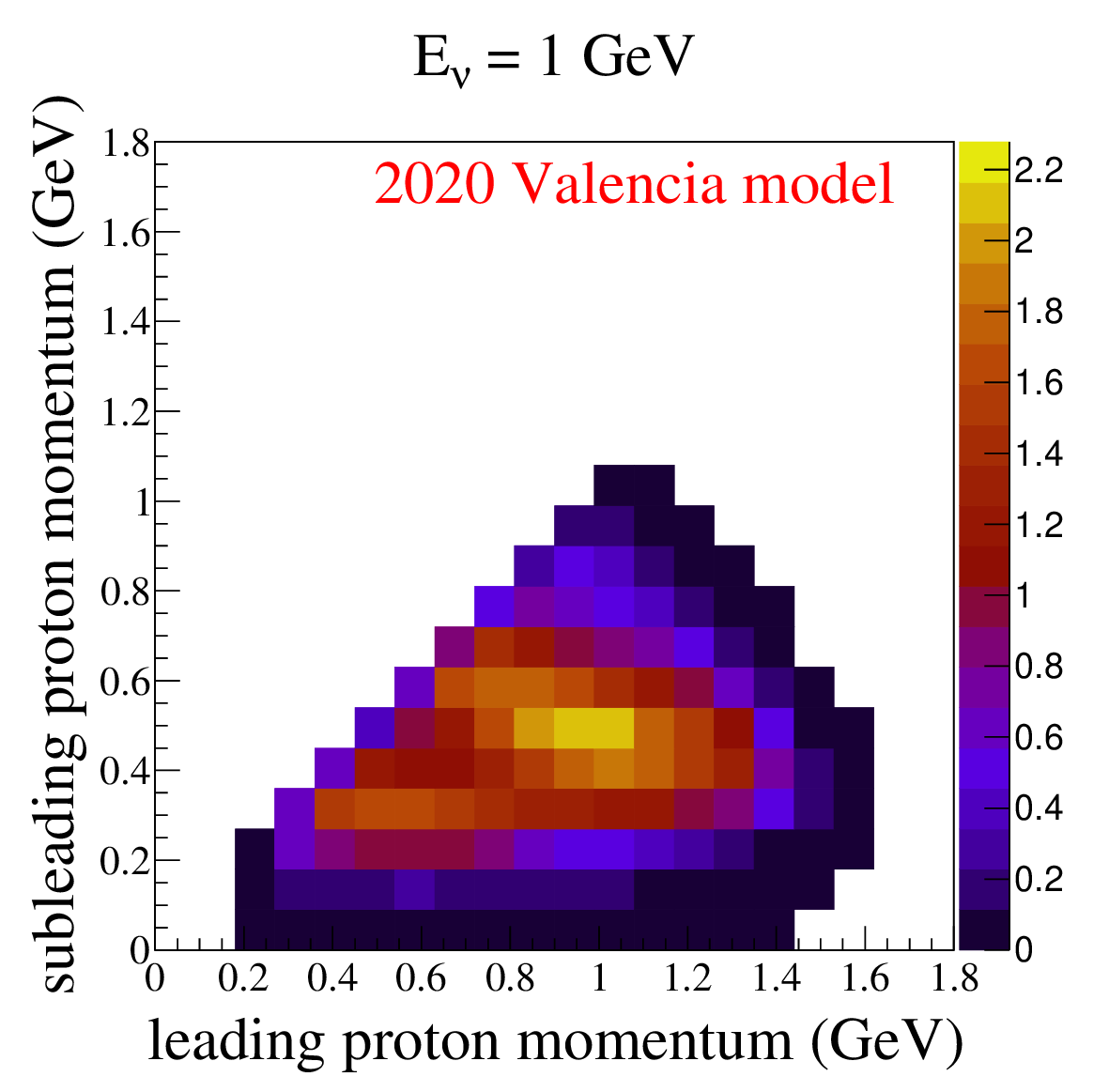}
        \end{adjustbox}
    \end{minipage}
        \begin{minipage}[b]{0.3\linewidth}
        \begin{adjustbox}{width=\linewidth}
            \includegraphics[width=\linewidth]{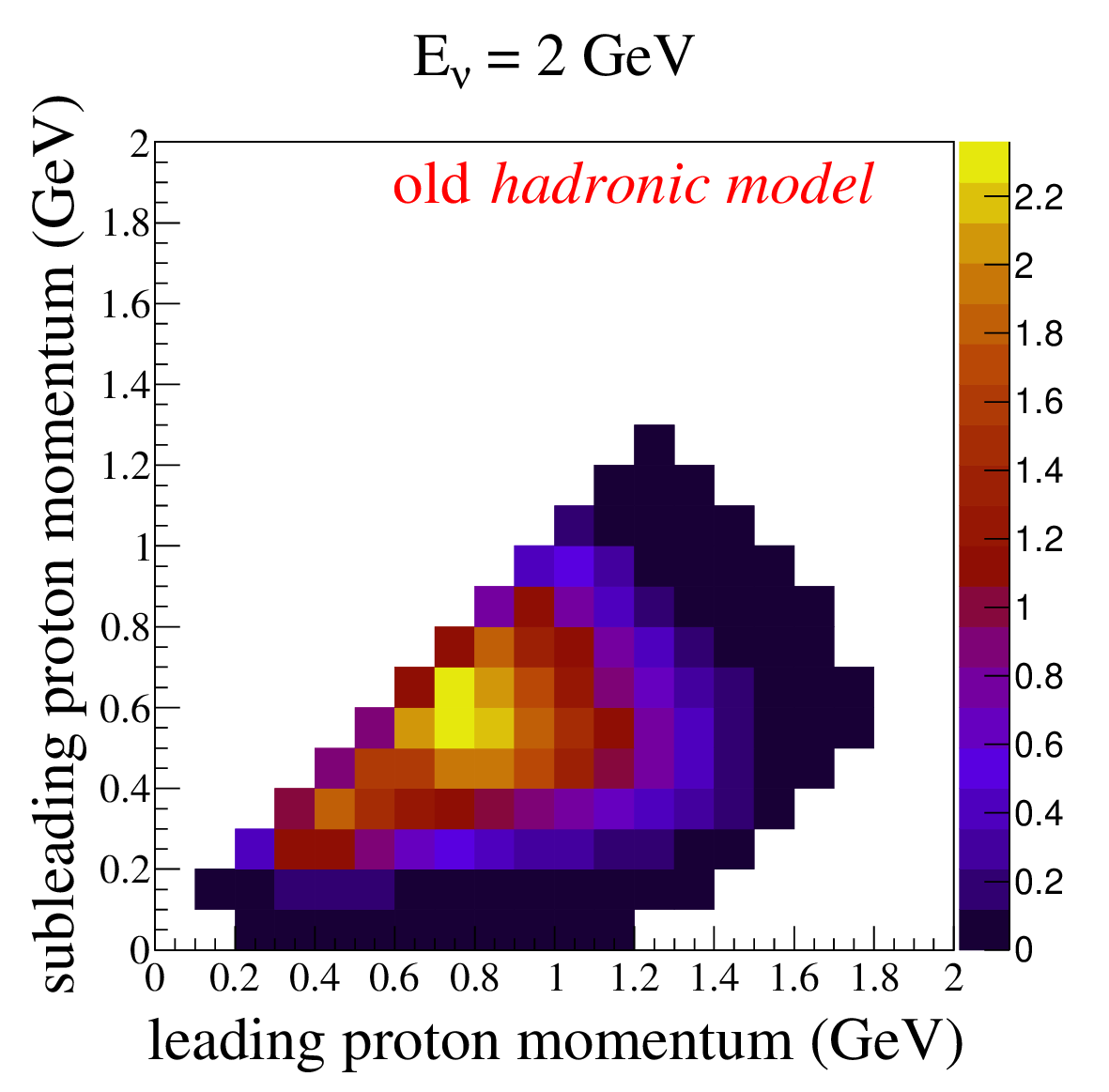}
        \end{adjustbox}
    \end{minipage}
    \begin{minipage}[b]{0.3\linewidth}
        \begin{adjustbox}{width=\linewidth}
            \includegraphics[width=\linewidth]{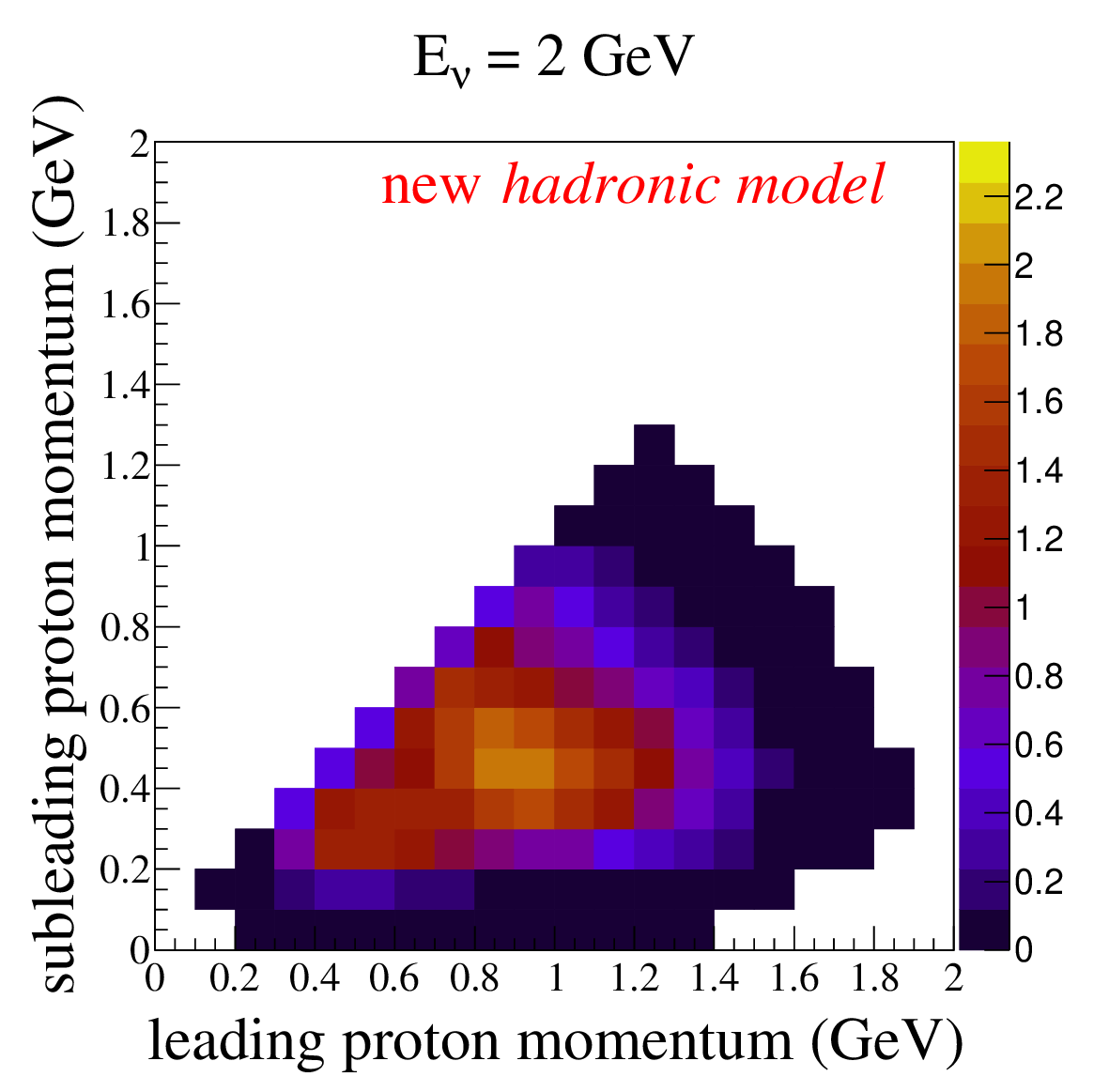}
        \end{adjustbox}
    \end{minipage}
    \begin{minipage}[b]{0.3\linewidth}
        \begin{adjustbox}{width=\linewidth}
            \includegraphics[width=\linewidth]{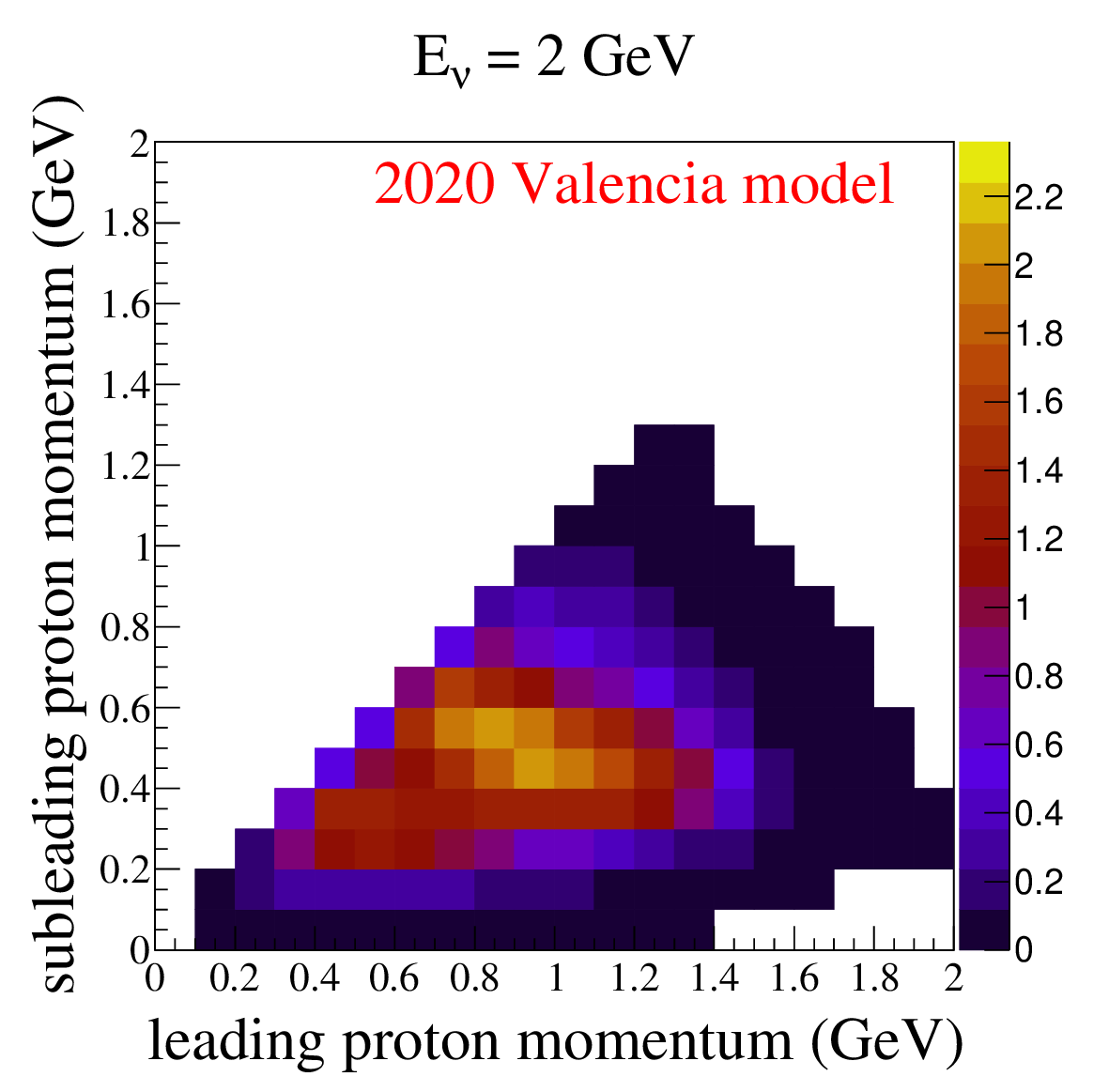}
        \end{adjustbox}
    \end{minipage}
    \caption{Outgoing nucleon distribution $d^2\sigma/d|\mathbf{p}_1| d|\mathbf{p}_2|$ ($10^{-39}$ cm$^2$/GeV$^2$) for two protons in the final state at three different neutrino energies. In all cases, the target nucleus is $\isotope[12][6]{C}$. The \textbf{left} panels show the nucleon phase space produced by the old \textit{hadronic model}, and the \textbf{middle} panels show the new \textit{hadronic model} as described in Sect.\ref{sec:New_MEC_model}, and the \textbf{right} panels represent the phase space from the 2020 Valencia model generated using the code provided by the authors of Ref.\cite{Sobczyk:2020dkn}.}
    \label{fig:nucleon-phasespace_pp}
\end{figure*}

The rightmost panel in Fig.~\ref{fig:nucleon-phasespace_pp}
 shows the $pp$ nucleon phase space from the 2020 Valencia model. The x-axis represents the proton with higher momentum denoted by $|\mathbf{p}_1|$ (the \textit{leading} proton), while the y-axis corresponds to the one with lower momentum denoted by $\mathbf{p}_2$ (the \textit{sub-leading} proton). At $E_{\nu} = 0.5$ GeV, we observe a peak for $|\mathbf{p}_1| \sim [0.4, 0.6]$ GeV and $|\mathbf{p}_2| \sim [0.25, 0.35]$ GeV, corresponding to the $N\Delta$ contribution. For $E_{\nu} \gtrsim 1$ GeV, two distinct peaks appear, one from the $N\Delta$ contribution and another from the $\Delta\Delta$ contribution at higher energies, located around $|\mathbf{p}_1| \sim [0.9, 1.2]$ GeV and $|\mathbf{p}_2| \sim [0.4, 0.6]$ GeV. The 2020 Valencia model assigns significantly different momenta to the two protons.\\

The left panel of Fig.~\ref{fig:nucleon-phasespace_pp} shows results obtained with the hadronic tables of the new model, but with old \textit{hadronic model} for nucleons (see \cite{Sobczyk:2012ms}). At $E_{\nu} = 0.5$ GeV, a single peak from the $N\Delta$ interference appears. For $E_{\nu} \geq 1$ GeV, both $N\Delta$ and $\Delta\Delta$ peaks are visible, with the peak situated at the diagonal in the nucleon phase space. This indicates that momentum differences between the final-state protons are less stringent as compared to the 2020 Valencia model. This behavior is expected because of our assumption \textemdash uniform polar angular distribution of nucleon pair in center-of-mass frame \textemdash which results in an equal number of nucleon pairs with either significant differences or with low differences in their momentum.\\

The middle panel in Fig.~\ref{fig:nucleon-phasespace_pp} represents the performance of the new NuWro hadronic model as described in Sect.~\ref{sec:New_MEC_model}. Here, too, we observe the $N\Delta$ peak at $E_{\nu}=0.5$ GeV. As we increase neutrino energy i.e. $E_{\nu} \geq 1$ GeV, both peaks become visible. Since $\hat{P}_{pp}^{\text{Global-fit}} > 0$, the peak shifts away from the diagonal to towards the lower-right, compared to the left panel. This is expected as the new \textit{hadronic model}  samples more nucleons that are (anti)parallel with moderate strength in the center-of-mass frame. Upon boosting back to the LAB frame, the \textit{forward} nucleon becomes more energetic due to its parallel component with the boost direction, while the \textit{backward} nucleon becomes less energetic. This results in a noticeable momentum difference between the two outgoing protons. The new NuWro hadronic model does not reproduce the results from the 2020 Valencia model exactly, but later on, we will demonstrate that the difference is largly annihilated by FSI effects not included in the analyzed distributions.

\subsubsection{\label{sec:nppn_nucleon_phase_space}$np+pn$ nucleon phase space} 
\begin{figure*}[htbp!]
    \begin{minipage}[b]{0.3\linewidth}
        \begin{adjustbox}{width=\linewidth}
            \includegraphics[width=\linewidth]{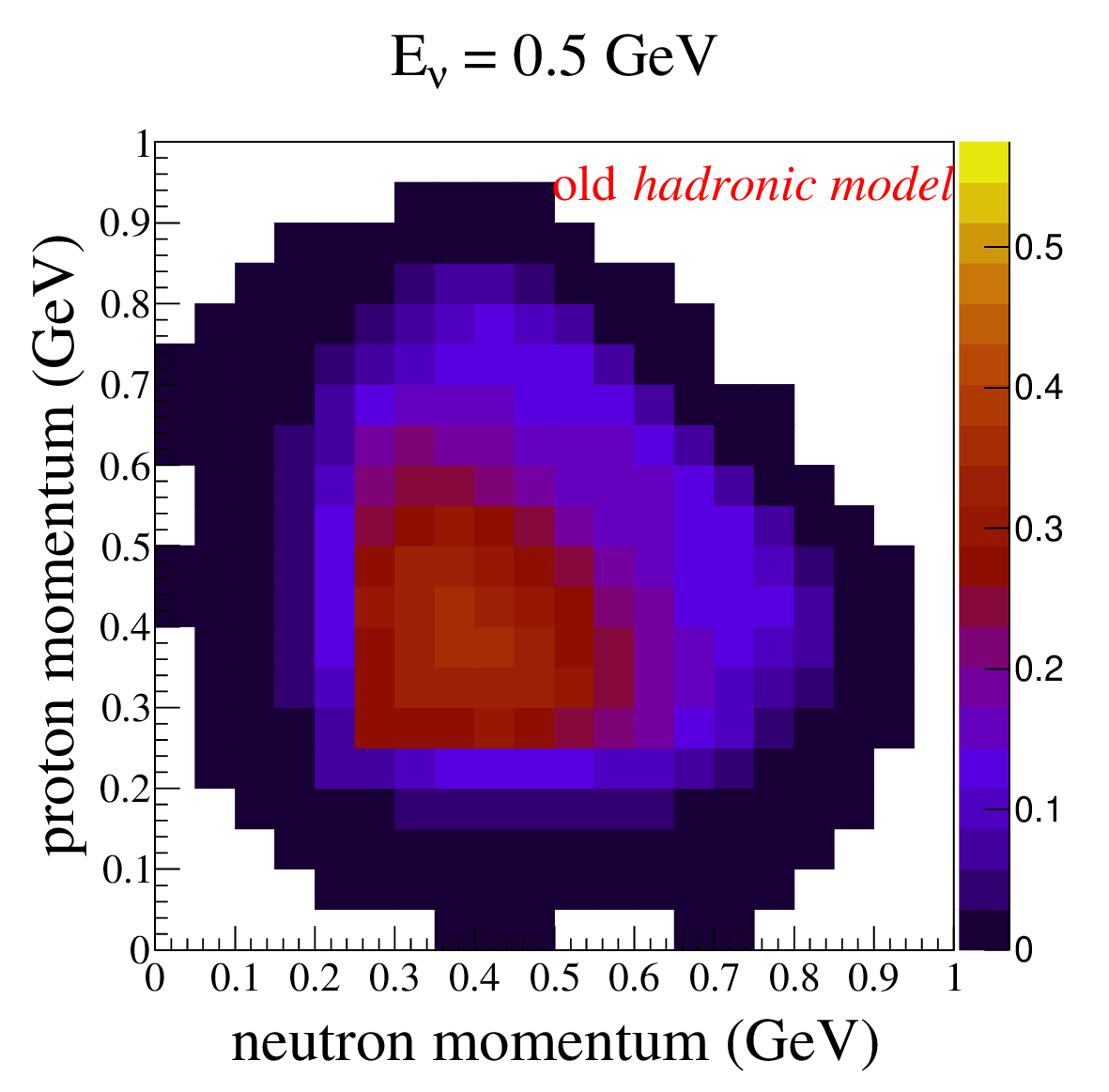}
        \end{adjustbox}
    \end{minipage}
    \begin{minipage}[b]{0.3\linewidth}
        \begin{adjustbox}{width=\linewidth}
            \includegraphics[width=\linewidth]{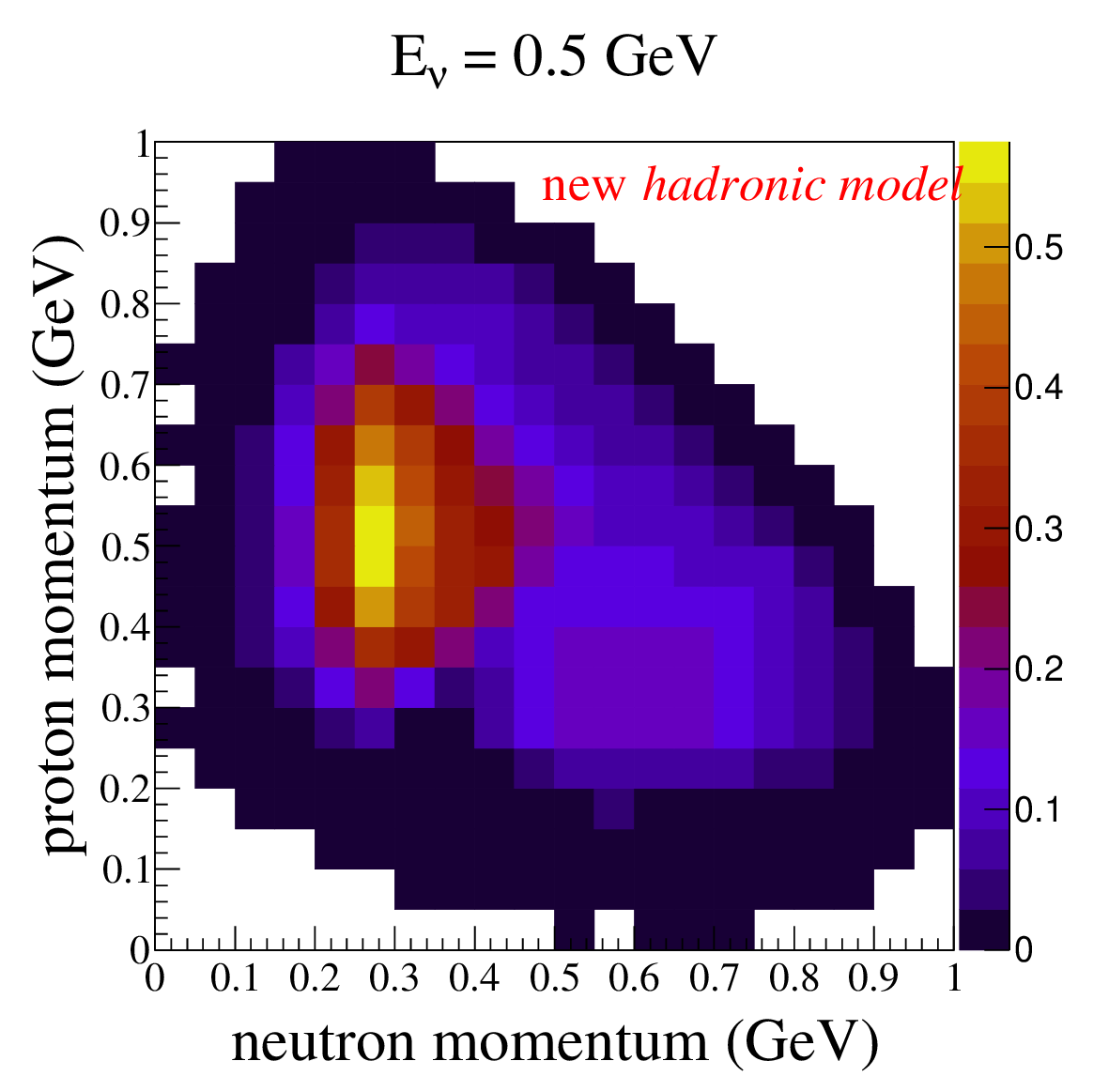}
        \end{adjustbox}
    \end{minipage}
    \begin{minipage}[b]{0.3\linewidth}
        \begin{adjustbox}{width=\linewidth}
            \includegraphics[width=\linewidth]{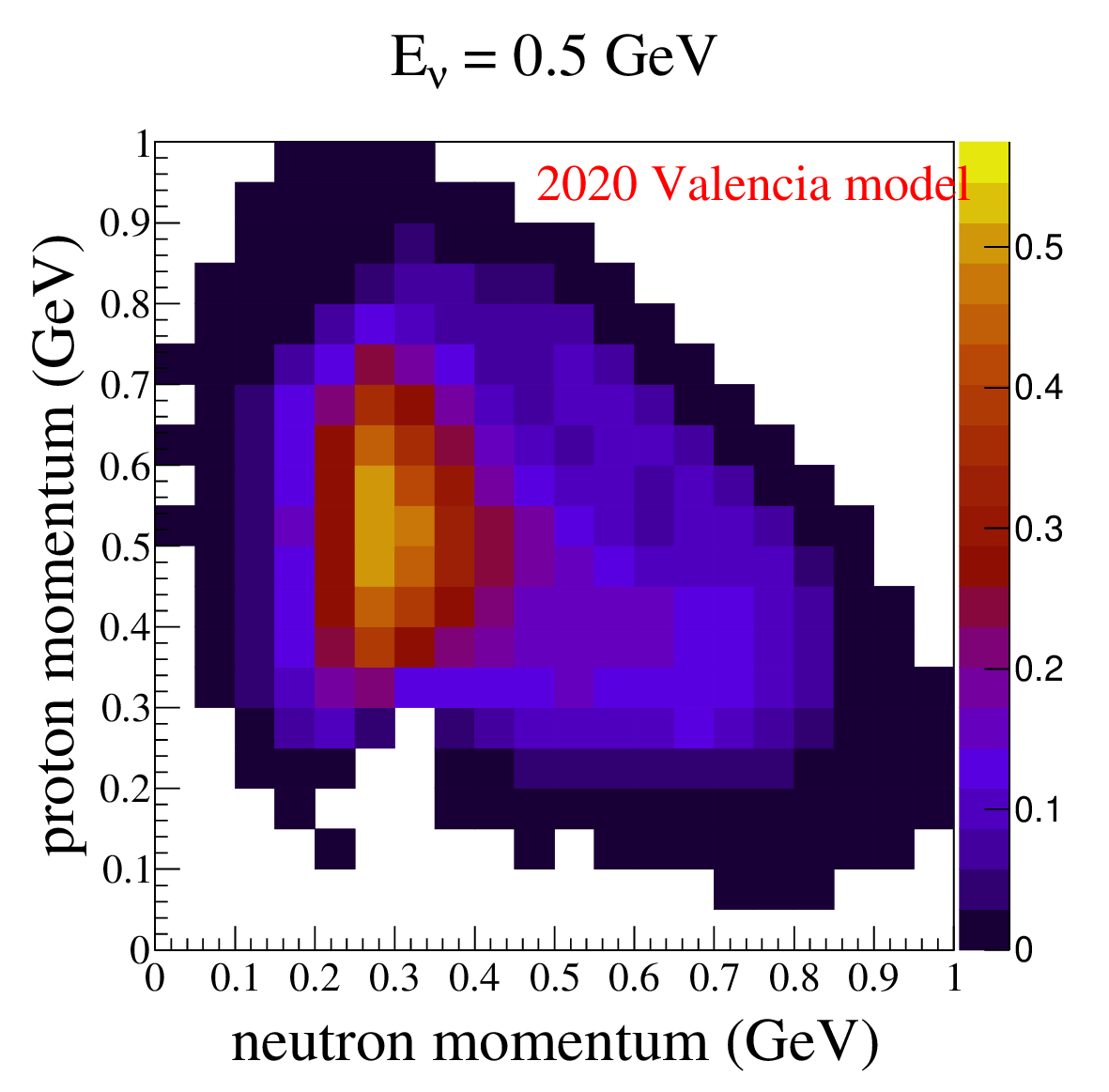}
        \end{adjustbox}
    \end{minipage}
    \begin{minipage}[b]{0.3\linewidth}
        \begin{adjustbox}{width=\linewidth}
            \includegraphics[width=\linewidth]{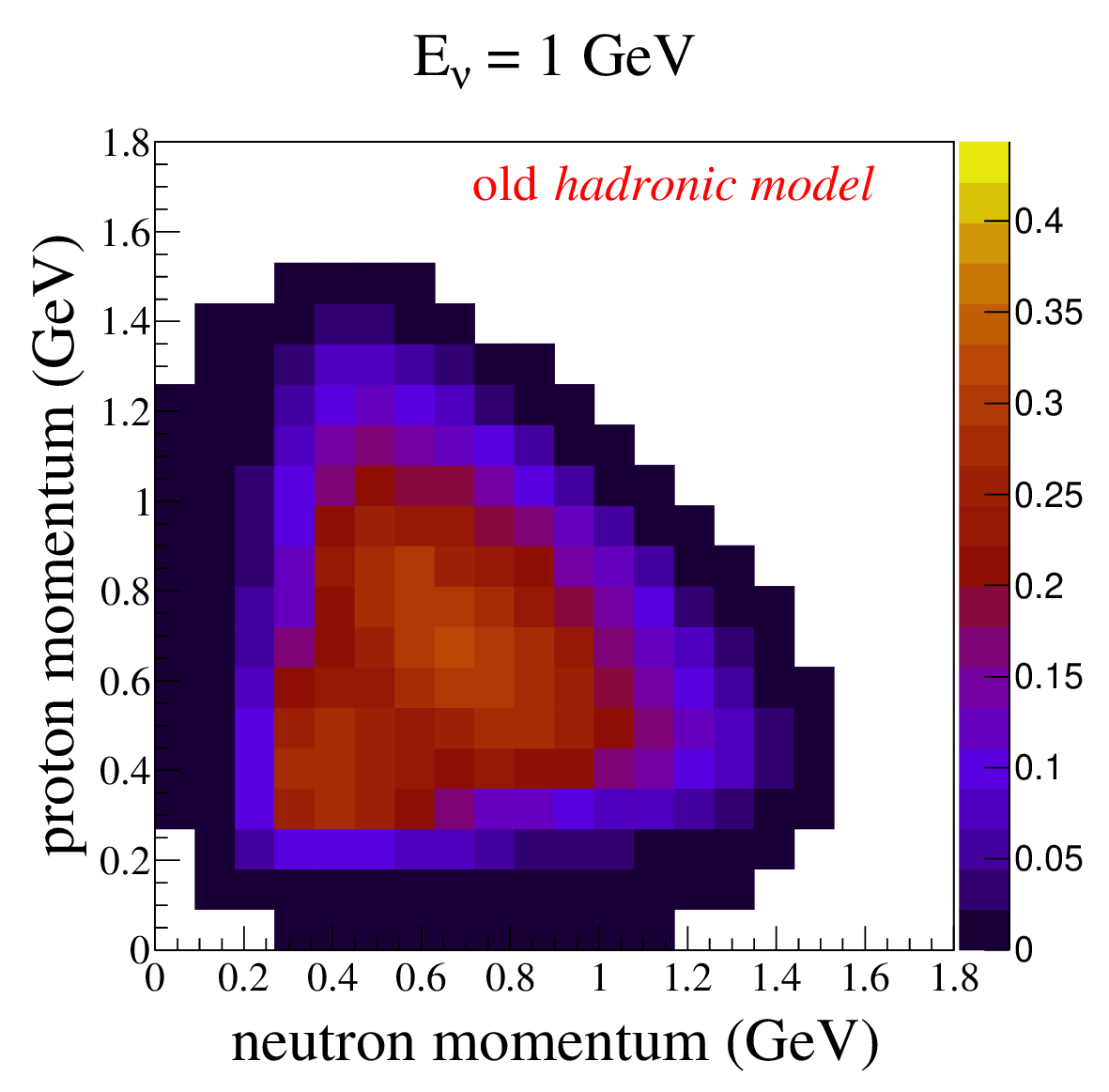}
        \end{adjustbox}
    \end{minipage}
    \begin{minipage}[b]{0.3\linewidth}
        \begin{adjustbox}{width=\linewidth}
            \includegraphics[width=\linewidth]{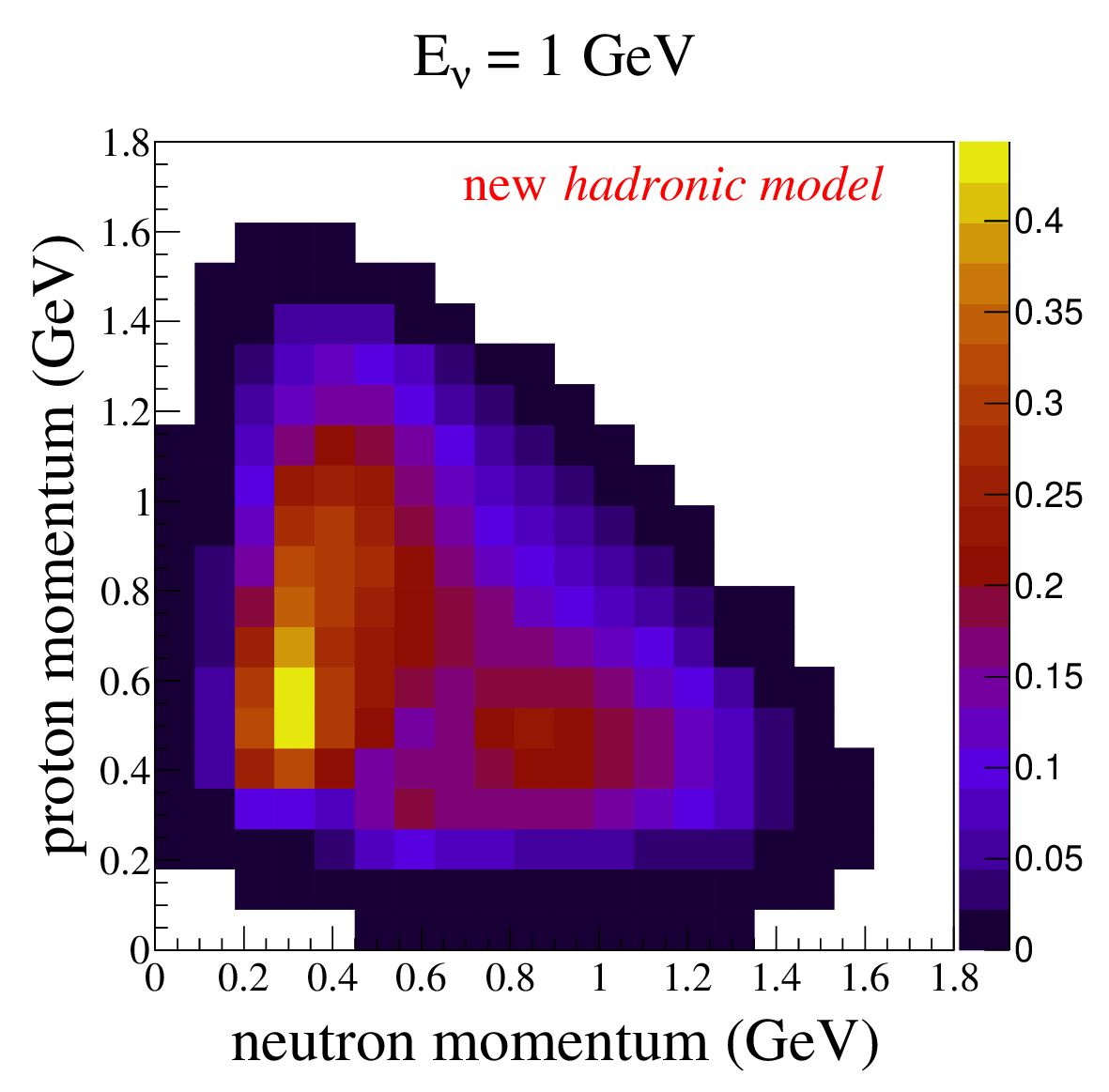}
        \end{adjustbox}
    \end{minipage}
    \begin{minipage}[b]{0.3\linewidth}
        \begin{adjustbox}{width=\linewidth}
            \includegraphics[width=\linewidth]{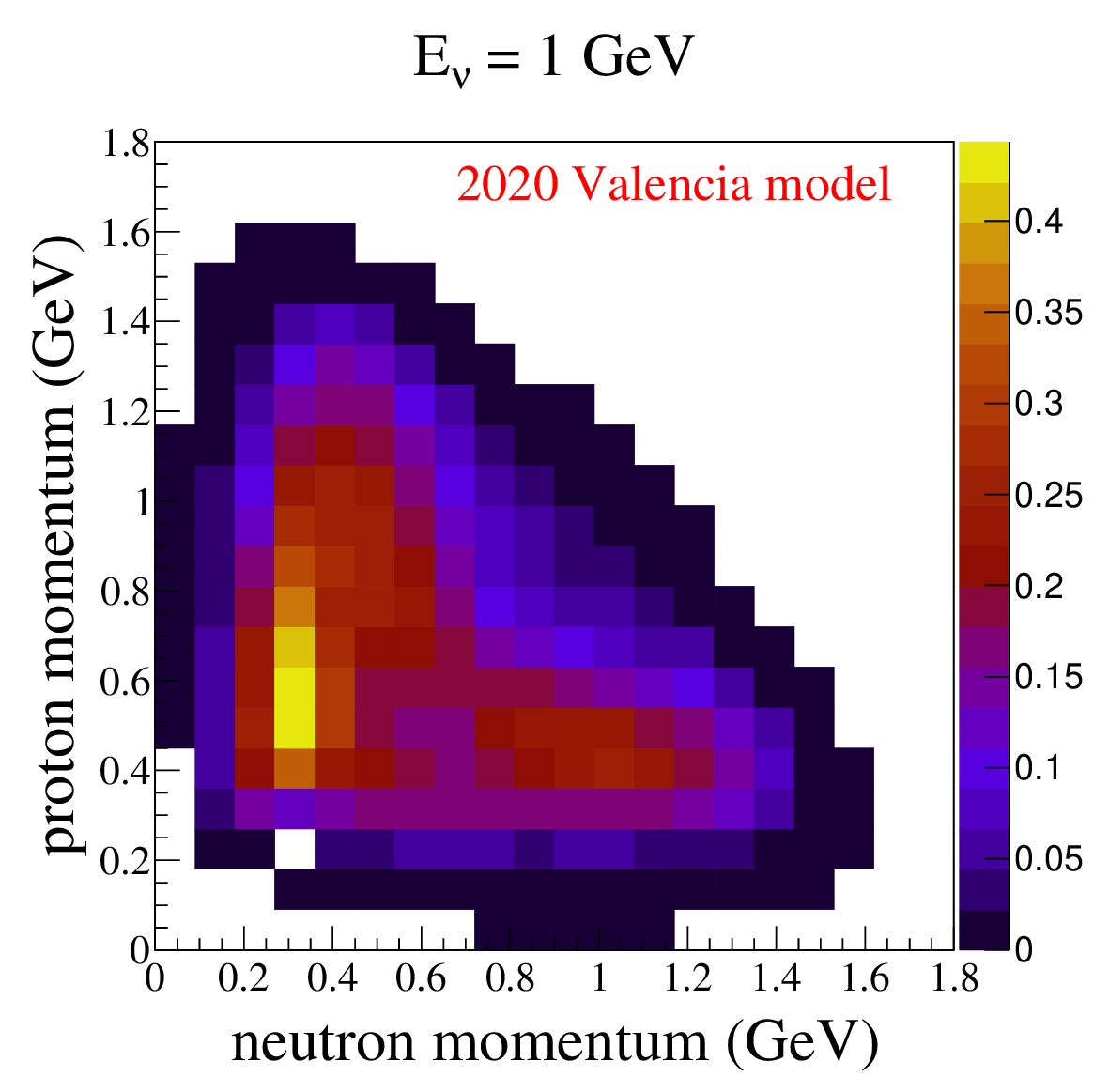}
        \end{adjustbox}
    \end{minipage}
        \begin{minipage}[b]{0.3\linewidth}
        \begin{adjustbox}{width=\linewidth}
            \includegraphics[width=\linewidth]{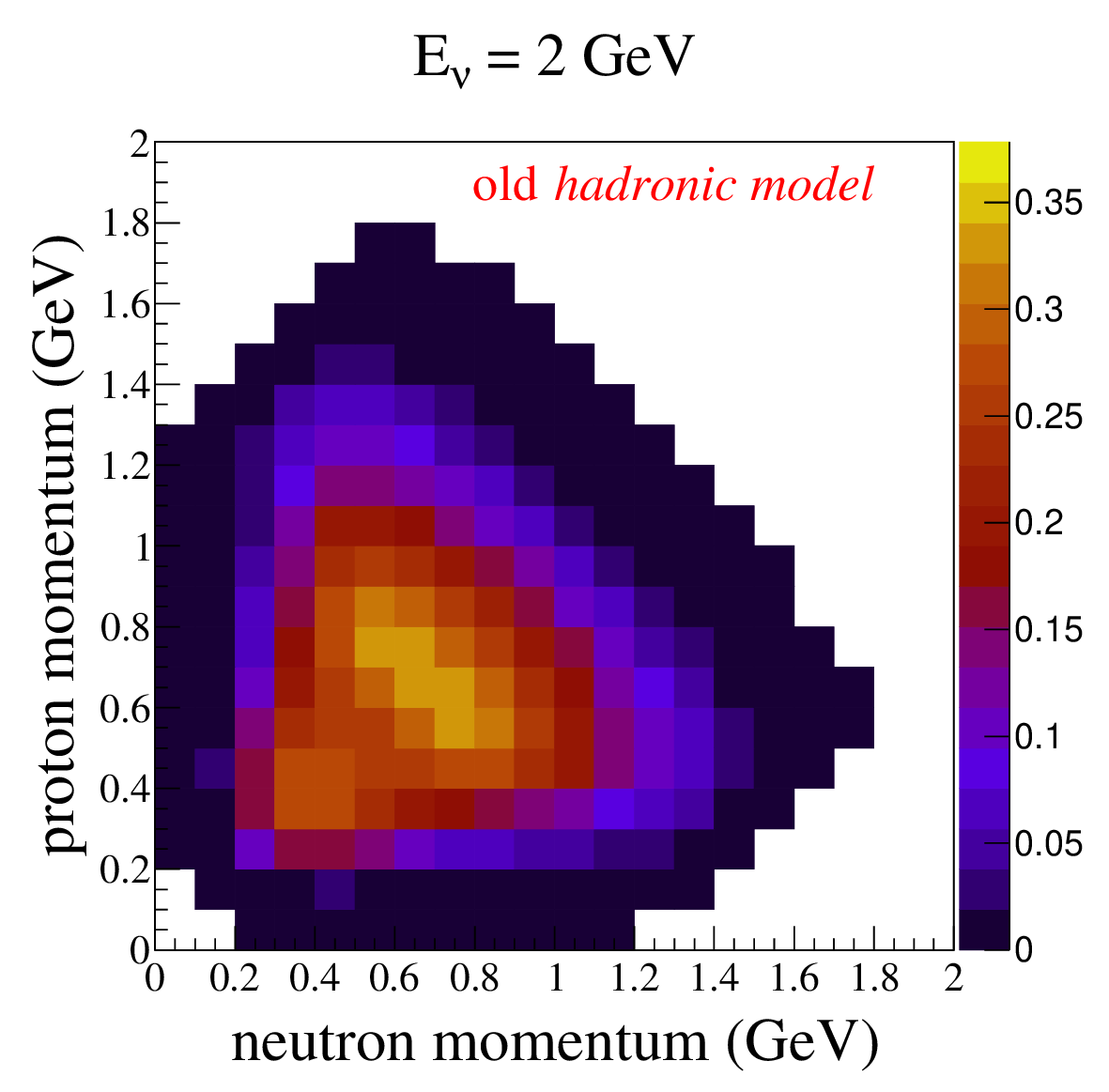}
        \end{adjustbox}
    \end{minipage}
    \begin{minipage}[b]{0.3\linewidth}
        \begin{adjustbox}{width=\linewidth}
            \includegraphics[width=\linewidth]{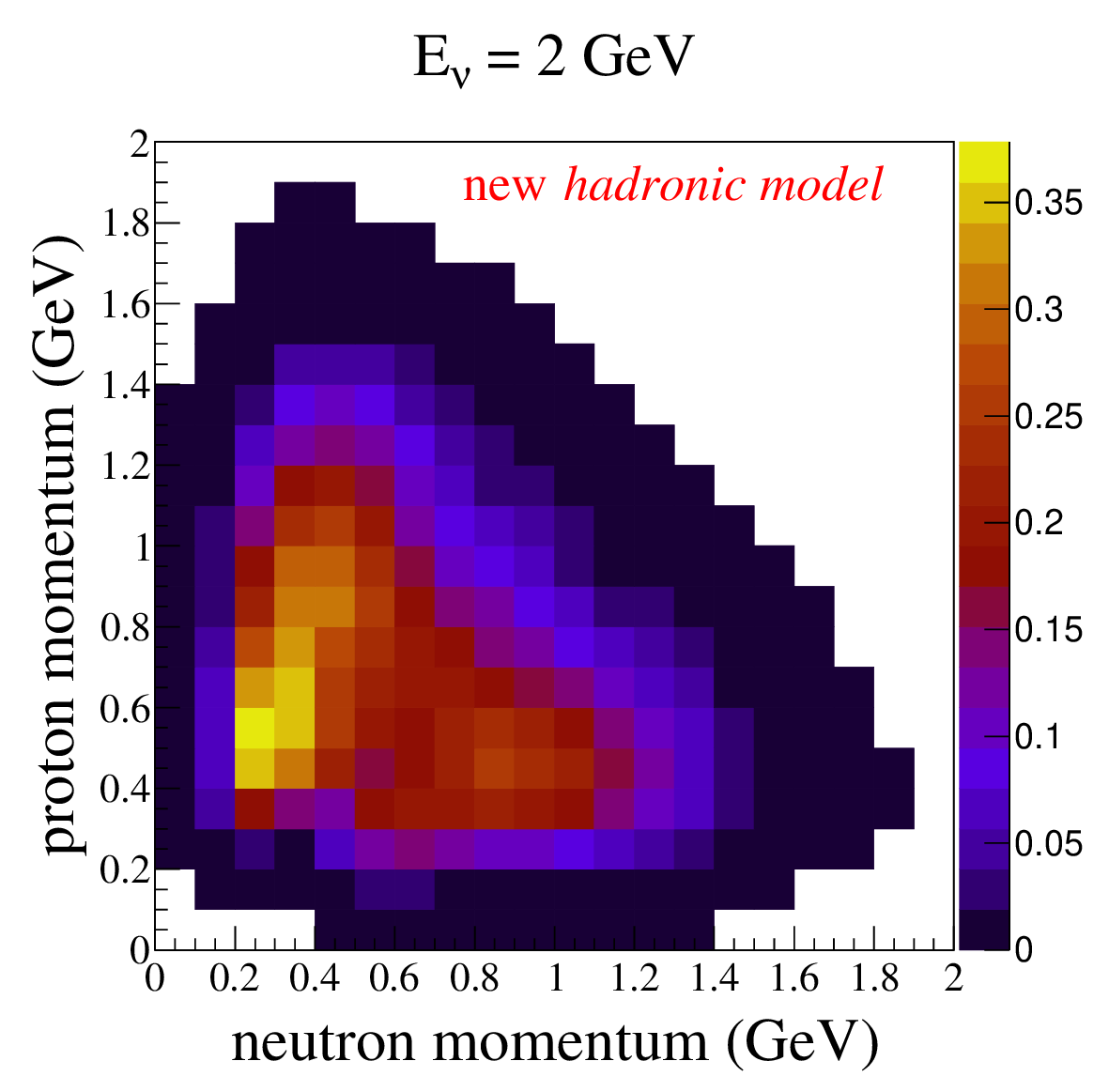}
        \end{adjustbox}
    \end{minipage}
    \begin{minipage}[b]{0.3\linewidth}
        \begin{adjustbox}{width=\linewidth}
            \includegraphics[width=\linewidth]{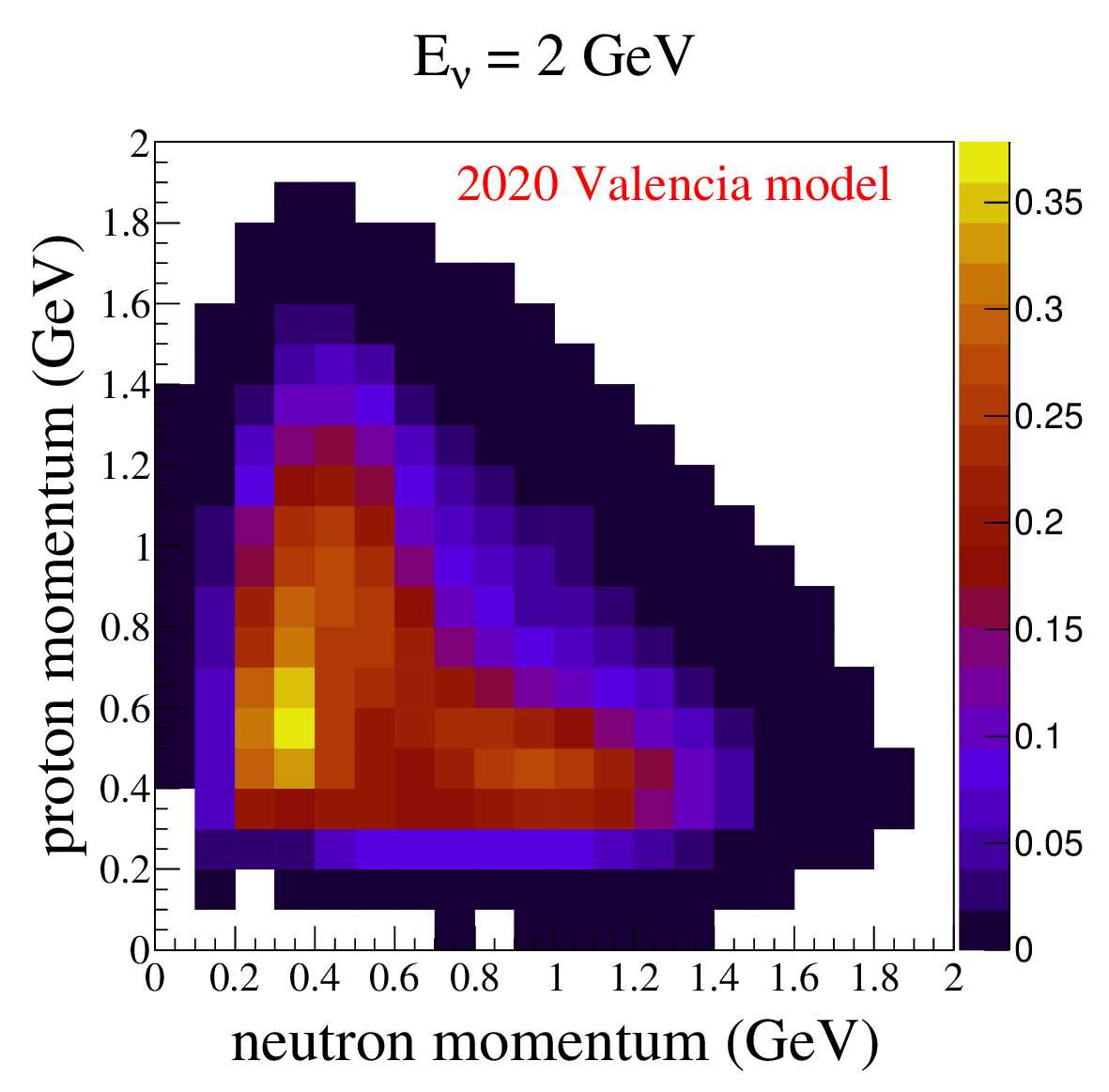}
        \end{adjustbox}
    \end{minipage}
    \caption{Outgoing nucleon distribution $d\sigma/d|\mathbf{p}_p| d|\mathbf{p}_n|$ ($10^{-39}$ cm$^2$/GeV$^2$) in the case of neutron-proton pair produced in the final state for three different neutrino energies. In all cases, the target nucleus is $\isotope[12][6]{C}$. The panel from \textbf{left} is for the old \textit{hadronic model} while the \textbf{middle} panel corresponds to the new \textit{hadronic model}. The \textbf{right} panel corresponds to phase space generated by the 2020 Valencia model obtained from the code provided by the authors of Ref. \cite{Sobczyk:2020dkn}.}
    \label{fig:nucleon-phasespace_nppn}
\end{figure*}

In this case, the contribution to the cross section is significantly smaller compared to the pp part, see Fig.~\ref{fig:inclusive_cros_section}. We first discuss the nucleon phase space from the 2020 Valencia model. The rightmost panel in Fig.~\ref{fig:nucleon-phasespace_nppn} shows the phase space for $E_{\nu} \in \{0.5, 1.0, 2.0\}$ GeV.
The x-axis corresponds to neutron momentum and the y-axis corresponds to the proton momentum. At $E_{\nu}=0.5$ GeV we see that phase space is mostly filled by $pn$ events. This is expected as the at low energy transfers $N\Delta$ mechanism has the dominant contribution to the total cross section and $pn$ events are more likely compared to $np$ events due to higher contribution to the total cross section. For higher neutrino energies i.e $E_{\nu} \geq 1$ GeV, pn events are more likely. \\

If a proton is produced in the \textit{ph} excitation directly connected to the $W^+$ boson, which is marked as a $pn$ event, then the probability of a higher energetic proton and a lower energetic neutron is quite high, as more energy is transferred to the proton. Conversely, if a neutron is produced in the \textit{ph} excitation directly connected to the $W^+$ boson, which is marked as a $np$ event, then the probability of a higher energetic neutron and a lower energetic proton is quite high, as more energy is transferred to the neutron. These two possible contributions have a similar strength and this causes the phase space to have a ``$\Lcorner$''-like shape.\\

The left panel of Fig.~\ref{fig:nucleon-phasespace_nppn} shows results obtained with the hadronic tables of the new model but with the \textit{old hadronic model}~\cite{Sobczyk:2012ms} for nucleons. We observe that the distribution is symmetric along the diagonal for all neutrino energies. This is because of two major assumptions in the old model. The first assumption is that $np$ and $pn$ events are equally likely. The second is that we assume a uniform polar angular distribution of the nucleon pair in the center-of-mass frame for all nucleon types. At $E_{\nu} = 0.5$ GeV, we observe a peak at $|\mathbf{p}_p|, |\mathbf{p}_n| \sim [0.35, 0.45]$ GeV, attributed to the $N\Delta$ mechanism. At higher neutrino energies, another peak appears at $|\mathbf{p}_p|, |\mathbf{p}_n| \sim [0.5, 0.8]$ GeV, corresponding to the $\Delta\Delta$ mechanism.\\

The middle panel in Fig.~\ref{fig:nucleon-phasespace_nppn} represents the performance of the new NuWro hadronic model (see Sect.\ref{sec:New_MEC_model}). We observe that due to our assumption \textemdash determination of the isospin of \textit{forward} and \textit{backward} nucleon \textemdash along with the positive global fit values $\hat{P}_{np}, \hat{P}_{pn} > 0$ the peak is split into two parts. We also observe that since $\hat{P}_{np} \approx \hat{P}_{pn}$ the strength of assigning higher energy is similar in both types of final state pair. We see that in this case an agreement between the new NuWro model and the 2020 Valencia model is very good, especially for larger neutrino energies.

\subsection{\label{sec:FSI_effects} FSI effects}
\begin{figure}[htbp!]
    \centering
    \includegraphics[width=1.1\linewidth]{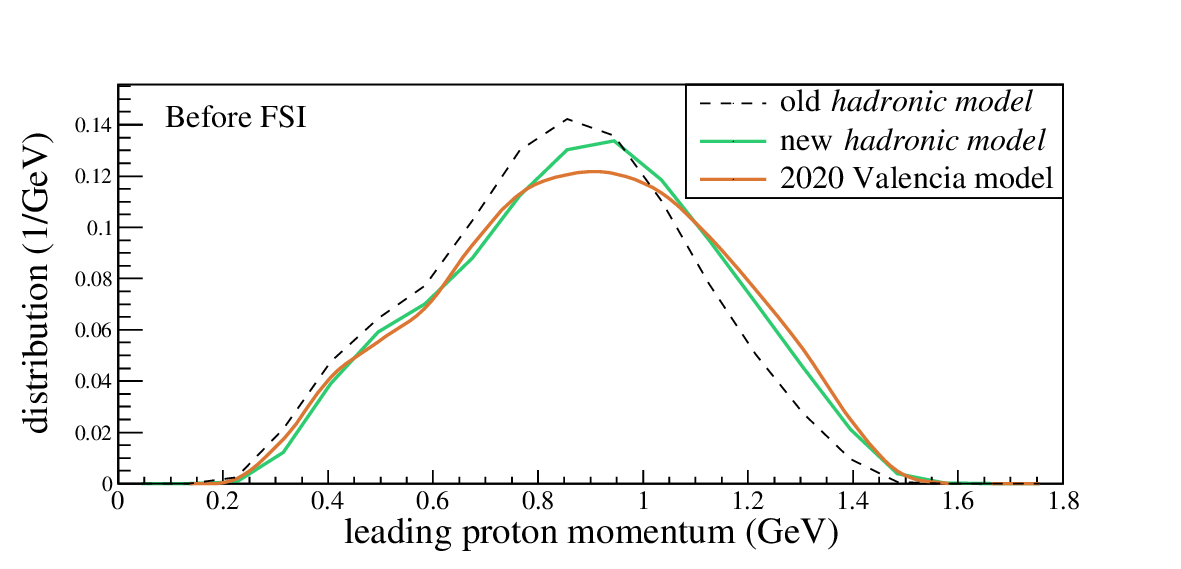}    \caption{\label{fig:maximal_nucleon_before_fsi} A distribution of events (normalized to unit area) from $pp$ nucleon phase space when projected on the x-axis ``leading proton momentum'' for $E_{\nu} = 1$ GeV (see middle row of Fig.~\ref{fig:nucleon-phasespace_pp}).}
\end{figure}

In this section, we quantify the impact of FSI effects on modeling the outgoing nucleons. We consider two proton final state events as it constitutes a much higher contribution to the total cross section. We look only at \textit{leading} protons since they are more likely to be detected. In Fig.~\ref{fig:maximal_nucleon_before_fsi} 
we show the distribution of the momentum of \textit{leading} proton from three models discussed in the previous subsection at neutrino energy $1$ GeV. The histogram with yellow, magenta, and green lines shows results from the new NuWro \textit{hadronic model}, 2020 Valencia model, and old NuWro hadronic model with new hadronic tables respectively. The new \textit{hadronic model} samples more highly energetic protons similar to the 2020 Valencia model (orange line). The old \textit{hadronic model} (dashed-black outline) however, produces high energetic nucleons at a significantly lower rate. We then pass the outgoing nucleons through NuWro's intranuclear cascade (INC) model~\cite{Niewczas:2019fro} to estimate the smearing effect. \\

Since the 2020 Valencia model does not account for the FSI effects, we must use the re-weighting technique to impose the FSI effects simulated by NuWro from the results produced by the 2020 Valencia model. For this purpose, we reproduce the nucleon phase space from the 2020 Valencia model within NuWro for any given neutrino energy. We do this by computing the scaling factor of each bin in the nucleon phase space which is the ratio of the number of events produced by the 2020 Valencia model to the number of events produced by the new \textit{hadronic model} of NuWro in that bin. We then scale the weights of the events that are within that specific bin by the scaling factor and analyze the maximal proton after FSI using NuWro's cascade model. This makes events with values of maximal proton after FSI among different bins more/less likely to occur based on the scaling factor within the nucleon phase space. In this way, the FSI effects are estimated for the 2020 Valencia model. \\

\begin{figure}[htbp!]
    \centering
    \includegraphics[width=1.1\linewidth]{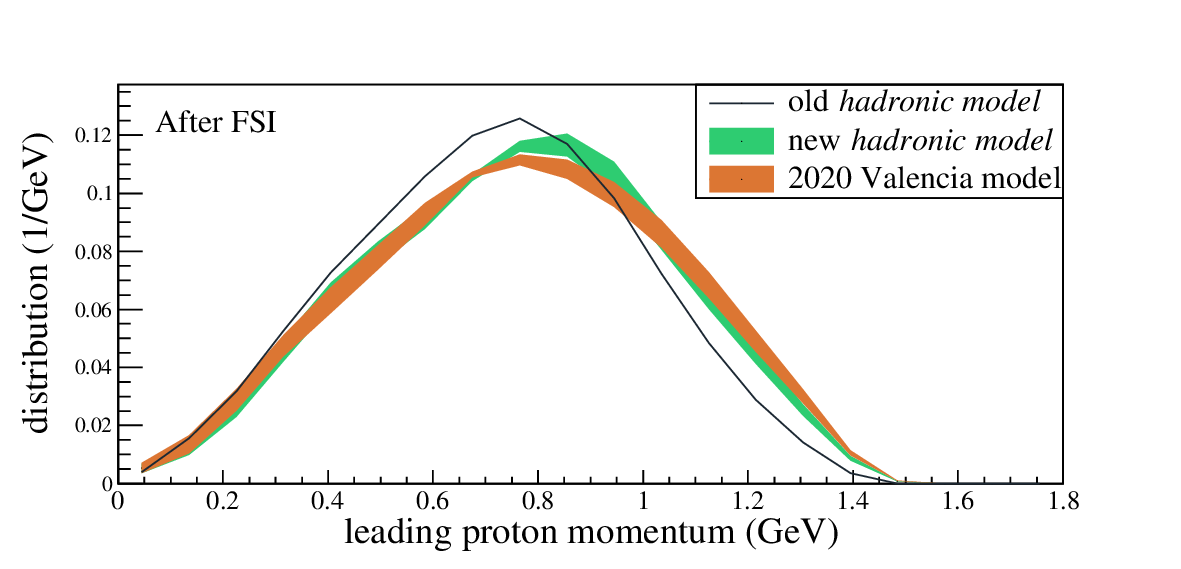}
    \caption{\label{fig:maximal_nucleon_after_fsi} Distribution of momentum of \textit{leading} proton after FSI for $E_{\nu} = 1$ GeV. The band around the curve represents the uncertainties during modeling FSI effects.}
\end{figure}

In Fig.~\ref{fig:maximal_nucleon_after_fsi}        
we show the distribution of momentum of \textit{leading} proton from three investigated models after FSI. The bands around the curves represent the uncertainties in modeling FSI effects estimated in Ref.~\cite{Niewczas:2019fro}. We observe that results from the 2020 Valencia model and its NuWro implementation coincide within FSI uncertainties. Also, both are significantly different from results from the old \textit{hadronic model}. This demonstrates that the treatment proposed in this paper even if it is an approximation, accounts for the most important features of this model.

\subsection{\label{sec:Nuwro_vs_data} Comparison to experimental data}

In this section, we discuss the impact of modeling correlations in outgoing momenta within NuWro by benchmarking NuWro's performance against recent cross section results from MINER$\ensuremath{\nu}$A experiment \cite{PhysRevD.101.092001, PhysRevLett.121.022504}
obtained with NuMI beam in low energy configuration (peak at $\sim 3$ GeV and a spread of $\sim 2.5$ GeV). The signal is defined as an event with no pions, one muon, and at least one proton, satisfying:

\begin{eqnarray}
    1.5\: \text{GeV}/c < p_{\mu} < 10\: \text{GeV}/c \quad\quad \theta_{\mu} < 20^\circ, \label{eqn:muon-cut} \\
    0.45\: \text{GeV}/c < p_{p} < 1.2\: \text{GeV}/c \quad\quad \theta_{p} < 70^\circ,
    \label{eqn:proton-cut}
\end{eqnarray}
where $p_{\mu}$ and $\theta_{\mu}$ ($p_{p}$ and $\theta_p$) are the muon (proton) momentum and polar angle upon exiting the nucleus with respect to the neutrino direction, see Ref.~\cite{PhysRevLett.121.022504}. 

We tested two different parametrizations of the axial form factor. In one case, we used the MINER$\ensuremath{\nu}$A parametrization ~\cite{osti_1923034, Banerjee:2023hub}, and in the other, dipole parametrization with the axial mass $M_A$ set to $M_A = 1.03$ GeV. We observed that varying the axial form factor parametrization led to $\sim 5\%$ variation in the differential cross section for quasi-elastic events, after applying the selection criteria given in Eqs.~(\ref{eqn:muon-cut}-\ref{eqn:proton-cut}). All results presented in this article are based on the MINER\ensuremath{\nu}A parametrization of the axial form factor. \\

In what follows we show results obtained with the new NuWro MEC model and compare them with results from the old NuWro implementation of the Valencia model.

\begin{figure}[htbp!]
    \centering
    \begin{minipage}[b]{\linewidth}
        \begin{adjustbox}{width=\linewidth}
            \includegraphics[width=\linewidth]{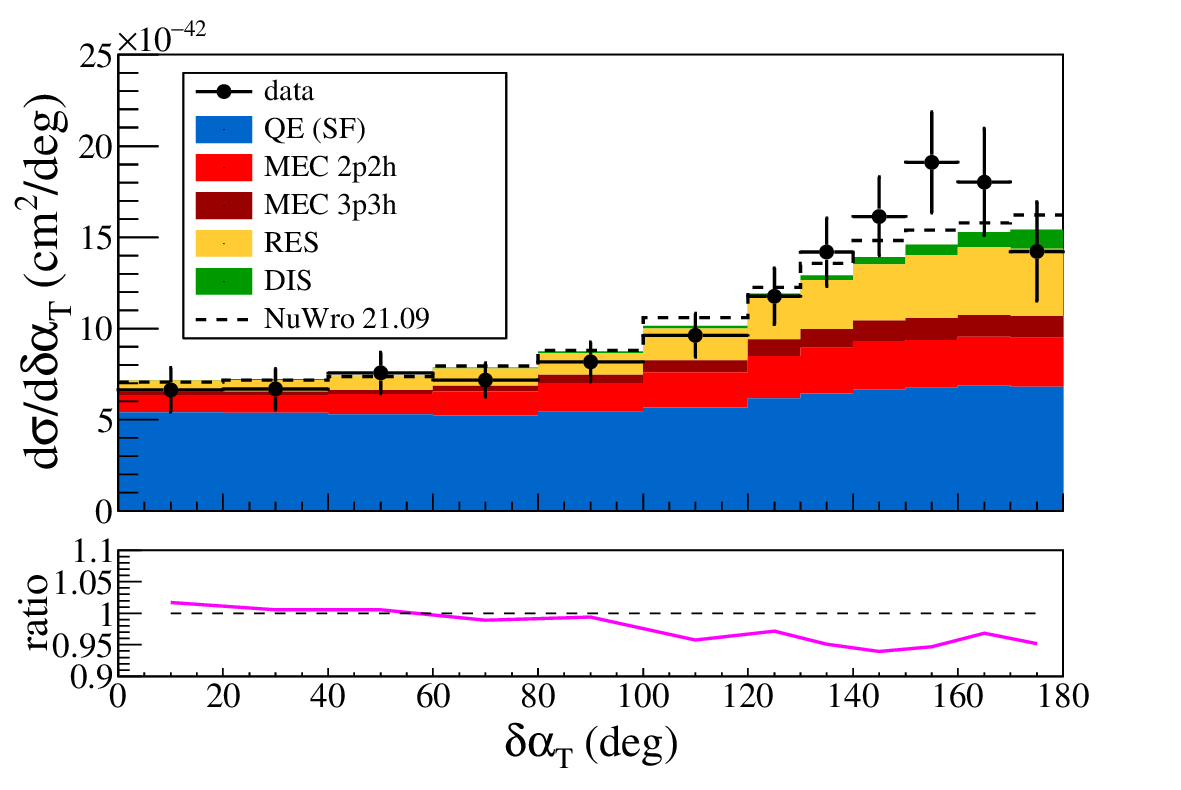}
        \end{adjustbox}
    \end{minipage}
    \begin{minipage}[b]{\linewidth}
        \begin{adjustbox}{width=\linewidth}
            \includegraphics[width=\linewidth]{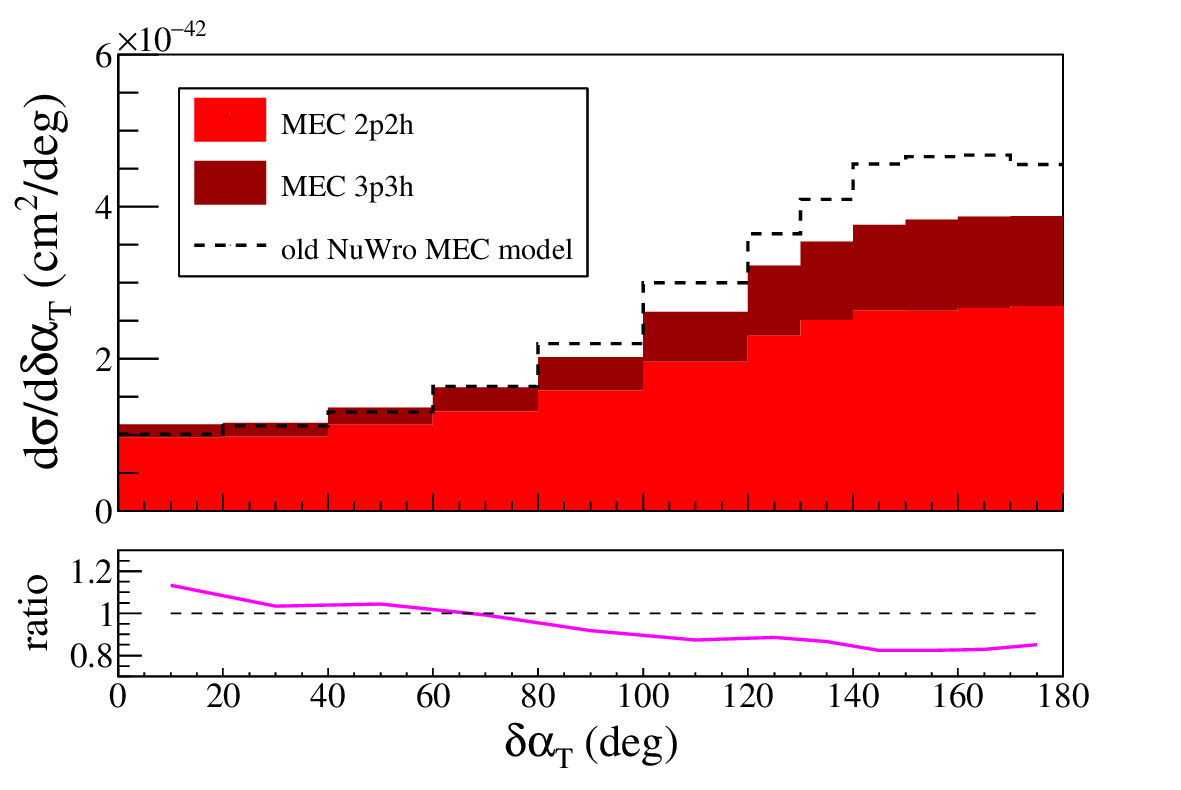}
        \end{adjustbox}
    \end{minipage}
    \caption{\label{fig:dAlphaT_model_comparision} (\textbf{Top}) MINER$\ensuremath{\nu}$A CC$1p0\pi$ differential cross section as a function of $\delta\alpha_{T}$ (single-TKI). For comparison, the old NuWro MEC model (dashed line histogram) is also shown. (\textbf{Bottom}) Contribution of MEC to the differential cross section, modeled using NuWro implementation of 2020 Valencia model. The experimental data are obtained from Ref.\cite{PhysRevD.101.092001} which is an updated version of Ref.\cite{PhysRevLett.121.022504}.  }
\end{figure}

In CC neutrino-nucleus interactions there is an imbalance between the initial neutrino momentum and the sum of final-state lepton and hadron momenta as a result of nuclear effects. This imbalance is denoted by  $\delta\mathbf{p}$. The transverse projection of this imbalance $\delta\mathbf{p}_{T}$ is used to define various useful observables. One of the observables defined using $\delta\mathbf{p}_{T}$ is its direction w.r.t the opposite of the transverse projection of the muon momentum $-\hat{\mathbf{p}}_{T}^{\mu}$ denoted by $\delta\alpha_{T}$. In Fig.~\ref{fig:dAlphaT_model_comparision} we present CC$1p0\pi$ differential cross section as a function of $\delta\alpha_{T}$ (single-TKI). NuWro's CCQE contribution is computed using spectral function formalism. On the top, contributions from individual NuWro interaction modes are shown separately. On the bottom, we show new and old MEC contributions.
In the top panel, the sub-plot (pink curve) shows the ratio of the differential cross section, including all channels (with MEC modeled as in Sect.~\ref{sec:New_MEC_model}), to that of the old NuWro model~\cite{Sobczyk:2012ms}. The bottom panel sub-plot represents the differential cross section ratio for only the MEC channel.\\

The variable $\delta\alpha_T$ is sensitive to Fermi motion (FM) and intranuclear momentum transfer (IMT)~\cite{PhysRevD.101.092001, PhysRevC.94.015503}. The IMT accounts for all kinds of nuclear effects, including correlations in the momenta of outgoing nucleons and the final state interactions. Without nuclear effects, the isotropic nature of FM would yield a uniform distribution of $\delta\alpha_T$~\cite{PhysRevD.101.092001}. Clearly, both the new and old models deviate from this uniformity. In the new MC MEC model, large angular values of $\delta\alpha_T$ are suppressed compared to old model (see the bottom panel of Fig.~\ref{fig:dAlphaT_model_comparision}). Based on the expression for $\delta\alpha_T$ (see~Ref.\cite{PhysRevLett.121.022504}), we infer that events with $|\mathbf{p}^p_T| < |\mathbf{p}^{\mu}_T|$ experience greater suppression in the new MC MEC model. When considering only MEC, the effect of this suppression is approximately $\sim 20\%$. However, this effect reduces to $\sim 5-7\%$ when all other channels are also considered.\\

\begin{figure}[htbp!]
    \centering
    \begin{minipage}[b]{\linewidth}
        \begin{adjustbox}{width=\linewidth}
            \includegraphics[width=\linewidth]{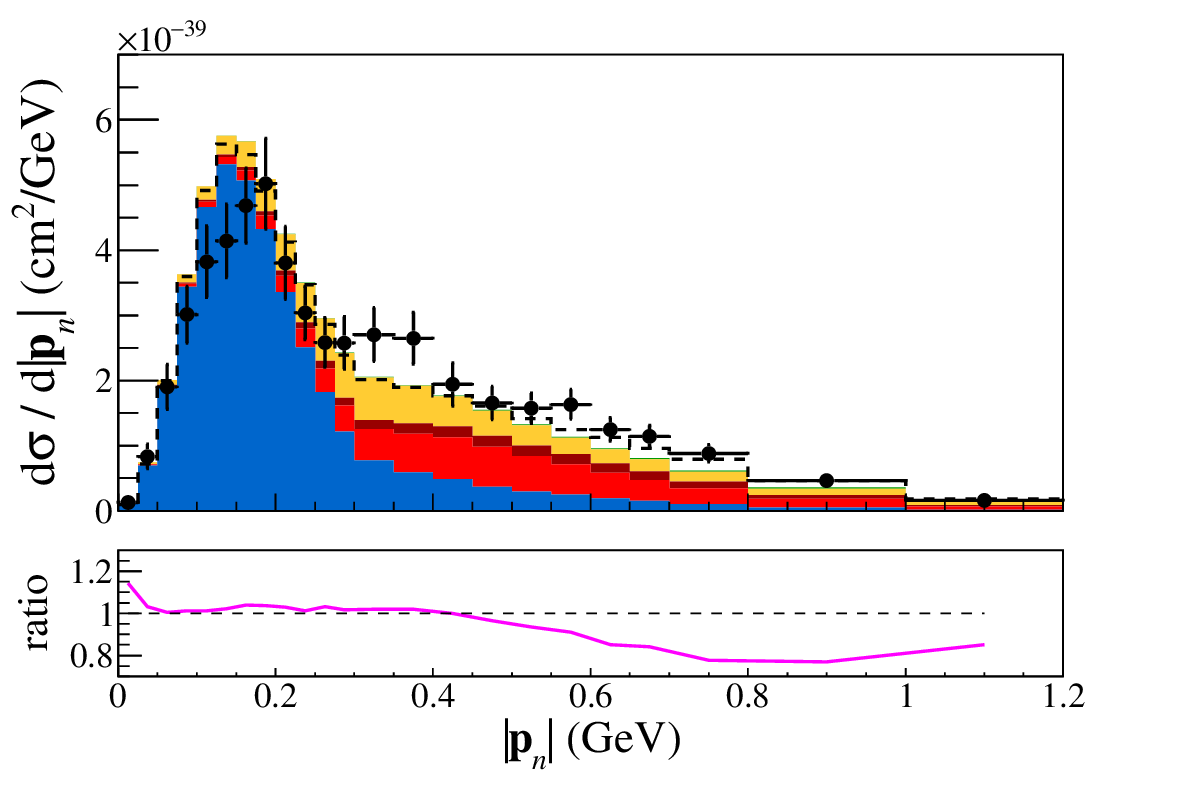}
        \end{adjustbox}
    \end{minipage}
    \begin{minipage}[b]{\linewidth}
        \begin{adjustbox}{width=\linewidth}
            \includegraphics[width=\linewidth]{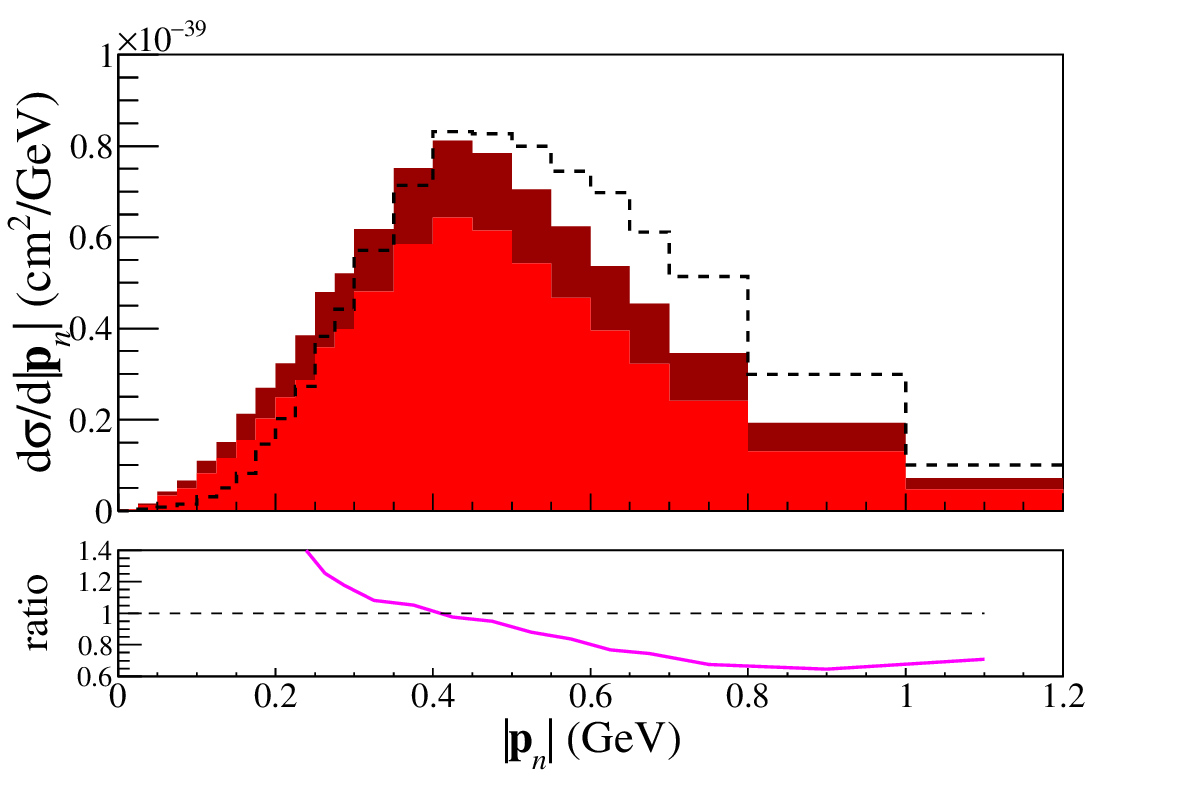}
        \end{adjustbox}
    \end{minipage}
    \caption{\label{fig:recoNeutron_model_comparision} Same as Fig.~\ref{fig:dAlphaT_model_comparision} but for reconstructed neutron momentum.}
\end{figure}

The separation between the IMT and the FM is more clear when one considers the \textit{reconstructed neutron momentum}
~\cite{Furmanski:2016wqo} exploring information also about the longitudinal component of the observed proton. Data points below $|\mathbf{p}_n|<0.25$ GeV constrain the modeling of FM. For $|\mathbf{p}_n|\gtrsim 0.4$ GeV, data points constrain modeling of IMT which arises from contributions like pion absorption and \textit{np-nh}, resulting in an extended tail of the distribution. The transition region from FM to IMT lies between this region. In Fig.~\ref{fig:recoNeutron_model_comparision} we show the differential cross section in the same format as Fig.~\ref{fig:dAlphaT_model_comparision}, but for the reconstructed neutron momentum. 
Here when we consider only MEC (see bottom panel of Fig.~\ref{fig:recoNeutron_model_comparision}, the effects of replacing the old MEC model with the new one produce two distinct results. The correlations in the momenta within the 2020 Valencia model are modeled such that the IMT contribution arising from \textit{2p2h} and \textit{3p3h} are heavily suppressed ($\sim 40\%$) at large values $|\mathbf{p}_n| \gtrsim 0.6$ GeV. If one includes other channels, where both MEC and RES have a dominant contribution, the effect of MEC nucleon correlations is still $\sim 20\%$. The effect is opposite in the FM and transition regions where the MEC cross section is enhanced, most importantly at $|\mathbf{p}_n| \lesssim 0.35$ GeV. However, the effect is insignificant when one adds contributions from other channels as well. \\

\begin{figure*}[htbp!]
    \centering
    \begin{minipage}[b]{0.32\linewidth}
        \begin{adjustbox}{width=1.1\linewidth}
            \includegraphics[width=1.1\linewidth]{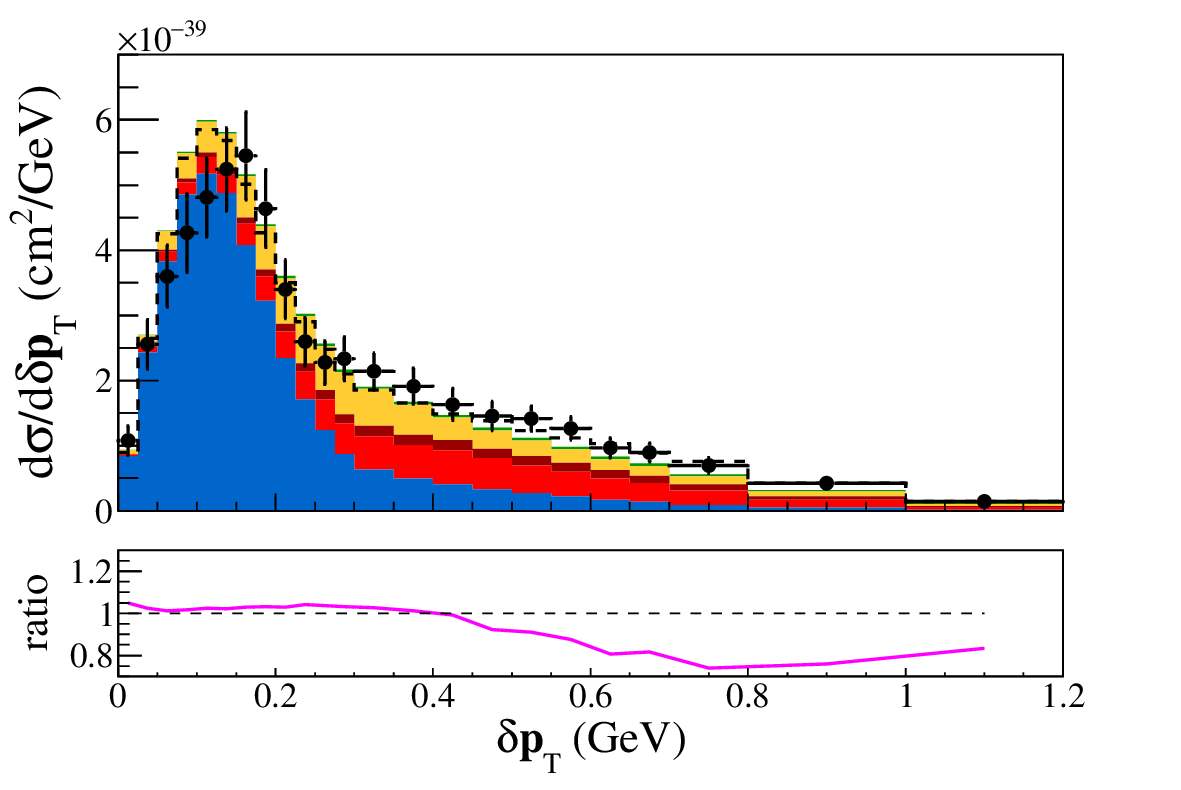}
        \end{adjustbox}
    \end{minipage}
    \begin{minipage}[b]{0.32\linewidth}
        \begin{adjustbox}{width=1.1\linewidth}
            \includegraphics[width=1.1\linewidth]{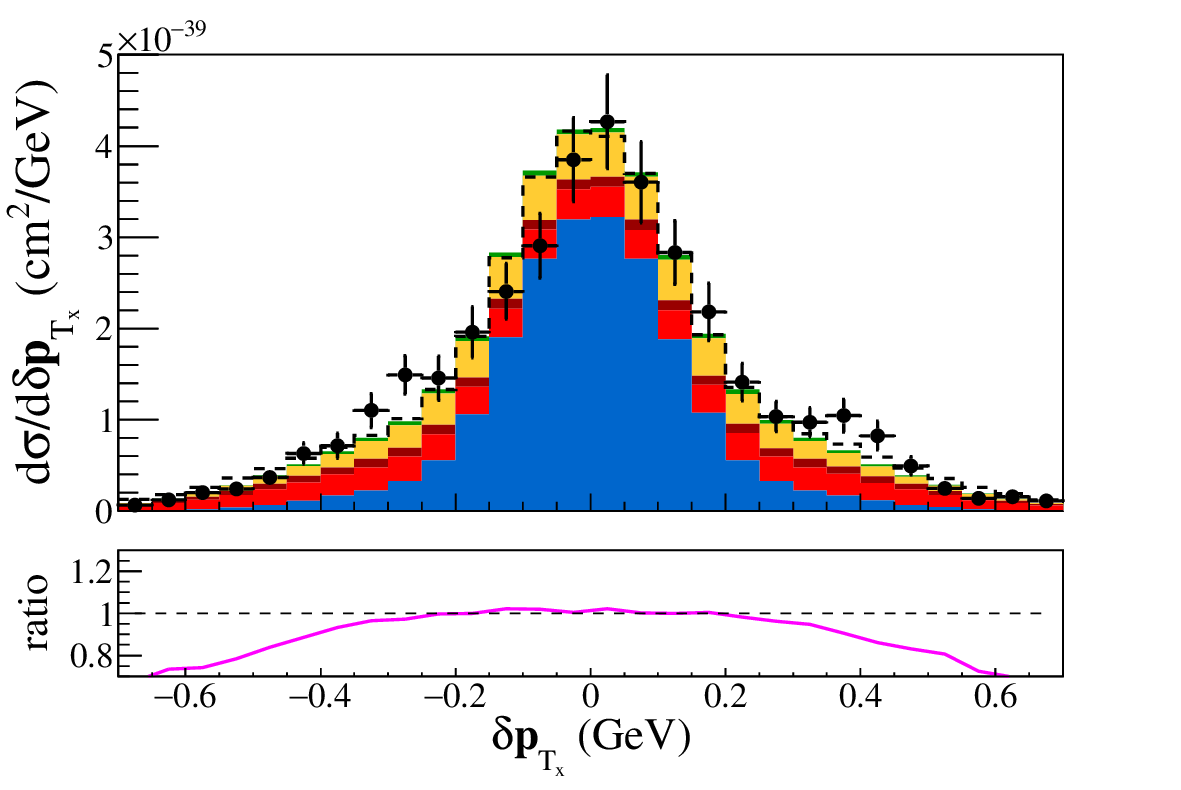}
        \end{adjustbox}
    \end{minipage}
        \begin{minipage}[b]{0.32\linewidth}
        \begin{adjustbox}{width=1.1\linewidth}
            \includegraphics[width=1.1\linewidth]{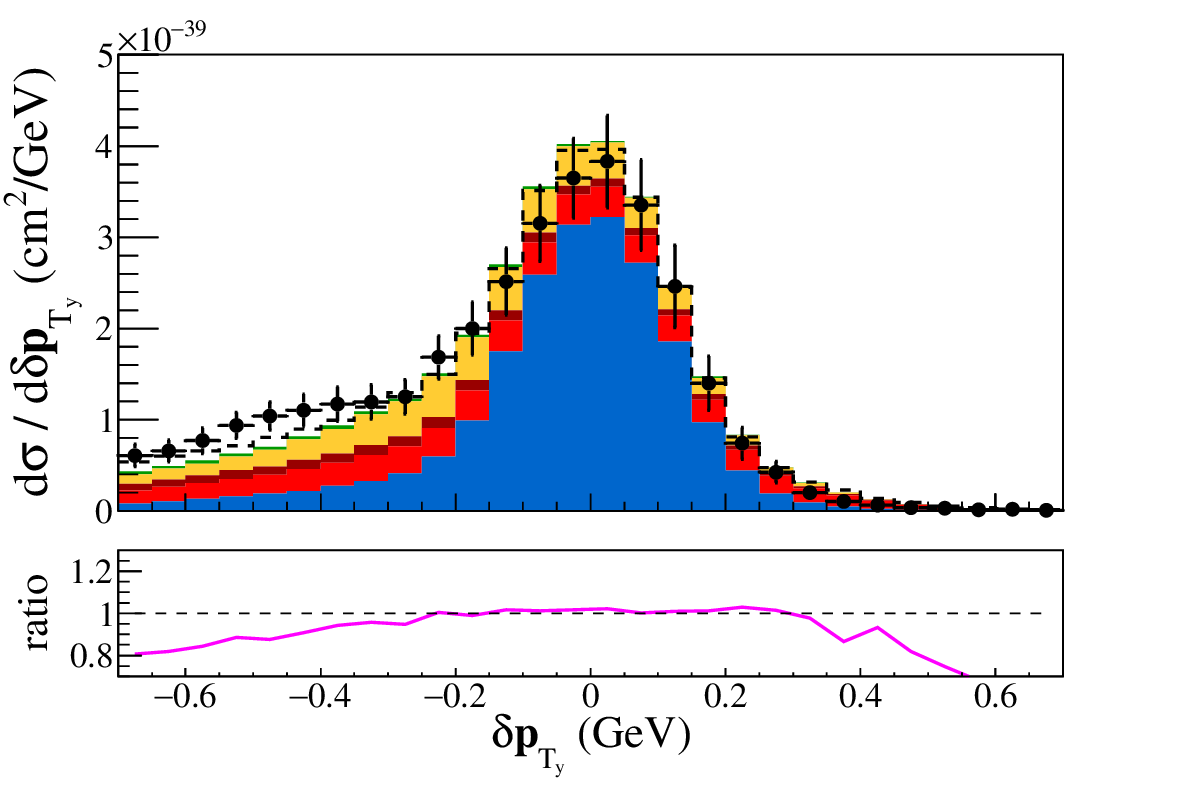}
        \end{adjustbox}
    \end{minipage}
    \begin{minipage}[b]{0.32\linewidth}
        \begin{adjustbox}{width=1.1\linewidth}
            \includegraphics[width=1.1\linewidth]{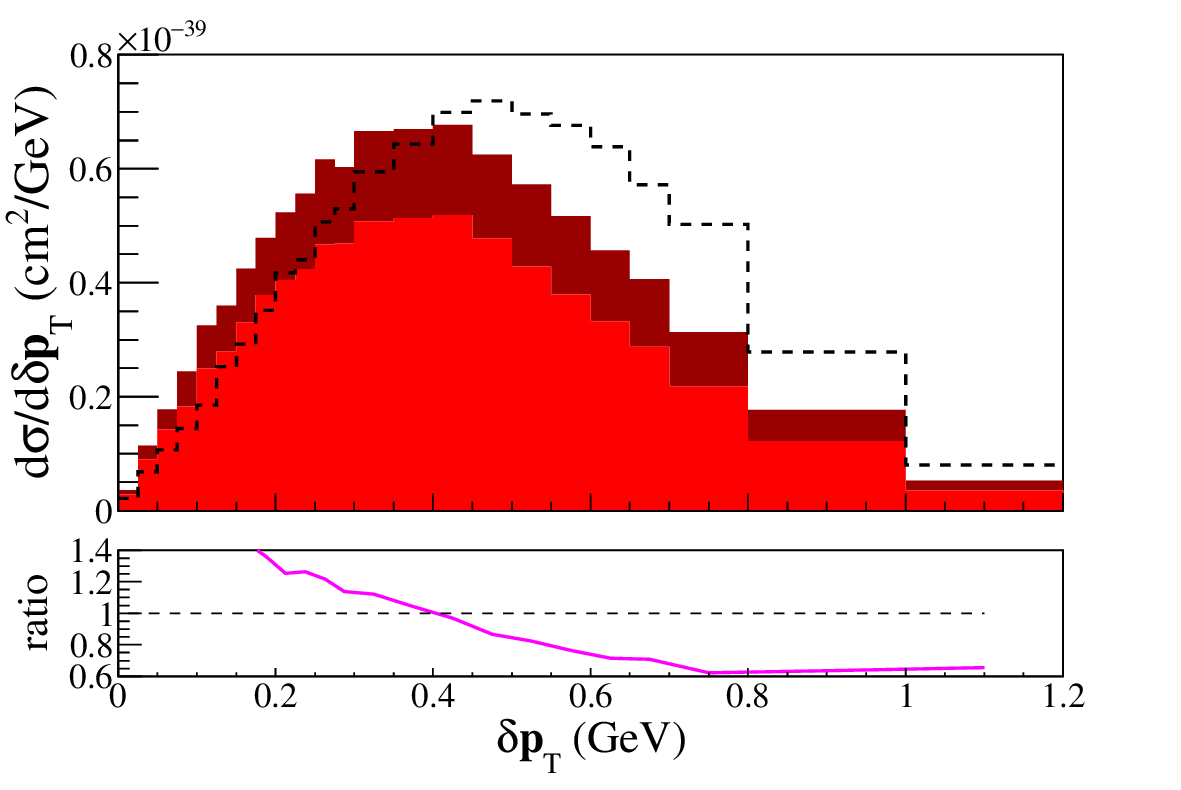}
        \end{adjustbox}
    \end{minipage}
    \begin{minipage}[b]{0.32\linewidth}
        \begin{adjustbox}{width=1.1\linewidth}
            \includegraphics[width=1.1\linewidth]{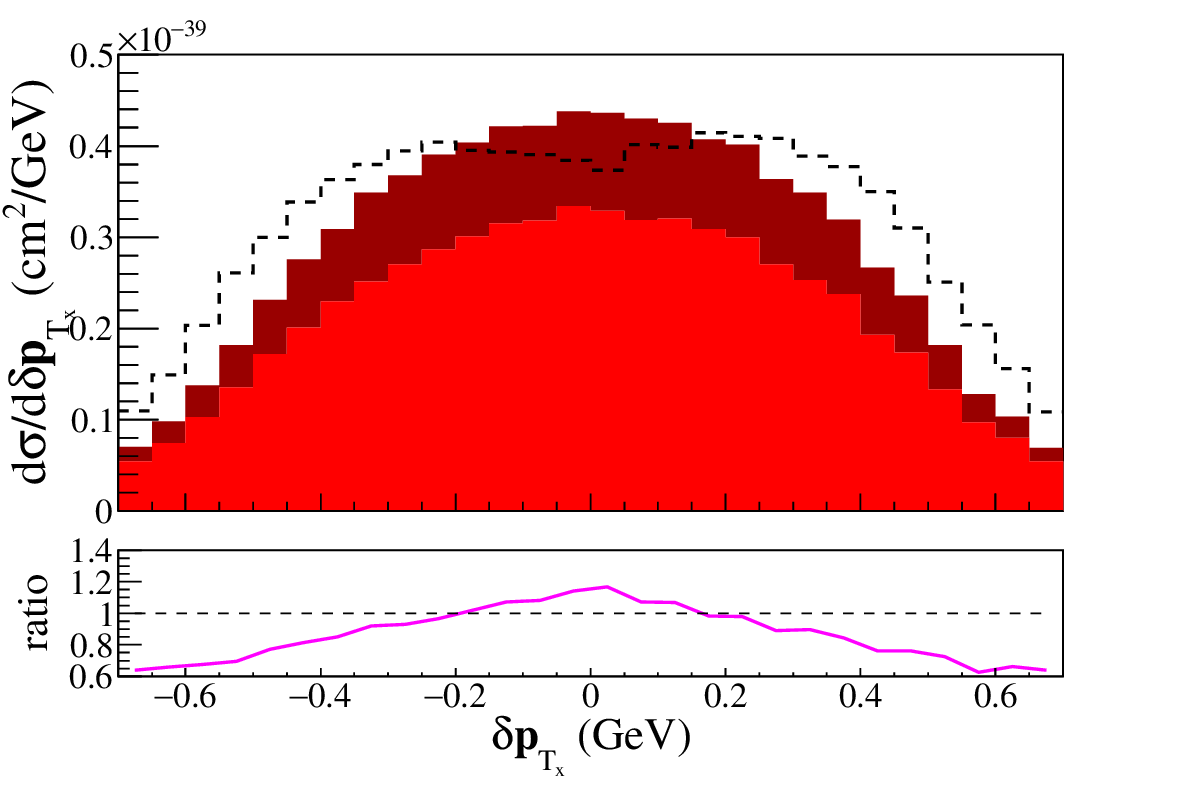}
        \end{adjustbox}
    \end{minipage}
        \begin{minipage}[b]{0.32\linewidth}
        \begin{adjustbox}{width=1.1\linewidth}
            \includegraphics[width=1.1\linewidth]{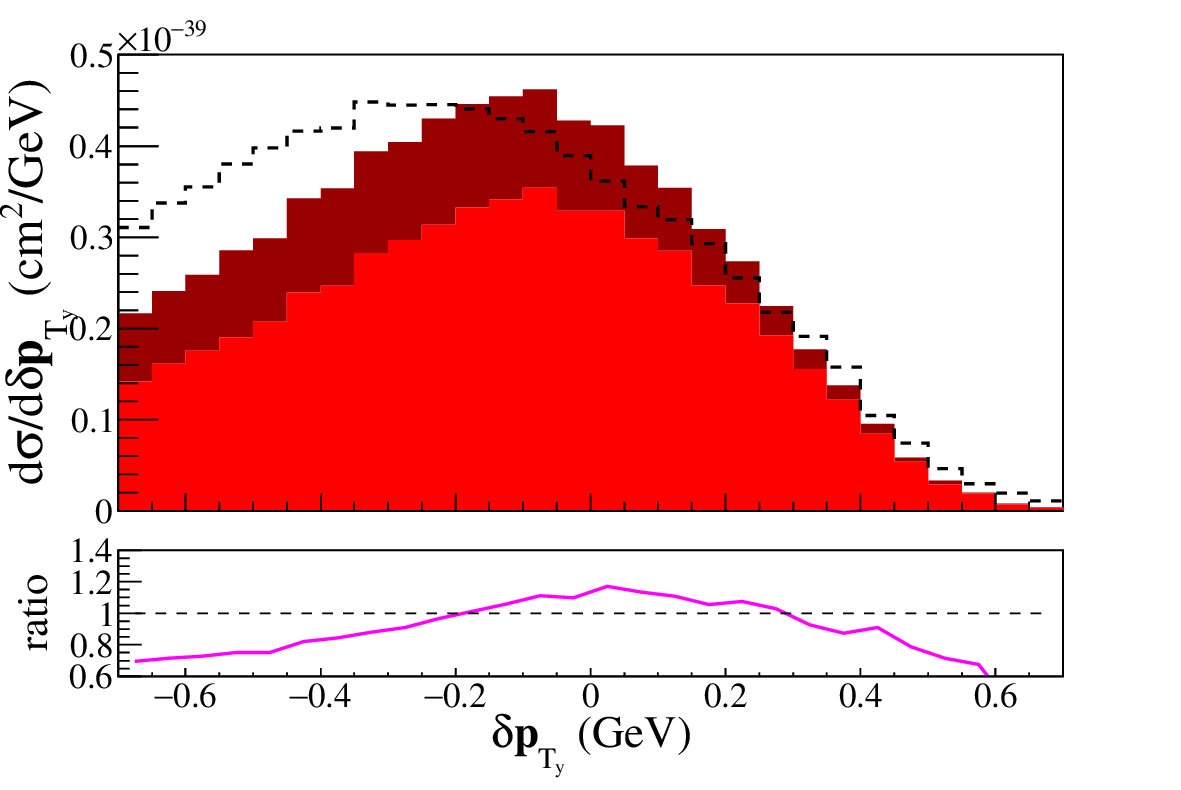}
        \end{adjustbox}
    \end{minipage}
    \caption{\label{fig:dpt_comparision} Same as the figure in  Fig.~\ref{fig:dAlphaT_model_comparision} but for $\delta \mathbf{p}_T$ (in the \textbf{left} panel) and x ,y projections of $\delta \mathbf{p}_T$ in the \textbf{middle} and \textbf{right} panel respectively. The \textbf{bottom} panel shows the MEC contributions within respective observable.}
\end{figure*}

In Fig.~\ref{fig:dpt_comparision} we present the differential cross section as a function of $\delta \mathbf{p}_{T}$ (single-TKI) with $x$,$y$ projections in the middle and bottom panel. Readers are encouraged to refer to Eq.(5) and Eq.(12) of Ref.\cite{PhysRevD.101.092001} for the definition of $\delta\mathbf{p}_{T}$ and its x,y-projections.
The peak region $|\delta\mathbf{p}_T| \leq 0.25$ GeV comes from the FM while the tail $|\delta\mathbf{p}_T| \gtrsim 0.4$ GeV arises from pion absorption and \textit{np-nh}. In Fig.~\ref{fig:dpt_comparision} (in the top left) we see that the tail of the distribution is suppressed. This effect is quite significant $\sim 40\%$ when considering only the MEC channel, while it is still around $\sim 20\%$ when considering the rest of the interaction channels. This is consistent with what we observe for $\delta\alpha_T$ where $|\mathbf{p}^p_T| < |\mathbf{p}^{\mu}_T|$ experience heavier suppression. The difference in modeling MEC is also visible in the x,y-projections of $\delta\mathbf{p}_{T}$. The effect is also seen in the projection of these variables. In right most panel of Fig.~\ref{fig:dpt_comparision} we show the differential cross section for $\delta\mathbf{p}_{T_y}$. 
We observe that the new model reduces the tails and enhances the peak of the distribution (see top and bottom rightmost panel in Fig~\ref{fig:dpt_comparision}). \\

The values of $\chi^2$ per degree of freedom for different observables are reported in the table \ref{tab:table_chi_square}.  
\begin{table}[htbp!]
\caption{\label{tab:table_chi_square} The values of $\chi^2$ per degree of freedom for the
agreement between the NuWro simulations and the MINER$\ensuremath{\nu}$A CC$1p0\pi$ data \cite{PhysRevD.101.092001},\cite{PhysRevLett.121.022504}. Quasi elastic events are generated using the spectral function approach. For RES, momenta assigned to initial nucleons are drawn from effective spectral function (ESF).}
\begin{ruledtabular}
\begin{tabular}{lcc}
\textrm{}&
\textrm{Old MEC model}&
\textrm{New MEC model}\\
\colrule
$d\sigma/d|\mathbf{p}_{p}|$ & 0.99 (24.76/25) & 0.79 (19.72/25) \\
$d\sigma/d\theta_{p}$ & 1.73 (44.88/26) & 1.62 (42.08/26) \\
$d\sigma/d\delta\Phi_{T}$ & 2.69 (61.78/23) & 1.95 (44.82/23) \\
$d\sigma/d\delta|\mathbf{p}_{T}|$ & 3.05 (73.10/24) & 2.82 (67.68/24) \\
$d\sigma/d\delta\alpha_{T}$ & 1.76 (21.13/12) & 1.63 (19.52/12) \\
$d\sigma/d|\mathbf{p}_n|$ & 2.70 (64.85/24) & 3.21 (77.09/24) \\
\end{tabular}
\end{ruledtabular}
\end{table}
We see that NuWro with both new and old MEC models produces very good results for the proton momentum. For most observables, NuWro with the new MEC model performs slightly better. The exception is $|\mathbf{p}_n|$ where the new model works poorly. It is important to remember that in the discussed comparison we test the NuWro model as a whole with its most important ingredients: CCQE, RES, MEC and FSI effects always play an important role. 

Next, we calculate $\chi^2$ per degree of freedom for the distributions of $\delta\mathbf{p}_{T_x}$ and $\delta\mathbf{p}_{T_y}$  in two regions namely $[0.2, 0.2]$ GeV and $[-0.7,0.7]$ GeV. The first region is dominated by CCQE interactions, however, the second region has a significant impact from different interaction mechanisms including MEC. The calculated values are provided in the table \ref{tab:table_chi_square2}.
\begin{table*}[htbp!]
    \centering
    \caption{\label{tab:table_chi_square2} The values of $\chi^2$ per degree of freedom for the
agreement between the NuWro simulations and the MINER$\ensuremath{\nu}$A CC$1p0\pi$ data \cite{PhysRevD.101.092001, PhysRevLett.121.022504} for $\delta\mathbf{p}_{T_x}$ and $\delta\mathbf{p}_{T_y}$.}
    \begin{ruledtabular}
    \begin{tabular}{lcc|cc}
        \textrm{}&
        \multicolumn{2}{c}{Old MEC model}&
        \multicolumn{2}{c}{New MEC model}\\
        \cline{1-5}
         & -0.2 - 0.2 & -0.7 - 0.7 & -0.2 - 0.2 & -0.7 - 0.7 \\
         & (GeV) & (GeV) & (GeV) & (GeV) \\
        \colrule
        $\delta\mathbf{p}_{T_x}$ & 2.21 (17.69/8)  & 2.29 (64.25/28) & 2.85 (22.77/8) & 1.95 (54.49/28) \\
        $\delta\mathbf{p}_{T_y}$ & 0.44 (3.51/8) & 3.23 (90.44/28) & 0.58 (4.63/8) & 2.05 (57.55/28) \\
    \end{tabular}
    \end{ruledtabular}
\end{table*}
We observe that for the wider region $[-0.7, 0.7]$ GeV which has a greater impact from MEC modeling, NuWro performs slightly better with the new MEC model for both $\delta\mathbf{p}_{T_x}$ and $\delta\mathbf{p}_{T_y}$. However, the accumulation of events at the origin of $\delta\mathbf{p}_{T_x}$ and $\delta\mathbf{p}_{T_y}$ observables within the new MEC model results in NuWro performing poorly in this narrow peak region $[-0.2, 0.2]$ GeV (see middle and bottom panel of Fig.~\ref{fig:dpt_comparision}).\\

In conclusion, this work highlights the sensitivity of hadronic observables to how correlations in the momenta of outgoing nucleons are modeled in multinucleon knockout models. It is important that theoretical exclusive models for the \textit{2p2h} and \textit{3p3h} mechanisms are implemented in MC generators used in neutrino oscillation experiments.

\section{\label{sec:outlook} OUTLOOK}
In this article, we introduce a new multinucleon knock-out model implemented in the NuWro Monte Carlo event generator for the MEC channel. The new model focuses on the different treatment of \textit{2p2h} and \textit{3p3h} mechanisms and is based on the theoretical framework developed by the Valencia group. The main goal of the new model is to add correlations in the momenta of outgoing nucleons based on different final-state nucleon pairs within \textit{2p2h} mechanism and asses its impact when compared with experimental data.
The construction of the new MC MEC model involves a nucleon sampling function that requires two adjustable free parameters to simulate correlations between the momenta of outgoing nucleons. We show that new MEC hadronic model is important for developing predictions of hadronic observables, which can influence experimental measurements of neutrino oscillation parameters. We would like to stress that the approach we propose in this paper is quite general and can be adopted in other neutrino Monte Carlo event generators.\\

During this study's final stage, we became aware of the most recent publication by the Valencia group \cite{Sobczyk:2024ecl}. We expect that the versatility of our model will allow us to incorporate these new changes into the Monte Carlo generators using the methodologies outlined in this work (see Sect. \ref{sec:New_MEC_model}). The work in this direction is already ongoing.

\begin{acknowledgments}
We would like to express our gratitude to J.E Sobczyk for sharing the code and meaningful discussions on the technicalities in the 2020 Valencia model. This work is fully (H.P., R.D.B, and B.E.K) or partly (J.T.S, A.M.A, and K.M.G) supported by the National Science Center under the Grant No.UMO-2021/41/B/ST2/02778. K.M.G is also partly supported by the {\sl Excellence Initiative – Research University,  2020-2026} at
the University of Wroclaw.
\end{acknowledgments}

\appendix
\section{\label{sec:derivation} Dynamic Pauli Blocking Condition Derivation}
Let $\beta$ and $\gamma$ be the boost parameters that transform the hadronic system (as described in section \ref{sec:New_MEC_model}) from the lab frame to the center-of-mass (CM) frame of the two outgoing nucleons. Let $(E_{b}^*, \mathbf{p}^*_b)$ and $(E_{b}^{\text{lab}}, \mathbf{p}^{\text{lab}}_b)$ represent the 4-momentum of the \textit{backward} outgoing nucleon in the CM frame and the lab frame, respectively. The Lorentz transformation between the two frames is given by:
\begin{equation}
    \begin{bmatrix}
        E_{b}^{\text{lab}} \\
        p^{\text{lab}}_{\parallel_b} \\
        p^{\text{lab}}_{\perp_b}
    \end{bmatrix} = 
    \begin{bmatrix}
        \gamma & \beta\gamma & 0 \\
        \beta\gamma & \gamma & 0 \\
        0 & 0 & 1
    \end{bmatrix}
    \begin{bmatrix}
        E_{b}^*\\
        p^*_{\parallel_b} \\
        p^*_{\perp_b} 
    \end{bmatrix}
    \label{eqn:lorentz_transformation}
\end{equation}
where $p^{\text{lab}}_{\parallel_b}$, $p^{\text{lab}}_{\perp_b}$, $p^*_{\parallel_b}$, and $p^*_{\perp_b}$ represent the parallel and perpendicular components of $\mathbf{p}^{\text{lab}}_b$ and $\mathbf{p}^*_b$, the 3-momentum of the \textit{backward} nucleon in the lab and CM frames, respectively. The expression for $E^{\text{lab}}_b$ in terms of CM frame quantities is given by
\begin{equation}
    E^{\text{lab}}_b = \gamma E_b^* + \beta\gamma p^*_{\parallel_b} 
\end{equation}

For the \textit{backward} nucleon, $p^*_{\parallel_b} = - p^*_b\cdot|\cos\theta^*|$ since $\cos\theta^* < 0$. The condition $E_{b}^{\text{lab}} > m_b + E_F$ can then be rewritten as 
\begin{eqnarray}
   && E_{b}^{\text{lab}} > m_b + E_F  \\
   \implies && \gamma E_b^* - \beta\gamma p^*_b\cdot|\cos\theta^*|  > m_b + E_F \\
   \implies &&  |\cos\theta^*| < \frac{\gamma E_b^* - m_b - E_{F}}{\beta\gamma p_b^*} 
\end{eqnarray}

Since $|\cos\theta^*| < 1$, we have
\begin{eqnarray}
 \cos\theta^* \in && [-\kappa, 0] \nonumber \\
&& \left\{ \kappa: \kappa = \min\left[1,  \frac{\gamma E_b^* - m_b - E_{F}}{\beta\gamma p_b^*}\right] \right\}
\end{eqnarray}

\begin{figure}[htb!]
    \centering
    \includegraphics[width=0.8\linewidth]{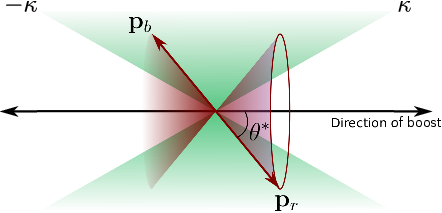}
    \caption{A schematic representation of the Pauli blocking condition in the center-of-mass frame.}
    \label{fig:schematic-pauli-blocking}
\end{figure}

A graphical representation of this restriction is shown in Fig.~\ref{fig:schematic-pauli-blocking}. The shaded green region in the figure illustrates the allowed range of the cosine angle of the outgoing nucleon in the CM frame. The momenta $\mathbf{p}_b$ and $\mathbf{p}_f$ represent the \textit{backward} and \textit{forward} nucleons, respectively. The red cone depicts vectors with the same polar angle $\theta^*$ but different azimuthal angles $\phi^*$. This condition can be used to restrict the \textit{cosine} angle of the nucleon pair in the CM frame due to Pauli blocking of outgoing nucleons.

\section{\label{sec:sampling} Nucleon sampling function}
\subsection{\label{subsec:two_parameter_nucleon_sampling_function}Function with two adjustable parameters}
A general form of the sampling function used in this paper is:

\begin{eqnarray}
            \label{eqn:nucleon-sampling-function}
        f(x)  = \left\{\begin{array}{lr}
        {\displaystyle {l+1\over\displaystyle 2 \kappa (l+1-nP)}\left(1 - P + P\left|\frac{x}{ \kappa}\right|^{l} \right)} &\quad P\in[0,1] \nonumber \\
        \nonumber \\
        {\displaystyle {l+1\over\displaystyle 2 \kappa (l+1+P)}\left(1 + P\left|\frac{x}{\kappa}\right|^{l}\right)} & P\in[-1,0] \nonumber \\
            \end{array}\right. && \\ 
\end{eqnarray}

\begin{figure}[htb]
\begin{minipage}[b]{0.8\linewidth}
\begin{adjustbox}{width=0.8\linewidth}
    \begin{tikzpicture}
        \tikzstyle{every node}=[font=\Large]
        \begin{axis}[xmin=-1, xmax=1, ymin=0, ymax=1.0,
                    axis x line=bottom,
                    axis y line=middle,
                    xlabel=$\cos\theta^*$,
                    ylabel=$f(\cos\theta^*)$]
            \addplot[line width=0.5mm, color=black, samples=50, domain=-1:1]{(0.8 + 0.2*(abs(x))^2)}
            node[right, pos=0.2]{$(0.2,2)$};
            \addplot[line width=0.5mm, color=WildStrawberry, dashed, samples=50, domain=-1:1]{(0.6 + 0.4*(abs(x))^2)}
            node[left, pos=0.7]{$(0.4,2)$};
            \addplot[line width=0.5mm, color=PineGreen, densely dotted, samples=50, domain=-1:1]{(0.3 + 0.7*(abs(x))^2)}
            node[right, pos=0.6]{$(0.7,2)$};
            \addplot[line width=0.5mm, color=blue, samples=50, domain=-1:1]{(0 + 1.0*(abs(x))^2)}
            node[left, pos=0.4]{$(1,2)$};
        \end{axis}
    \end{tikzpicture}
\end{adjustbox}
\end{minipage}
\begin{minipage}[b]{0.8\linewidth}
\begin{adjustbox}{width=0.8\linewidth}
    \begin{tikzpicture}
        \tikzstyle{every node}=[font=\Large]
        \begin{axis}[xmin=-1, xmax=1, ymin=0, ymax=1.0,
                    axis x line=bottom,
                    axis y line=middle,
                    xlabel=$\cos\theta^*$,
                    ylabel=$f(\cos\theta^*)$]
            \addplot[line width=0.5mm, color=black, samples=50, domain=-1:1]{(1 - 0.3*(abs(x))^3)}
            node[right, pos=0.1]{$(-0.3,3)$};
            \addplot[line width=0.5mm, color=WildStrawberry, dashed, samples=50, domain=-1:1]{(1 - 0.6*(abs(x))^3)}
            node[right, pos=0.1]{$(-0.6,3)$};
            \addplot[line width=0.5mm, color=PineGreen, densely dotted, samples=50, domain=-1:1]{(1 - 0.9*(abs(x))^3)}
            node[left, pos=0.8]{$(-0.9,3)$};
        \end{axis}       
    \end{tikzpicture}
\end{adjustbox}
\end{minipage}
    \caption{Nucleon-sampling function (not normalized) as represented in Eq.\eqref{eqn:nucleon-sampling-function} obtained for for various values of $(P,l)$ at $\kappa =1$. \textbf{Top} panel shows $f(\cos\theta^*)$ for different  values of $P>0$ 
    and $l=2$. \textbf{Bottom} panel shows $f(\cos\theta^*)$ for some different $P<0$
    and $n=3$. } 
    \label{fig:nucleon-sampling-function1}
\end{figure}
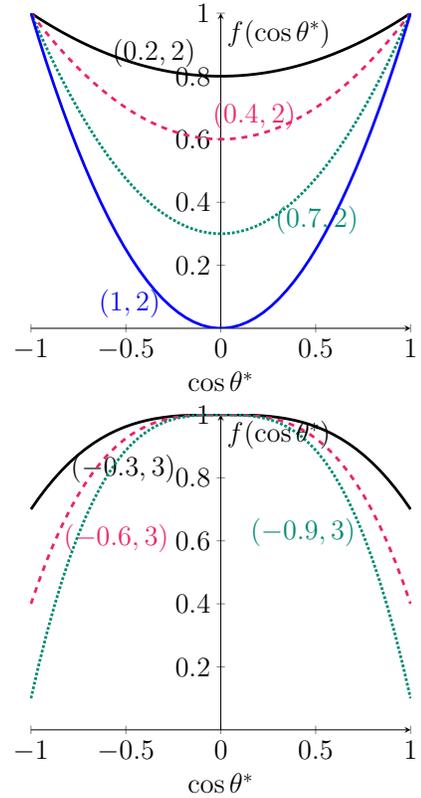

The shape of the function is visualized in Fig.\ref{fig:nucleon-sampling-function1}. 
The function takes two parameters, $P$ and $l$, corresponding to \verb+MEC_cm_direction+ and \verb+MEC_cm_strength+, respectively. The parameter $P \in [-1, 1]$ controls the direction of the outgoing nucleons relative to the boost direction: for $P = -1$, nucleons perpendicular to the boost are preferred, while for $P = 1$, more parallel nucleons are selected. As $P$ increases, the probability of selecting parallel nucleons increases at the expense of perpendicular ones, with $P = 0$ resulting in a flat distribution. The parameter $l \in \mathbb{N}$ controls the severity of the selection: for $P > 0$, increasing $l$ suppresses perpendicular nucleons and enhances parallel ones, while for $P < 0$, it suppresses parallel nucleons and favors perpendicular or slightly angled ones. In the lab frame, for $P \to 1$ and $l > 1$, nucleons parallel to the boost in the center-of-mass frame lead to one high-momentum and one low-momentum nucleon. For $P \to -1$ and $l > 1$, nucleons perpendicular to the boost result in two nucleons with similar momenta in the lab frame.
\subsection{\label{subsec:Alternatives_to_nucleon_samping_function}Function with five adjustable parameters}

In the course of our study
we also tried other methods of defining the nucleon-sampling function. For example, we considered a family of functions with five free parameters, namely $a,b,c,r$ and $s$ with $r < s$. The functional form of such a function is given as 
\begin{eqnarray}
     \label{eqn:five-parameter-nucleon-sampling-function} 
    && f(x)  = N(\kappa)\cdot \left\{ \begin{array}{lr}
        {\displaystyle a + x{b-a\over\displaystyle r}} &\quad  0 \leq x < r  \\
        & \nonumber \\
        {\displaystyle b + (x-r){c-b\over\displaystyle s-r}} &\quad  r \leq x < s \\ 
        & \nonumber \\
        {\displaystyle c + (x-s){1-c \over\displaystyle \kappa-s} }&\quad s \leq x \leq \kappa 
    \end{array} \right.  \\
    && \nonumber \\
    && \text{where}\: N(\kappa) = \frac{2}{(a+b)r + (b+c)(s-r) + (c+1)(\kappa-s)} \nonumber 
\end{eqnarray}

The function in Eq.\eqref{eqn:five-parameter-nucleon-sampling-function} allows to model ``hat''-shaped type distribution which is not possible for any values of $(P,n)$ in Eq.\eqref{eqn:nucleon-sampling-function}. The parameter space was defined for the three final state pairs are \(\{a,b,c\in [0,1] \: \text{and} \: r,s (r<s) \in [0, \kappa]. \} \) 
We found that this type of nucleon-sampling function produces a $\tilde{\chi}^2_{\text{Best-fit}}$ very close to that produced by two-parameters nucleon-sampling function for $pp$ nucleon phase space. The shape of $f(x)$ produced by the optimal parameters is very similar to the shape produced by the two-parameter nucleon-sampling function.

\bibliography{main}

\providecommand{\noopsort}[1]{}\providecommand{\singleletter}[1]{#1}%
\begin{thebibliography}{43}%
\makeatletter
\providecommand \@ifxundefined [1]{%
 \@ifx{#1\undefined}
}%
\providecommand \@ifnum [1]{%
 \ifnum #1\expandafter \@firstoftwo
 \else \expandafter \@secondoftwo
 \fi
}%
\providecommand \@ifx [1]{%
 \ifx #1\expandafter \@firstoftwo
 \else \expandafter \@secondoftwo
 \fi
}%
\providecommand \natexlab [1]{#1}%
\providecommand \enquote  [1]{``#1''}%
\providecommand \bibnamefont  [1]{#1}%
\providecommand \bibfnamefont [1]{#1}%
\providecommand \citenamefont [1]{#1}%
\providecommand \href@noop [0]{\@secondoftwo}%
\providecommand \href [0]{\begingroup \@sanitize@url \@href}%
\providecommand \@href[1]{\@@startlink{#1}\@@href}%
\providecommand \@@href[1]{\endgroup#1\@@endlink}%
\providecommand \@sanitize@url [0]{\catcode `\\12\catcode `\$12\catcode `\&12\catcode `\#12\catcode `\^12\catcode `\_12\catcode `\%12\relax}%
\providecommand \@@startlink[1]{}%
\providecommand \@@endlink[0]{}%
\providecommand \url  [0]{\begingroup\@sanitize@url \@url }%
\providecommand \@url [1]{\endgroup\@href {#1}{\urlprefix }}%
\providecommand \urlprefix  [0]{URL }%
\providecommand \Eprint [0]{\href }%
\providecommand \doibase [0]{https://doi.org/}%
\providecommand \selectlanguage [0]{\@gobble}%
\providecommand \bibinfo  [0]{\@secondoftwo}%
\providecommand \bibfield  [0]{\@secondoftwo}%
\providecommand \translation [1]{[#1]}%
\providecommand \BibitemOpen [0]{}%
\providecommand \bibitemStop [0]{}%
\providecommand \bibitemNoStop [0]{.\EOS\space}%
\providecommand \EOS [0]{\spacefactor3000\relax}%
\providecommand \BibitemShut  [1]{\csname bibitem#1\endcsname}%
\let\auto@bib@innerbib\@empty
\bibitem [{\citenamefont {Abe}\ \emph {et~al.}(2015)\citenamefont {Abe} \emph {et~al.}}]{Hyper-KamiokandeProto-:2015xww}%
  \BibitemOpen
  \bibfield  {author} {\bibinfo {author} {\bibfnamefont {K.}~\bibnamefont {Abe}} \emph {et~al.} (\bibinfo {collaboration} {Hyper-Kamiokande Proto-}),\ }\bibfield  {title} {\bibinfo {title} {{Physics potential of a long-baseline neutrino oscillation experiment using a J-PARC neutrino beam and Hyper-Kamiokande}},\ }\href {https://doi.org/10.1093/ptep/ptv061} {\bibfield  {journal} {\bibinfo  {journal} {PTEP}\ }\textbf {\bibinfo {volume} {2015}},\ \bibinfo {pages} {053C02} (\bibinfo {year} {2015})},\ \Eprint {https://arxiv.org/abs/1502.05199} {arXiv:1502.05199 [hep-ex]} \BibitemShut {NoStop}%
\bibitem [{\citenamefont {Acciarri}\ \emph {et~al.}(2016)\citenamefont {Acciarri} \emph {et~al.}}]{DUNE:2016hlj}%
  \BibitemOpen
  \bibfield  {author} {\bibinfo {author} {\bibfnamefont {R.}~\bibnamefont {Acciarri}} \emph {et~al.} (\bibinfo {collaboration} {DUNE}),\ }\bibfield  {title} {\bibinfo {title} {{Long-Baseline Neutrino Facility (LBNF) and Deep Underground Neutrino Experiment (DUNE)}: {Conceptual Design Report, Volume 1: The LBNF and DUNE Projects}},\ }\href@noop {} {\  (\bibinfo {year} {2016})},\ \Eprint {https://arxiv.org/abs/1601.05471} {arXiv:1601.05471 [physics.ins-det]} \BibitemShut {NoStop}%
\bibitem [{\citenamefont {Alvarez-Ruso}\ \emph {et~al.}(2018)\citenamefont {Alvarez-Ruso} \emph {et~al.}}]{NuSTEC:2017hzk}%
  \BibitemOpen
  \bibfield  {author} {\bibinfo {author} {\bibfnamefont {L.}~\bibnamefont {Alvarez-Ruso}} \emph {et~al.} (\bibinfo {collaboration} {NuSTEC}),\ }\bibfield  {title} {\bibinfo {title} {{NuSTEC White Paper: Status and challenges of neutrino\textendash{}nucleus scattering}},\ }\href {https://doi.org/10.1016/j.ppnp.2018.01.006} {\bibfield  {journal} {\bibinfo  {journal} {Prog. Part. Nucl. Phys.}\ }\textbf {\bibinfo {volume} {100}},\ \bibinfo {pages} {1} (\bibinfo {year} {2018})},\ \Eprint {https://arxiv.org/abs/1706.03621} {arXiv:1706.03621 [hep-ph]} \BibitemShut {NoStop}%
\bibitem [{\citenamefont {Gallagher}\ and\ \citenamefont {Hayato}(2024)}]{GallagherHayato}%
  \BibitemOpen
  \bibfield  {author} {\bibinfo {author} {\bibfnamefont {H.}~\bibnamefont {Gallagher}}\ and\ \bibinfo {author} {\bibfnamefont {Y.}~\bibnamefont {Hayato}} (\bibinfo {collaboration} {in Particle Data Group}),\ }\bibfield  {title} {\bibinfo {title} {{Review of particle physics}},\ }\href {https://doi.org/10.1103/PhysRevD.110.030001} {\bibfield  {journal} {\bibinfo  {journal} {Phys. Rev. D}\ }\textbf {\bibinfo {volume} {110}},\ \bibinfo {pages} {030001} (\bibinfo {year} {2024})}\BibitemShut {NoStop}%
\bibitem [{\citenamefont {Gonzalez-Garcia}\ and\ \citenamefont {Yokoyama}(2024)}]{Gonzalez-GarciaYokoyama}%
  \BibitemOpen
  \bibfield  {author} {\bibinfo {author} {\bibfnamefont {M.}~\bibnamefont {Gonzalez-Garcia}}\ and\ \bibinfo {author} {\bibfnamefont {M.}~\bibnamefont {Yokoyama}} (\bibinfo {collaboration} {in Particle Data Group}),\ }\bibfield  {title} {\bibinfo {title} {{Review of particle physics}},\ }\href {https://doi.org/10.1103/PhysRevD.110.030001} {\bibfield  {journal} {\bibinfo  {journal} {Phys. Rev. D}\ }\textbf {\bibinfo {volume} {110}},\ \bibinfo {pages} {030001} (\bibinfo {year} {2024})}\BibitemShut {NoStop}%
\bibitem [{\citenamefont {Abe}\ \emph {et~al.}(2022)\citenamefont {Abe} \emph {et~al.}}]{Super-Kamiokande:2021the}%
  \BibitemOpen
  \bibfield  {author} {\bibinfo {author} {\bibfnamefont {K.}~\bibnamefont {Abe}} \emph {et~al.} (\bibinfo {collaboration} {Super-Kamiokande}),\ }\bibfield  {title} {\bibinfo {title} {{First gadolinium loading to Super-Kamiokande}},\ }\href {https://doi.org/10.1016/j.nima.2021.166248} {\bibfield  {journal} {\bibinfo  {journal} {Nucl. Instrum. Meth. A}\ }\textbf {\bibinfo {volume} {1027}},\ \bibinfo {pages} {166248} (\bibinfo {year} {2022})},\ \Eprint {https://arxiv.org/abs/2109.00360} {arXiv:2109.00360 [physics.ins-det]} \BibitemShut {NoStop}%
\bibitem [{\citenamefont {Llewellyn~Smith}(1972)}]{LlewellynSmith:1971uhs}%
  \BibitemOpen
  \bibfield  {author} {\bibinfo {author} {\bibfnamefont {C.~H.}\ \bibnamefont {Llewellyn~Smith}},\ }\bibfield  {title} {\bibinfo {title} {{Neutrino Reactions at Accelerator Energies}},\ }\href {https://doi.org/10.1016/0370-1573(72)90010-5} {\bibfield  {journal} {\bibinfo  {journal} {Phys. Rept.}\ }\textbf {\bibinfo {volume} {3}},\ \bibinfo {pages} {261} (\bibinfo {year} {1972})}\BibitemShut {NoStop}%
\bibitem [{\citenamefont {Collaboration}(2021)}]{ETM2021}%
  \BibitemOpen
  \bibfield  {author} {\bibinfo {author} {\bibfnamefont {E.}~\bibnamefont {Collaboration}},\ }\bibfield  {title} {\bibinfo {title} {Axial form factors of the nucleon from lattice qcd},\ }\href {https://doi.org/10.1007/JHEP06(2021)123} {\bibfield  {journal} {\bibinfo  {journal} {Journal of High Energy Physics}\ }\textbf {\bibinfo {volume} {2021}},\ \bibinfo {pages} {123} (\bibinfo {year} {2021})}\BibitemShut {NoStop}%
\bibitem [{\citenamefont {Aguilar-Arevalo}\ \emph {et~al.}(2010)\citenamefont {Aguilar-Arevalo} \emph {et~al.}}]{MiniBooNE:2010bsu}%
  \BibitemOpen
  \bibfield  {author} {\bibinfo {author} {\bibfnamefont {A.~A.}\ \bibnamefont {Aguilar-Arevalo}} \emph {et~al.} (\bibinfo {collaboration} {MiniBooNE}),\ }\bibfield  {title} {\bibinfo {title} {{First Measurement of the Muon Neutrino Charged Current Quasielastic Double Differential Cross Section}},\ }\href {https://doi.org/10.1103/PhysRevD.81.092005} {\bibfield  {journal} {\bibinfo  {journal} {Phys. Rev. D}\ }\textbf {\bibinfo {volume} {81}},\ \bibinfo {pages} {092005} (\bibinfo {year} {2010})},\ \Eprint {https://arxiv.org/abs/1002.2680} {arXiv:1002.2680 [hep-ex]} \BibitemShut {NoStop}%
\bibitem [{\citenamefont {McFarland}\ and\ \citenamefont {Cai}(2020)}]{McFarland2020}%
  \BibitemOpen
  \bibfield  {author} {\bibinfo {author} {\bibfnamefont {K.~S.}\ \bibnamefont {McFarland}}\ and\ \bibinfo {author} {\bibfnamefont {T.}~\bibnamefont {Cai}},\ }\bibfield  {title} {\bibinfo {title} {Neutrino–nucleus cross sections for oscillation experiments},\ }\href {https://doi.org/10.1146/annurev-nucl-030220-065259} {\bibfield  {journal} {\bibinfo  {journal} {Annual Review of Nuclear and Particle Science}\ }\textbf {\bibinfo {volume} {70}},\ \bibinfo {pages} {269} (\bibinfo {year} {2020})}\BibitemShut {NoStop}%
\bibitem [{\citenamefont {Marteau}(1999)}]{Marteau:1999kt}%
  \BibitemOpen
  \bibfield  {author} {\bibinfo {author} {\bibfnamefont {J.}~\bibnamefont {Marteau}},\ }\bibfield  {title} {\bibinfo {title} {{Effects of the nuclear correlations on the neutrino oxygen interactions}},\ }\href {https://doi.org/10.1007/s100500050274} {\bibfield  {journal} {\bibinfo  {journal} {Eur. Phys. J. A}\ }\textbf {\bibinfo {volume} {5}},\ \bibinfo {pages} {183} (\bibinfo {year} {1999})},\ \Eprint {https://arxiv.org/abs/hep-ph/9902210} {arXiv:hep-ph/9902210} \BibitemShut {NoStop}%
\bibitem [{\citenamefont {Martini}\ \emph {et~al.}(2009)\citenamefont {Martini}, \citenamefont {Ericson}, \citenamefont {Chanfray},\ and\ \citenamefont {Marteau}}]{Martini:2009uj}%
  \BibitemOpen
  \bibfield  {author} {\bibinfo {author} {\bibfnamefont {M.}~\bibnamefont {Martini}}, \bibinfo {author} {\bibfnamefont {M.}~\bibnamefont {Ericson}}, \bibinfo {author} {\bibfnamefont {G.}~\bibnamefont {Chanfray}},\ and\ \bibinfo {author} {\bibfnamefont {J.}~\bibnamefont {Marteau}},\ }\bibfield  {title} {\bibinfo {title} {{A Unified approach for nucleon knock-out, coherent and incoherent pion production in neutrino interactions with nuclei}},\ }\href {https://doi.org/10.1103/PhysRevC.80.065501} {\bibfield  {journal} {\bibinfo  {journal} {Phys. Rev. C}\ }\textbf {\bibinfo {volume} {80}},\ \bibinfo {pages} {065501} (\bibinfo {year} {2009})},\ \Eprint {https://arxiv.org/abs/0910.2622} {arXiv:0910.2622 [nucl-th]} \BibitemShut {NoStop}%
\bibitem [{\citenamefont {Nieves}\ \emph {et~al.}(2011)\citenamefont {Nieves}, \citenamefont {Ruiz~Simo},\ and\ \citenamefont {Vicente~Vacas}}]{Nieves:2011pp}%
  \BibitemOpen
  \bibfield  {author} {\bibinfo {author} {\bibfnamefont {J.}~\bibnamefont {Nieves}}, \bibinfo {author} {\bibfnamefont {I.}~\bibnamefont {Ruiz~Simo}},\ and\ \bibinfo {author} {\bibfnamefont {M.~J.}\ \bibnamefont {Vicente~Vacas}},\ }\bibfield  {title} {\bibinfo {title} {{Inclusive Charged--Current Neutrino--Nucleus Reactions}},\ }\href {https://doi.org/10.1103/PhysRevC.83.045501} {\bibfield  {journal} {\bibinfo  {journal} {Phys. Rev. C}\ }\textbf {\bibinfo {volume} {83}},\ \bibinfo {pages} {045501} (\bibinfo {year} {2011})},\ \Eprint {https://arxiv.org/abs/1102.2777} {arXiv:1102.2777 [hep-ph]} \BibitemShut {NoStop}%
\bibitem [{\citenamefont {Megias}\ \emph {et~al.}(2016)\citenamefont {Megias}, \citenamefont {Amaro}, \citenamefont {Barbaro}, \citenamefont {Caballero}, \citenamefont {Donnelly},\ and\ \citenamefont {Ruiz~Simo}}]{Megias:2016fjk}%
  \BibitemOpen
  \bibfield  {author} {\bibinfo {author} {\bibfnamefont {G.~D.}\ \bibnamefont {Megias}}, \bibinfo {author} {\bibfnamefont {J.~E.}\ \bibnamefont {Amaro}}, \bibinfo {author} {\bibfnamefont {M.~B.}\ \bibnamefont {Barbaro}}, \bibinfo {author} {\bibfnamefont {J.~A.}\ \bibnamefont {Caballero}}, \bibinfo {author} {\bibfnamefont {T.~W.}\ \bibnamefont {Donnelly}},\ and\ \bibinfo {author} {\bibfnamefont {I.}~\bibnamefont {Ruiz~Simo}},\ }\bibfield  {title} {\bibinfo {title} {{Charged-current neutrino-nucleus reactions within the superscaling meson-exchange current approach}},\ }\href {https://doi.org/10.1103/PhysRevD.94.093004} {\bibfield  {journal} {\bibinfo  {journal} {Phys. Rev. D}\ }\textbf {\bibinfo {volume} {94}},\ \bibinfo {pages} {093004} (\bibinfo {year} {2016})},\ \Eprint {https://arxiv.org/abs/1607.08565} {arXiv:1607.08565 [nucl-th]} \BibitemShut {NoStop}%
\bibitem [{\citenamefont {Sobczyk}(2012)}]{Sobczyk:2012ms}%
  \BibitemOpen
  \bibfield  {author} {\bibinfo {author} {\bibfnamefont {J.~T.}\ \bibnamefont {Sobczyk}},\ }\bibfield  {title} {\bibinfo {title} {{Multinucleon Ejection Model for Meson Exchange Current Neutrino Interactions}},\ }\href {https://doi.org/10.1103/PhysRevC.86.015504} {\bibfield  {journal} {\bibinfo  {journal} {Phys. Rev. C}\ }\textbf {\bibinfo {volume} {86}},\ \bibinfo {pages} {015504} (\bibinfo {year} {2012})},\ \Eprint {https://arxiv.org/abs/1201.3673} {arXiv:1201.3673 [hep-ph]} \BibitemShut {NoStop}%
\bibitem [{\citenamefont {Hayato}(2009)}]{Hayato:2009zz}%
  \BibitemOpen
  \bibfield  {author} {\bibinfo {author} {\bibfnamefont {Y.}~\bibnamefont {Hayato}},\ }\bibfield  {title} {\bibinfo {title} {{A neutrino interaction simulation program library NEUT}},\ }\href@noop {} {\bibfield  {journal} {\bibinfo  {journal} {Acta Phys. Polon. B}\ }\textbf {\bibinfo {volume} {40}},\ \bibinfo {pages} {2477} (\bibinfo {year} {2009})}\BibitemShut {NoStop}%
\bibitem [{\citenamefont {Andreopoulos}\ \emph {et~al.}(2009)\citenamefont {Andreopoulos}, \citenamefont {Agafonova}, \citenamefont {Albrecht} \emph {et~al.}}]{andreopoulos2009genie}%
  \BibitemOpen
  \bibfield  {author} {\bibinfo {author} {\bibfnamefont {C.}~\bibnamefont {Andreopoulos}}, \bibinfo {author} {\bibfnamefont {M.}~\bibnamefont {Agafonova}}, \bibinfo {author} {\bibfnamefont {G.}~\bibnamefont {Albrecht}}, \emph {et~al.},\ }\bibfield  {title} {\bibinfo {title} {Genie: A neutrino monte carlo generator},\ }\href {https://doi.org/10.1016/j.nima.2008.10.047} {\bibfield  {journal} {\bibinfo  {journal} {Nucl. Instrum. Meth. A}\ }\textbf {\bibinfo {volume} {600}},\ \bibinfo {pages} {415} (\bibinfo {year} {2009})}\BibitemShut {NoStop}%
\bibitem [{\citenamefont {Sobczyk}\ \emph {et~al.}(2020)\citenamefont {Sobczyk}, \citenamefont {Nieves},\ and\ \citenamefont {S\'anchez}}]{Sobczyk:2020dkn}%
  \BibitemOpen
  \bibfield  {author} {\bibinfo {author} {\bibfnamefont {J.~E.}\ \bibnamefont {Sobczyk}}, \bibinfo {author} {\bibfnamefont {J.}~\bibnamefont {Nieves}},\ and\ \bibinfo {author} {\bibfnamefont {F.}~\bibnamefont {S\'anchez}},\ }\bibfield  {title} {\bibinfo {title} {{Exclusive-final-state hadron observables from neutrino-nucleus multinucleon knockout}},\ }\href {https://doi.org/10.1103/PhysRevC.102.024601} {\bibfield  {journal} {\bibinfo  {journal} {Phys. Rev. C}\ }\textbf {\bibinfo {volume} {102}},\ \bibinfo {pages} {024601} (\bibinfo {year} {2020})},\ \Eprint {https://arxiv.org/abs/2002.08302} {arXiv:2002.08302 [nucl-th]} \BibitemShut {NoStop}%
\bibitem [{\citenamefont {Sobczyk}\ and\ \citenamefont {Nieves}(2024)}]{Sobczyk:2024ecl}%
  \BibitemOpen
  \bibfield  {author} {\bibinfo {author} {\bibfnamefont {J.~E.}\ \bibnamefont {Sobczyk}}\ and\ \bibinfo {author} {\bibfnamefont {J.}~\bibnamefont {Nieves}},\ }\bibfield  {title} {\bibinfo {title} {{Neutrino and antineutrino charged-current multi-nucleon cross sections revisited}},\ }\href@noop {} {\  (\bibinfo {year} {2024})},\ \Eprint {https://arxiv.org/abs/2407.21587} {arXiv:2407.21587 [nucl-th]} \BibitemShut {NoStop}%
\bibitem [{\citenamefont {Juszczak}\ \emph {et~al.}(2006)\citenamefont {Juszczak}, \citenamefont {Nowak},\ and\ \citenamefont {Sobczyk}}]{Juszczak:2005zs}%
  \BibitemOpen
  \bibfield  {author} {\bibinfo {author} {\bibfnamefont {C.}~\bibnamefont {Juszczak}}, \bibinfo {author} {\bibfnamefont {J.~A.}\ \bibnamefont {Nowak}},\ and\ \bibinfo {author} {\bibfnamefont {J.~T.}\ \bibnamefont {Sobczyk}},\ }\bibfield  {title} {\bibinfo {title} {{Simulations from a new neutrino event generator}},\ }\href {https://doi.org/10.1016/j.nuclphysbps.2006.08.069} {\bibfield  {journal} {\bibinfo  {journal} {Nucl. Phys. B Proc. Suppl.}\ }\textbf {\bibinfo {volume} {159}},\ \bibinfo {pages} {211} (\bibinfo {year} {2006})},\ \Eprint {https://arxiv.org/abs/hep-ph/0512365} {arXiv:hep-ph/0512365} \BibitemShut {NoStop}%
\bibitem [{\citenamefont {Golan}\ \emph {et~al.}(2012)\citenamefont {Golan}, \citenamefont {Juszczak},\ and\ \citenamefont {Sobczyk}}]{Golan:2012wx}%
  \BibitemOpen
  \bibfield  {author} {\bibinfo {author} {\bibfnamefont {T.}~\bibnamefont {Golan}}, \bibinfo {author} {\bibfnamefont {C.}~\bibnamefont {Juszczak}},\ and\ \bibinfo {author} {\bibfnamefont {J.~T.}\ \bibnamefont {Sobczyk}},\ }\bibfield  {title} {\bibinfo {title} {{Final State Interactions Effects in Neutrino-Nucleus Interactions}},\ }\href {https://doi.org/10.1103/PhysRevC.86.015505} {\bibfield  {journal} {\bibinfo  {journal} {Phys. Rev. C}\ }\textbf {\bibinfo {volume} {86}},\ \bibinfo {pages} {015505} (\bibinfo {year} {2012})},\ \Eprint {https://arxiv.org/abs/1202.4197} {arXiv:1202.4197 [nucl-th]} \BibitemShut {NoStop}%
\bibitem [{\citenamefont {Benhar}\ \emph {et~al.}(1994)\citenamefont {Benhar}, \citenamefont {Fabrocini}, \citenamefont {Fantoni},\ and\ \citenamefont {Sick}}]{Benhar:1999bg}%
  \BibitemOpen
  \bibfield  {author} {\bibinfo {author} {\bibfnamefont {O.}~\bibnamefont {Benhar}}, \bibinfo {author} {\bibfnamefont {A.}~\bibnamefont {Fabrocini}}, \bibinfo {author} {\bibfnamefont {S.}~\bibnamefont {Fantoni}},\ and\ \bibinfo {author} {\bibfnamefont {I.}~\bibnamefont {Sick}},\ }\bibfield  {title} {\bibinfo {title} {{Spectral function of finite nuclei and scattering of GeV electrons}},\ }\href {https://doi.org/10.1016/0375-9474(94)90920-2} {\bibfield  {journal} {\bibinfo  {journal} {Nucl. Phys. A}\ }\textbf {\bibinfo {volume} {579}},\ \bibinfo {pages} {493} (\bibinfo {year} {1994})}\BibitemShut {NoStop}%
\bibitem [{\citenamefont {Ankowski}\ and\ \citenamefont {Sobczyk}(2006)}]{Ankowski:2005wi}%
  \BibitemOpen
  \bibfield  {author} {\bibinfo {author} {\bibfnamefont {A.~M.}\ \bibnamefont {Ankowski}}\ and\ \bibinfo {author} {\bibfnamefont {J.~T.}\ \bibnamefont {Sobczyk}},\ }\bibfield  {title} {\bibinfo {title} {{Argon spectral function and neutrino interactions}},\ }\href {https://doi.org/10.1103/PhysRevC.74.054316} {\bibfield  {journal} {\bibinfo  {journal} {Phys. Rev. C}\ }\textbf {\bibinfo {volume} {74}},\ \bibinfo {pages} {054316} (\bibinfo {year} {2006})},\ \Eprint {https://arxiv.org/abs/nucl-th/0512004} {arXiv:nucl-th/0512004} \BibitemShut {NoStop}%
\bibitem [{\citenamefont {Juszczak}\ \emph {et~al.}(2005)\citenamefont {Juszczak}, \citenamefont {Nowak},\ and\ \citenamefont {Sobczyk}}]{Juszczak:2005wk}%
  \BibitemOpen
  \bibfield  {author} {\bibinfo {author} {\bibfnamefont {C.}~\bibnamefont {Juszczak}}, \bibinfo {author} {\bibfnamefont {J.~A.}\ \bibnamefont {Nowak}},\ and\ \bibinfo {author} {\bibfnamefont {J.~T.}\ \bibnamefont {Sobczyk}},\ }\bibfield  {title} {\bibinfo {title} {{Spectrum of recoil nucleons in quasi-elastic neutrino nucleus interactions}},\ }\href {https://doi.org/10.1140/epjc/s2004-02086-9} {\bibfield  {journal} {\bibinfo  {journal} {Eur. Phys. J. C}\ }\textbf {\bibinfo {volume} {39}},\ \bibinfo {pages} {195} (\bibinfo {year} {2005})}\BibitemShut {NoStop}%
\bibitem [{\citenamefont {Yariv}\ \emph {et~al.}(2008)\citenamefont {Yariv} \emph {et~al.}}]{Yariv2007}%
  \BibitemOpen
  \bibfield  {author} {\bibinfo {author} {\bibfnamefont {Y.}~\bibnamefont {Yariv}} \emph {et~al.},\ }\bibfield  {title} {\bibinfo {title} {{Intra-nuclear cascade models at low energy?}},\ }in\ \href {https://doi.org/10.1051/ndata:07738} {\emph {\bibinfo {booktitle} {Bersillon, O., et al. (eds.)}}},\ \bibinfo {series and number} {\bibinfo {number} {p.1125}}\ (\bibinfo  {publisher} {EDP Sciences, Paris},\ \bibinfo {year} {2008})\BibitemShut {NoStop}%
\bibitem [{\citenamefont {Metropolis}\ \emph {et~al.}(1958{\natexlab{a}})\citenamefont {Metropolis}, \citenamefont {Bivins}, \citenamefont {Storm}, \citenamefont {Turkevich}, \citenamefont {Miller},\ and\ \citenamefont {Friedlander}}]{Metropolis:1958wvo}%
  \BibitemOpen
  \bibfield  {author} {\bibinfo {author} {\bibfnamefont {N.}~\bibnamefont {Metropolis}}, \bibinfo {author} {\bibfnamefont {R.}~\bibnamefont {Bivins}}, \bibinfo {author} {\bibfnamefont {M.}~\bibnamefont {Storm}}, \bibinfo {author} {\bibfnamefont {A.}~\bibnamefont {Turkevich}}, \bibinfo {author} {\bibfnamefont {J.~M.}\ \bibnamefont {Miller}},\ and\ \bibinfo {author} {\bibfnamefont {G.}~\bibnamefont {Friedlander}},\ }\bibfield  {title} {\bibinfo {title} {{Monte Carlo Calculations on Intranuclear Cascades. I. Low-Energy Studies}},\ }\href {https://doi.org/10.1103/PhysRev.110.185} {\bibfield  {journal} {\bibinfo  {journal} {Phys. Rev.}\ }\textbf {\bibinfo {volume} {110}},\ \bibinfo {pages} {185} (\bibinfo {year} {1958}{\natexlab{a}})}\BibitemShut {NoStop}%
\bibitem [{\citenamefont {Metropolis}\ \emph {et~al.}(1958{\natexlab{b}})\citenamefont {Metropolis}, \citenamefont {Bivins}, \citenamefont {Storm}, \citenamefont {Miller}, \citenamefont {Friedlander},\ and\ \citenamefont {Turkevich}}]{Metropolis:1958sb}%
  \BibitemOpen
  \bibfield  {author} {\bibinfo {author} {\bibfnamefont {N.}~\bibnamefont {Metropolis}}, \bibinfo {author} {\bibfnamefont {R.}~\bibnamefont {Bivins}}, \bibinfo {author} {\bibfnamefont {M.}~\bibnamefont {Storm}}, \bibinfo {author} {\bibfnamefont {J.~M.}\ \bibnamefont {Miller}}, \bibinfo {author} {\bibfnamefont {G.}~\bibnamefont {Friedlander}},\ and\ \bibinfo {author} {\bibfnamefont {A.}~\bibnamefont {Turkevich}},\ }\bibfield  {title} {\bibinfo {title} {{Monte Carlo Calculations on Intranuclear Cascades. 2. High-Energy Studies and Pion Processes}},\ }\href {https://doi.org/10.1103/PhysRev.110.204} {\bibfield  {journal} {\bibinfo  {journal} {Phys. Rev.}\ }\textbf {\bibinfo {volume} {110}},\ \bibinfo {pages} {204} (\bibinfo {year} {1958}{\natexlab{b}})}\BibitemShut {NoStop}%
\bibitem [{\citenamefont {Pandharipande}\ and\ \citenamefont {Pieper}(1992)}]{Pandharipande:1992zz}%
  \BibitemOpen
  \bibfield  {author} {\bibinfo {author} {\bibfnamefont {V.~R.}\ \bibnamefont {Pandharipande}}\ and\ \bibinfo {author} {\bibfnamefont {S.~C.}\ \bibnamefont {Pieper}},\ }\bibfield  {title} {\bibinfo {title} {{Nuclear transparency to intermediate-energy nucleons from (e, e'p) reactions}},\ }\href {https://doi.org/10.1103/PhysRevC.45.791} {\bibfield  {journal} {\bibinfo  {journal} {Phys. Rev.}\ }\textbf {\bibinfo {volume} {C45}},\ \bibinfo {pages} {791} (\bibinfo {year} {1992})}\BibitemShut {NoStop}%
\bibitem [{\citenamefont {Salcedo}\ \emph {et~al.}(1988)\citenamefont {Salcedo}, \citenamefont {Oset}, \citenamefont {Vicente-Vacas},\ and\ \citenamefont {Garcia-Recio}}]{Salcedo:1987md}%
  \BibitemOpen
  \bibfield  {author} {\bibinfo {author} {\bibfnamefont {L.~L.}\ \bibnamefont {Salcedo}}, \bibinfo {author} {\bibfnamefont {E.}~\bibnamefont {Oset}}, \bibinfo {author} {\bibfnamefont {M.~J.}\ \bibnamefont {Vicente-Vacas}},\ and\ \bibinfo {author} {\bibfnamefont {C.}~\bibnamefont {Garcia-Recio}},\ }\bibfield  {title} {\bibinfo {title} {Computer simulation of inclusive pion nuclear reactions},\ }\href@noop {} {\bibfield  {journal} {\bibinfo  {journal} {Nucl.\ Phys.\ A}\ }\textbf {\bibinfo {volume} {484}},\ \bibinfo {pages} {557} (\bibinfo {year} {1988})}\BibitemShut {NoStop}%
\bibitem [{\citenamefont {Dytman}\ \emph {et~al.}(2021)\citenamefont {Dytman}, \citenamefont {Hayato}, \citenamefont {Raboanary}, \citenamefont {Sobczyk}, \citenamefont {Tena~Vidal},\ and\ \citenamefont {Vololoniaina}}]{Dytman:2021ohr}%
  \BibitemOpen
  \bibfield  {author} {\bibinfo {author} {\bibfnamefont {S.}~\bibnamefont {Dytman}}, \bibinfo {author} {\bibfnamefont {Y.}~\bibnamefont {Hayato}}, \bibinfo {author} {\bibfnamefont {R.}~\bibnamefont {Raboanary}}, \bibinfo {author} {\bibfnamefont {J.~T.}\ \bibnamefont {Sobczyk}}, \bibinfo {author} {\bibfnamefont {J.}~\bibnamefont {Tena~Vidal}},\ and\ \bibinfo {author} {\bibfnamefont {N.}~\bibnamefont {Vololoniaina}},\ }\bibfield  {title} {\bibinfo {title} {{Comparison of validation methods of simulations for final state interactions in hadron production experiments}},\ }\href {https://doi.org/10.1103/PhysRevD.104.053006} {\bibfield  {journal} {\bibinfo  {journal} {Phys. Rev. D}\ }\textbf {\bibinfo {volume} {104}},\ \bibinfo {pages} {053006} (\bibinfo {year} {2021})},\ \Eprint {https://arxiv.org/abs/2103.07535} {arXiv:2103.07535 [hep-ph]} \BibitemShut {NoStop}%
\bibitem [{\citenamefont {Bodek}\ \emph {et~al.}(2011)\citenamefont {Bodek}, \citenamefont {Budd},\ and\ \citenamefont {Christy}}]{Bodek:2011ps}%
  \BibitemOpen
  \bibfield  {author} {\bibinfo {author} {\bibfnamefont {A.}~\bibnamefont {Bodek}}, \bibinfo {author} {\bibfnamefont {H.~S.}\ \bibnamefont {Budd}},\ and\ \bibinfo {author} {\bibfnamefont {M.~E.}\ \bibnamefont {Christy}},\ }\bibfield  {title} {\bibinfo {title} {{Neutrino Quasielastic Scattering on Nuclear Targets: Parametrizing Transverse Enhancement (Meson Exchange Currents)}},\ }\href {https://doi.org/10.1140/epjc/s10052-011-1726-y} {\bibfield  {journal} {\bibinfo  {journal} {Eur. Phys. J. C}\ }\textbf {\bibinfo {volume} {71}},\ \bibinfo {pages} {1726} (\bibinfo {year} {2011})},\ \Eprint {https://arxiv.org/abs/1106.0340} {arXiv:1106.0340 [hep-ph]} \BibitemShut {NoStop}%
\bibitem [{\citenamefont {Golan}\ \emph {et~al.}(2013)\citenamefont {Golan}, \citenamefont {Graczyk}, \citenamefont {Juszczak},\ and\ \citenamefont {Sobczyk}}]{Golan:2013jtj}%
  \BibitemOpen
  \bibfield  {author} {\bibinfo {author} {\bibfnamefont {T.}~\bibnamefont {Golan}}, \bibinfo {author} {\bibfnamefont {K.~M.}\ \bibnamefont {Graczyk}}, \bibinfo {author} {\bibfnamefont {C.}~\bibnamefont {Juszczak}},\ and\ \bibinfo {author} {\bibfnamefont {J.~T.}\ \bibnamefont {Sobczyk}},\ }\bibfield  {title} {\bibinfo {title} {{Extraction of Axial Mass and Strangeness Values from the MiniBooNE Neutral Current Elastic Cross Section Measurement}},\ }\href {https://doi.org/10.1103/PhysRevC.88.024612} {\bibfield  {journal} {\bibinfo  {journal} {Phys. Rev. C}\ }\textbf {\bibinfo {volume} {88}},\ \bibinfo {pages} {024612} (\bibinfo {year} {2013})},\ \Eprint {https://arxiv.org/abs/1302.3890} {arXiv:1302.3890 [hep-ph]} \BibitemShut {NoStop}%
\bibitem [{\citenamefont {Abe}\ \emph {et~al.}(2023)\citenamefont {Abe} \emph {et~al.}}]{KamLAND:2022ptk}%
  \BibitemOpen
  \bibfield  {author} {\bibinfo {author} {\bibfnamefont {S.}~\bibnamefont {Abe}} \emph {et~al.} (\bibinfo {collaboration} {KamLAND}),\ }\bibfield  {title} {\bibinfo {title} {{First measurement of the strange axial coupling constant using neutral-current quasielastic interactions of atmospheric neutrinos at KamLAND}},\ }\href {https://doi.org/10.1103/PhysRevD.107.072006} {\bibfield  {journal} {\bibinfo  {journal} {Phys. Rev. D}\ }\textbf {\bibinfo {volume} {107}},\ \bibinfo {pages} {072006} (\bibinfo {year} {2023})},\ \Eprint {https://arxiv.org/abs/2211.13911} {arXiv:2211.13911 [hep-ex]} \BibitemShut {NoStop}%
\bibitem [{\citenamefont {Gran}\ \emph {et~al.}(2013)\citenamefont {Gran}, \citenamefont {Nieves}, \citenamefont {Sanchez},\ and\ \citenamefont {Vicente~Vacas}}]{Gran:2013kda}%
  \BibitemOpen
  \bibfield  {author} {\bibinfo {author} {\bibfnamefont {R.}~\bibnamefont {Gran}}, \bibinfo {author} {\bibfnamefont {J.}~\bibnamefont {Nieves}}, \bibinfo {author} {\bibfnamefont {F.}~\bibnamefont {Sanchez}},\ and\ \bibinfo {author} {\bibfnamefont {M.~J.}\ \bibnamefont {Vicente~Vacas}},\ }\bibfield  {title} {\bibinfo {title} {{Neutrino-nucleus quasi-elastic and 2p2h interactions up to 10 GeV}},\ }\href {https://doi.org/10.1103/PhysRevD.88.113007} {\bibfield  {journal} {\bibinfo  {journal} {Phys. Rev. D}\ }\textbf {\bibinfo {volume} {88}},\ \bibinfo {pages} {113007} (\bibinfo {year} {2013})},\ \Eprint {https://arxiv.org/abs/1307.8105} {arXiv:1307.8105 [hep-ph]} \BibitemShut {NoStop}%
\bibitem [{\citenamefont {Garrow}\ \emph {et~al.}(2001)\citenamefont {Garrow}, \citenamefont {Offermann}, \citenamefont {Arrington} \emph {et~al.}}]{garrow2001nuclear}%
  \BibitemOpen
  \bibfield  {author} {\bibinfo {author} {\bibfnamefont {K.}~\bibnamefont {Garrow}}, \bibinfo {author} {\bibfnamefont {E.~A. J.~M.}\ \bibnamefont {Offermann}}, \bibinfo {author} {\bibfnamefont {J.}~\bibnamefont {Arrington}}, \emph {et~al.},\ }\bibfield  {title} {\bibinfo {title} {Nuclear transparency from quasielastic a(e,e'p) reactions up to q²=8.1 (gev/c)²},\ }\href {https://doi.org/10.1103/PhysRevC.64.014602} {\bibfield  {journal} {\bibinfo  {journal} {Physical Review C}\ }\textbf {\bibinfo {volume} {64}},\ \bibinfo {pages} {014602} (\bibinfo {year} {2001})}\BibitemShut {NoStop}%
\bibitem [{\citenamefont {Niewczas}\ and\ \citenamefont {Sobczyk}(2019)}]{Niewczas:2019fro}%
  \BibitemOpen
  \bibfield  {author} {\bibinfo {author} {\bibfnamefont {K.}~\bibnamefont {Niewczas}}\ and\ \bibinfo {author} {\bibfnamefont {J.~T.}\ \bibnamefont {Sobczyk}},\ }\bibfield  {title} {\bibinfo {title} {{Nuclear Transparency in Monte Carlo Neutrino Event Generators}},\ }\href {https://doi.org/10.1103/PhysRevC.100.015505} {\bibfield  {journal} {\bibinfo  {journal} {Phys. Rev. C}\ }\textbf {\bibinfo {volume} {100}},\ \bibinfo {pages} {015505} (\bibinfo {year} {2019})},\ \Eprint {https://arxiv.org/abs/1902.05618} {arXiv:1902.05618 [hep-ex]} \BibitemShut {NoStop}%
\bibitem [{\citenamefont {Pickering}\ \emph {et~al.}(2017)\citenamefont {Pickering}, \citenamefont {Stowell},\ and\ \citenamefont {Sobczyk}}]{Pickering_2017}%
  \BibitemOpen
  \bibfield  {author} {\bibinfo {author} {\bibfnamefont {L.}~\bibnamefont {Pickering}}, \bibinfo {author} {\bibfnamefont {P.}~\bibnamefont {Stowell}},\ and\ \bibinfo {author} {\bibfnamefont {J.}~\bibnamefont {Sobczyk}},\ }\bibfield  {title} {\bibinfo {title} {Event reweighting with the nuwro neutrino interaction generator},\ }\href {https://doi.org/10.1088/1742-6596/888/1/012175} {\bibfield  {journal} {\bibinfo  {journal} {Journal of Physics: Conference Series}\ }\textbf {\bibinfo {volume} {888}},\ \bibinfo {pages} {012175} (\bibinfo {year} {2017})}\BibitemShut {NoStop}%
\bibitem [{\citenamefont {Cai}\ \emph {et~al.}(2020)\citenamefont {Cai}, \citenamefont {Lu}, \citenamefont {Harewood} \emph {et~al.}}]{PhysRevD.101.092001}%
  \BibitemOpen
  \bibfield  {author} {\bibinfo {author} {\bibfnamefont {T.}~\bibnamefont {Cai}}, \bibinfo {author} {\bibfnamefont {X.-G.}\ \bibnamefont {Lu}}, \bibinfo {author} {\bibfnamefont {L.~A.}\ \bibnamefont {Harewood}}, \emph {et~al.} (\bibinfo {collaboration} {The MINER\ensuremath{\nu}A Collaboration}),\ }\bibfield  {title} {\bibinfo {title} {Nucleon binding energy and transverse momentum imbalance in neutrino-nucleus reactions},\ }\href {https://doi.org/10.1103/PhysRevD.101.092001} {\bibfield  {journal} {\bibinfo  {journal} {Phys. Rev. D}\ }\textbf {\bibinfo {volume} {101}},\ \bibinfo {pages} {092001} (\bibinfo {year} {2020})}\BibitemShut {NoStop}%
\bibitem [{\citenamefont {Lu}\ \emph {et~al.}(2018)\citenamefont {Lu}, \citenamefont {Betancourt}, \citenamefont {Walton} \emph {et~al.}}]{PhysRevLett.121.022504}%
  \BibitemOpen
  \bibfield  {author} {\bibinfo {author} {\bibfnamefont {X.-G.}\ \bibnamefont {Lu}}, \bibinfo {author} {\bibfnamefont {M.}~\bibnamefont {Betancourt}}, \bibinfo {author} {\bibfnamefont {T.}~\bibnamefont {Walton}}, \emph {et~al.} (\bibinfo {collaboration} {MINERvA Collaboration}),\ }\bibfield  {title} {\bibinfo {title} {Measurement of final-state correlations in neutrino muon-proton mesonless production on hydrocarbon at $⟨{E}_{\ensuremath{\nu}}⟩=3\text{ }\text{ }\mathrm{GeV}$},\ }\href {https://doi.org/10.1103/PhysRevLett.121.022504} {\bibfield  {journal} {\bibinfo  {journal} {Phys. Rev. Lett.}\ }\textbf {\bibinfo {volume} {121}},\ \bibinfo {pages} {022504} (\bibinfo {year} {2018})}\BibitemShut {NoStop}%
\bibitem [{\citenamefont {Cai}\ \emph {et~al.}(2023)\citenamefont {Cai}, \citenamefont {Moore}, \citenamefont {Olivier} \emph {et~al.}}]{osti_1923034}%
  \BibitemOpen
  \bibfield  {author} {\bibinfo {author} {\bibfnamefont {T.}~\bibnamefont {Cai}}, \bibinfo {author} {\bibfnamefont {M.~L.}\ \bibnamefont {Moore}}, \bibinfo {author} {\bibfnamefont {A.}~\bibnamefont {Olivier}}, \emph {et~al.},\ }\bibfield  {title} {\bibinfo {title} {Measurement of the axial vector form factor from antineutrino–proton scattering},\ }\bibfield  {journal} {\bibinfo  {journal} {Nature (London)}\ }\textbf {\bibinfo {volume} {614}},\ \href {https://doi.org/10.1038/s41586-022-05478-3} {10.1038/s41586-022-05478-3} (\bibinfo {year} {2023})\BibitemShut {NoStop}%
\bibitem [{\citenamefont {Banerjee}\ \emph {et~al.}(2024)\citenamefont {Banerjee}, \citenamefont {Ankowski}, \citenamefont {Graczyk}, \citenamefont {Kowal}, \citenamefont {Prasad},\ and\ \citenamefont {Sobczyk}}]{Banerjee:2023hub}%
  \BibitemOpen
  \bibfield  {author} {\bibinfo {author} {\bibfnamefont {R.~D.}\ \bibnamefont {Banerjee}}, \bibinfo {author} {\bibfnamefont {A.~M.}\ \bibnamefont {Ankowski}}, \bibinfo {author} {\bibfnamefont {K.~M.}\ \bibnamefont {Graczyk}}, \bibinfo {author} {\bibfnamefont {B.~E.}\ \bibnamefont {Kowal}}, \bibinfo {author} {\bibfnamefont {H.}~\bibnamefont {Prasad}},\ and\ \bibinfo {author} {\bibfnamefont {J.~T.}\ \bibnamefont {Sobczyk}},\ }\bibfield  {title} {\bibinfo {title} {{JLab spectral functions of argon in nuwro and their implications for MicroBooNE}},\ }\href {https://doi.org/10.1103/PhysRevD.109.073004} {\bibfield  {journal} {\bibinfo  {journal} {Phys. Rev. D}\ }\textbf {\bibinfo {volume} {109}},\ \bibinfo {pages} {073004} (\bibinfo {year} {2024})},\ \Eprint {https://arxiv.org/abs/2312.13369} {arXiv:2312.13369 [hep-ph]} \BibitemShut {NoStop}%
\bibitem [{\citenamefont {Lu}\ \emph {et~al.}(2016)\citenamefont {Lu}, \citenamefont {Pickering}, \citenamefont {Dolan}, \citenamefont {Barr}, \citenamefont {Coplowe}, \citenamefont {Uchida}, \citenamefont {Wark}, \citenamefont {Wascko}, \citenamefont {Weber},\ and\ \citenamefont {Yuan}}]{PhysRevC.94.015503}%
  \BibitemOpen
  \bibfield  {author} {\bibinfo {author} {\bibfnamefont {X.-G.}\ \bibnamefont {Lu}}, \bibinfo {author} {\bibfnamefont {L.}~\bibnamefont {Pickering}}, \bibinfo {author} {\bibfnamefont {S.}~\bibnamefont {Dolan}}, \bibinfo {author} {\bibfnamefont {G.}~\bibnamefont {Barr}}, \bibinfo {author} {\bibfnamefont {D.}~\bibnamefont {Coplowe}}, \bibinfo {author} {\bibfnamefont {Y.}~\bibnamefont {Uchida}}, \bibinfo {author} {\bibfnamefont {D.}~\bibnamefont {Wark}}, \bibinfo {author} {\bibfnamefont {M.~O.}\ \bibnamefont {Wascko}}, \bibinfo {author} {\bibfnamefont {A.}~\bibnamefont {Weber}},\ and\ \bibinfo {author} {\bibfnamefont {T.}~\bibnamefont {Yuan}},\ }\bibfield  {title} {\bibinfo {title} {Measurement of nuclear effects in neutrino interactions with minimal dependence on neutrino energy},\ }\href {https://doi.org/10.1103/PhysRevC.94.015503} {\bibfield  {journal} {\bibinfo  {journal} {Phys. Rev. C}\ }\textbf {\bibinfo {volume} {94}},\ \bibinfo {pages} {015503} (\bibinfo {year} {2016})}\BibitemShut {NoStop}%
\bibitem [{\citenamefont {Furmanski}\ and\ \citenamefont {Sobczyk}(2017)}]{Furmanski:2016wqo}%
  \BibitemOpen
  \bibfield  {author} {\bibinfo {author} {\bibfnamefont {A.~P.}\ \bibnamefont {Furmanski}}\ and\ \bibinfo {author} {\bibfnamefont {J.~T.}\ \bibnamefont {Sobczyk}},\ }\bibfield  {title} {\bibinfo {title} {{Neutrino energy reconstruction from one muon and one proton events}},\ }\href {https://doi.org/10.1103/PhysRevC.95.065501} {\bibfield  {journal} {\bibinfo  {journal} {Phys. Rev. C}\ }\textbf {\bibinfo {volume} {95}},\ \bibinfo {pages} {065501} (\bibinfo {year} {2017})},\ \Eprint {https://arxiv.org/abs/1609.03530} {arXiv:1609.03530 [hep-ex]} \BibitemShut {NoStop}%
\end{thebibliography}%

\end{document}